\documentclass{article}
\usepackage{amsmath, amssymb, amsthm, amsfonts, txfonts, pxfonts}
\usepackage{fullpage}
\usepackage{graphics}
\usepackage{amssymb}  
\usepackage{amsmath}
\usepackage{amscd}
\usepackage{euscript}
\usepackage{enumerate}
\usepackage[all]{xy}
\usepackage{paralist}
\usepackage{hyperref}

\newtheorem{Theorem}{Theorem}
\newtheorem{Corollary}[Theorem]{Corollary}
\newtheorem{Example}[Theorem]{Example}
\newtheorem{Definition}[Theorem]{Definition}
\newtheorem{Remark}[Theorem]{Remark}

\newtheorem{Lemma}[Theorem]{Lemma}
\newtheorem{Proposition}[Theorem]{Proposition}

\usepackage{color}

\newcommand{\dd}{\partial}

\def \coker {\rm coker}

\def \s {\sigma}
\def \btn {\,\overline{\otimes}\,}

\def \S {\EuScript{S}}
\def \GL {\mathsf{gl}}

\def \U {\EuScript{U}}
\def \s {\sigma}
\def \a {\mathfrak{a}}

\def \C {\mathbb{C}}

\def \e {\mathfrak{e}}
\def \w {\omega}
\def \id {\mathrm{id}}
\def \g {\mathfrak {g}}

\def \R {\mathbb{R}}

\def \f {\phi}
\def \dd {\partial}
\def \M {\EuScript{M}}

\def \P {\EuScript{P}}
\def \b {\beta}
\def \g {\mathfrak{g}}

\def \tn {\otimes}
\def \t {\triangleright}

\def \ra {\xrightarrow}
\def \rra {\rightarrow}
\def \lra {\longrightarrow} 
\def \le {\mathfrak{h}}

\def \diff {\mathrm{diff}}

\def \Z {\mathbb{Z}}

\def \gl {\mathfrak{gl}}

\def \H {\EuScript{F}}

\def \Hom {\mathrm{Hom}}

\def\be{\begin{equation}}
\def\ee{\end{equation}}
\def\bea{\begin{eqnarray}}
\def\eea{\end{eqnarray}}

\def \Aut {\mathrm{Aut}}

\def \act {\triangleright}

\def \g {\gamma}
\def \lg {\mathfrak{g} }
\def \lh {\mathfrak{h}}
\def \lG {\mathfrak{G}}
\def \cG {\mathcal{G}}
\def \V {\EuScript{V}}
\def \U {\EuScript{U}}
\def \cW {\EuScript{W}}
\def \gl {\mathfrak{gl}}
\def \cHom {\mathcal{HOM}}
\def \gg {\mathfrak{g}}
\def \H {\mathcal{H}}
\def \ot {\otimes}
\def \ii {\mathrm{i}}
\def \H {\mathrm{H}}

\def \ep {e_1}
\def \eo {e_0}
\def \em {e_{-1}}
\def \sl {\mathfrak{sl}_2}
\def \Fo {\mathbb{F}_0}
\def \Fu {\mathbb{F}_1}
\def \fsl {\Fu\rtimes_{\alpha}\sl}
\newcommand{\zp}[1]{(0,#1)}
\newcommand{\al}[2]{\alpha(#1,#2)}
\def \d {\mathrm{d}}
\def \rb {\bar{r}}
\def \eeo {(0,\eo)}
\def \eep {(0,\ep)}
\def \eem {(0,\em)}
\def \lU {\mathfrak{U}}
\def \B {\mathcal{B}}
\def \la {\mathfrak{a}}
\def \lk {\mathfrak{k}}

\def \F {\mathbb{F}}
\def \R {\mathcal{R}}
\def \st {\mathfrak{String}}
\def \ufo {1_{\Fo}}

\begin{document}

\title{{Categorifying the $\mathfrak{sl}(2,\C)$ Knizhnik-Zamolodchikov Connection via an Infinitesimal 2-Yang-Baxter Operator in the String Lie-2-Algebra} }

\author{ Lucio Simone Cirio
\footnote{Present address: Mathematisches Institut, Universit\"at M\"unster, Einsteinstrasse 62, 48149 M\"unster, Deutschland} \\ 
{\footnotesize Mathematics Research Unit, University of Luxembourg}\\
{\footnotesize 6, rue Richard Coudenhove-Kalergi, L-1359 Luxembourg, Luxembourg}\\
{\it \small lucio.cirio@uni-muenster.de} \\[.2cm]
Jo\~{a}o  Faria Martins  \\ 
{\footnotesize Departamento de Matem\'{a}tica  and Centro de Matem\'{a}tica e Aplica\c{c}\~{o}es,} \\ 
{\footnotesize Faculdade de Ci\^{e}ncias e Tecnologia,   Universidade Nova de Lisboa} \\
{\footnotesize  Quinta da Torre, 2829-516 Caparica, Portugal } \\
{\it \small jn.martins@fct.unl.pt } }
\date{}
\maketitle

\begin{abstract}
\noindent
{We construct a flat (and fake-flat)}  2-connection in the configuration space of $n$ indistinguishable  particles in the complex plane, which categorifies the  $\mathfrak{sl}(2,\C)$-Knizhnik-Zamolodchikov connection obtained from the adjoint  representation of $\mathfrak{sl}(2,\C)$. This will be done by considering the adjoint categorical representation of the string Lie 2-algebra and the notion of an infinitesimal {2-Yang-Baxter operator in a differential crossed module}. {Specifically, we find an infinitesimal 2-Yang-Baxter operator in the string Lie 2-algebra, proving that any (strict) categorical representation of the string Lie-2-algebra, in a chain-complex of vector spaces, yields a flat and (fake flat) 2-connection in the configuration space, categorifying the $\mathfrak{sl}(2,\C)$-Knizhnik-Zamolodchikov connection.}
{We will give very detailed explanation of all concepts involved, in particular discussing the relevant theory of 2-connections and their two dimensional holonomy, in the specific case of 2-groups derived from chain complexes of vector spaces.}

\end{abstract}

\noindent {\it Keywords:} {Higher Gauge Theory, two-dimensional holonomy, Categorification, crossed module, braid group, braided surface, configuration spaces, Knizhnik-Zamolodchikov equations, Zamolodchikov tetrahedron equation, infinitesimal braid group relations, infinitesimal relations for braid cobordisms, {categorical representation}.}
\\[.25cm]
{\it MSC2010:} {16T25, 
20F36 
(principal);  18D05, 
17B37, 
53C29, 
57M25, 
57Q45 
(secondary).}

\section{Introduction}

{Let $I=[-1,1]$. Given a positive integer $n$, recall that a braid \cite{Bi,BiBr,Ka,Art}, with $n$-strands is, by definition, a neat embedding $B$ of the manifold  ${\sqcup}_{i=1}^n I$ into $\C \times I$ (where $\C$  is the complex plane) such that the projection on the last variable (which unusually we take to be the horizontal one) is monotone. Moreover, we suppose that $B \cap (\C \times \{\pm 1\})=\{1,\dots ,n\}\times \{0\} \times \{\pm 1\}.$ Two braids $B$ and $B'$  are said to be isotopic (or equivalent) if there exists a level preserving (with respect to the last variable)  isotopy of $\C \times I$ , relative to the boundary of $\C \times I$,  sending $B$ to $B'$.} Isotopy classes of braids with $n$-strands form a group, the Artin braid group $\B_n$, where two braids $B$ and $B'$ are multiplied, defining $BB'$, by putting $B'$ on the right hand side of $B$. 

Isotopy classes of braids are, {canonically}, in one-to-one correspondence with elements of the fundamental group of the manifold $\C(n)/S_n$,  the configuration space of $n$ indistinguishable particles in the complex plane. In fact $\B_n\cong \pi_1(\C(n)/S_n)$, as groups. Here: 
$$\C(n)=\{(z_1,\dots,z_n) \in \C^n: i \neq j \implies z_i \neq z_j\} $$
is the configuration space of $n$ distinguishable particles in the complex plane, with the obvious (properly discontinuous) action of the symmetric group $S_n$ by diffeomorphisms.

Consider a vector bundle, with typical fibre $V$, over the configuration space $\C(n)/S_n$, with a connection. If the connection is flat, then the holonomy of it (which in this  case is  invariant under homotopy) will yield a homomorphism $\B_n \cong \pi_1(\C(n)/S_n)\ra{\H} {\rm GL}(V)$, where ${\rm GL}(V)$ is  the Lie group of invertible linear maps $V \to V$. An example of this is given by the well known 
 Knizhnik-Zamolodchikov connection \cite{KZ,BN,Ko,Ka}. Let us explain its construction. Choose a Lie algebra $\lg$ and a symmetric tensor $r=\sum_{i} x_i \tn y_i \in \lg \tn \lg$ such that the following relation, called the 4-term relation, is satisfied:
\begin{equation}\label{4termR}
 [r_{12}+r_{13},r_{23}]=0.
\end{equation}
Explicitly \eqref{4termR} means that in $\lg \tn \lg \tn \lg$ we have:
$$\sum_{i,j} x_i \tn [y_i,x_j]\tn y_j+x_i \tn x_j \tn [y_i,y_j]=0.$$
{Such element  $r \in \lg \tn \lg$ will be called an infinitesimal {Yang-Baxter operator} (or infinitesimal $\R$-matrix) in $\lg$.}

Let $\rho\colon \lg \to \gl(V)$ be a Lie algebra representation of $\lg$ in the vector space $V$. Here $\gl(V)$ is the Lie-algebra of linear maps $V \to V$, with the usual commutator $[f,g]=fg-gf$. Denote the tensor product $V\tn \dots \tn V$ of $V$ with itself $n$ times as $V^{\tn n}$. Consider the trivial vector bundle $\C(n) \times V^{\tn n}$. Given $1 \leq a <b \leq n$, define a linear map $\overline{\phi}_{ab}^\rho(r) \colon V^{\tn n} \to V^{\tn n}$ (called insertion map) as: 
\begin{equation}\label{imap}
\bar{\phi}_{ab}^\rho(r)(v_1 \tn  \dots \tn v_a \tn \dots \tn v_b\tn \dots\tn v_n)=\sum_{i} v_1 \tn  \dots \tn x_i\t v_a \tn \dots \tn y_i \t v_b\tn \dots\tn v_n,
\end{equation}
where  $\rho(X){(v)}=X \t v$, for $v \in V$ and $X \in \lg$.
We therefore inserted $(\rho \tn \rho)(r)$ in the positions $a$ and $b$ of $V^{\tn n}$.

The Knizhnik-Zamolodchikov connection   is given by the following $\gl(V^{\tn n})$-valued 1-form in the configuration space $\C(n)$:
\begin{equation}\label{kzi}
A=\frac{h}{2 \pi i}\sum_{1\leq a <b \leq n} \w_{ab} \bar{\phi}^\rho_{ab}(r),
\end{equation}
where   $$\w_{ab}=\frac{d z_a -d z_b}{z_a-z_b}.$$
This connection  appeared originally in the context of conformal field theory \cite{KZ}, and while calculating  Chern-Simons path integrals \cite{W}; see also the reference \cite{Kh2}. It follows easily that the Knizhnik-Zamolodchikov  connection is flat, by using the 4-term relation  \eqref{4termR}; see for example \cite{BN,Ka,Kh1}.

We have obvious left actions of $S_n$ on $\C(n)$ and on $V^{\tn n}$. Clearly, the product action of $S_n$ on $\C(n) \times V^{\tn n}$ is an action by vector bundle maps, so we can consider the  quotient vector bundle $(\C(n) \times V^{\tn n})/S_n$, over the configuration space  $\C(n)/S_n$; see \cite{Kh1}. Since the Knizhnik-Zamolodchikov connection is invariant under this action, we have a quotient Knizhnik-Zamolodchikov connection, also called  $A$, in the vector bundle $(\C(n) \times V^{\tn n})/S_n$. 

There is a standard way to find tensors $r \in \lg\tn \lg$ satisfying the 4-term relation: consider a $\lg$-invariant non-degenerate symmetric bilinear form $\langle, \rangle$ in $\lg$. Let $\{x_i\}$ be a basis of $\lg$. Let $\{y^i\}$ be the dual basis of $\lg^*$, identified with $\lg$ by using the bilinear form $\langle, \rangle$. {Then $r=\sum_i x_i \tn {y_i}$ is symmetric and  $\lg$-invariant, namely:}
\begin{equation}\label{infBraiding}
[\Delta(X),r]=0, \textrm{ for each } X \in \lg, \textrm{ in  } {\cal U}(\lg)\otimes {\cal U}(\lg)\, ; \
\end{equation}
{where ${\cal U}(\lg)$ is the universal enveloping algebra of $\lg$, with the usual co-product $\Delta\colon {\cal U}(\lg)\to  {\cal U}(\lg)\otimes {\cal U}(\lg).$ From \eqref{infBraiding} we can easily see \cite{BN,CV,Ka} that $r$ satisfies 
 the 4-term relation \eqref{4termR}, thus $r$ is an infinitesimal  Yang-Baxter operator in $\lg$. Relation \eqref{infBraiding}, {in other words the $\lg$-invariance of a tensor $r \in \lg \tn \lg$,} is  not implied by the 4-term relation \eqref{4termR}. A symmetric $\lg$-invariant tensor $r \in \lg \tn \lg$ will be called an infinitesimal braiding \cite{Ka} in $\lg$.}

Given a semisimple Lie algebra $\lg$, let $r$ be the infinitesimal {Yang-Baxter operator} coming from the Cartan-Killing form in $\lg$, a $\lg$-invariant, symmetric, non-degenerate, bilinear form. In the particular case of $\mathfrak{sl}(2,\C)=\sl$, with the standard generators $\{\em,\eo,\ep\}$, satisfying the commutation relations:
$$ [\eo , \em ] = - \, \em \, , \qquad\qquad [\em , \ep] = 2\eo \, , \qquad\qquad [\eo , \ep] = \ep \, ,$$
 the associated {infinitesimal Yang-Baxter operator} has the form:
\begin{equation}\label{rm}
 r = 2 \, \em\tn\ep + 2 \, \ep\tn\em - 4 \, \eo\tn\eo \, . 
\end{equation}
Given a finite dimensional representation $\rho$ of $\lg$, the representation of the braid group constructed  via  the holonomy of the  Knizhnik-Zamolodchikov connection defined from $r$ and $\rho$, see \eqref{kzi}, is equivalent to the representation of the braid group coming from the $\R$-matrix of the  quantum group $U_h(\lg)$, considering the  representation $\rho_h$ of $U_h(\lg)$ which quantizes $\rho$. This  beautiful result is due to Kohno \cite{Kh1}; see also \cite{Ka,D}.

The holonomy of the Knizhnik-Zamolodchikov connection does not immediately yield invariants of links in $S^3$,  given that the forms $\w_{ab}$ explode at maximal and minimal points. However, there exist well defined regularisation methods for the holonomy at the extreme points, albeit forcing us to introduce knot framings. For this see \cite{AF,LM}, in the identical case of the Kontsevich integral, also defined from the holonomy of a universal Knizhnik-Zamolodchikov connection. This construction gives knot invariants which coincide with the usual quantum group knot invariants; see \cite{Ka,Dr}.

This paper, which is a sequel of \cite{CFM11} and solves one of its main open problems,  has the objective of addressing how to extend this holonomy  approach for defining invariants of braids to the case of  braided surfaces. 
A (simple) braided surface  $B \ra{\S} B'$ \cite{CKS} (which is called a braid cobordism in \cite{KT}), connecting the braids $B$ and $B'$,  both being embedded 1-manifolds in {$\C  \times I$, and not considered to be up to isotopy, is a  2-manifold with corners $\S$,  neatly embedded in $\C \times I \times I$, defining an embedded cobordism between $B$ and $B'$. One also requires that the projection of $\S$ onto ${\{0\}} \times {I}^2$ is a simple branched cover, with a finite number of branch points, and moreover  that  the restriction  of $\S$ to $\C \times \{\pm 1\} \times I$ does not depend on the last variable. Two braided surfaces $B \ra{\S_1} B'$  and  $B \ra{\S_2} B'$ are said to be equivalent if they are isotopic, relative to the boundary of $\C \times I \times I$.} There exists a (non-strict) monoidal category whose objects are the braids with $n$-strands (not considered to be up to isotopy), the morphisms $B \to B'$ are the braided surfaces $\S$ connecting the braids $B$ and $B'$, up to isotopy, with the obvious vertical composition, and the tensor product $(B_1 \ra{\S_1} B_1') \tn  (B_2 \ra{\S_2} B_2')$  is given by the obvious horizontal juxtaposition $B_1 B_2 \ra{\S_1 \S_2} B_1'B_2'$.

A {braided surface} $B \ra{\S} B'$, without branch points, defines a map $\S'\colon I^2 \to \C(n)/S_n$,  \cite[1.6]{CKS}, restricting to $B\colon I\to \C(n)/S_n$ and to $B'\colon I \to \C(n)/S_n$ on the top and bottom of $I^2$, with $\S'$ being constant on the left and right sides of $I^2$. If $\S$ has branch points,  then $\S'$ is defined on $I^2$ minus the set of branch points of the projection of $\S$ onto ${\{0\}} \times I^2$. In each branch point $\S'$ has a very particular type of singularities; see \cite{CFM11}.

Let $M$ be a smooth manifold. The theory of Lie 2-group 2-connections over $M$ was initiated in \cite{BrMe,ACJ,BS1,BS2}; see also the excellent review \cite{BaH}. Their (in general non-abelian) two-dimensional (surface) holonomy was addressed in \cite{BS1,BS2,SW1,SW2}; see also \cite{FMP1,FMP2,FMP3}. We will always see a {strict} Lie 2-group \cite{BL,BHS,BM} as being represented by its associated Lie crossed module $(\beta \colon H \to G, \t)$, where $G$ and $H$ are Lie groups and $\t$ is a smooth left action of $G$ on $H$ by automorphisms, satisfying the well known Peiffer relations, \cite{BHS}. Analogously, we will always consider, instead of a {strict} Lie 2-algebra \cite{BC,B1},  its associated differential crossed module $\lG=(\beta \colon \lh \to \lg,\t)$, where $\lg$ and $\lh$ are Lie algebras and $\t$ is a left action of $\lg$ on $\lh$ by derivations. These are to satisfy the following natural compatibility conditions, called Peiffer relations:
\begin{equation}
\label{peiffer}
\partial(u) \t v=[u,v] \quad \textrm{ and } \quad \partial(X \t v)=[X,\partial{(v)}] \, , \qquad \forall \, u,v \in \lh \, , \forall \, X \in \lg.
\end{equation}

Let us briefly explain the (local) differential geometry of 2-connections and their two-dimensional holonomy, in the particular case when the underlying Lie 2-group is obtained from a chain complex of vector spaces $\V$; \cite{BC,BL,KP,FMP3,FMM,CFM11}. We will provide full definitions and  detailed calculations in this paper. {We will not make use of the general theory of 2-vector bundles, since everything we use can be presented locally, and the natural global setting for the examples we present fits nicely inside the theory of vector bundles whose typical fibre is a chain-complex of vector spaces (differential graded vector bundles \cite{AC}).}

Let $\V$ be a chain complex of vector spaces. Let $\Aut(\V)$ be the 2-category, with a single object, whose 1-morphisms are the (degree 0) chain-maps $f\colon \V \to \V$, with composition (in the reverse order) as horizontal composition.  Given chain maps $f,f'\colon \V \to \V$, the 2-morphisms $f \stackrel{s}{\implies} f'$ in $\Aut(\V)$ are given by (equivalence classes of) the  usual chain complex homotopies (degree-1 maps $s\colon \V \to \V$), connecting $f$ and $f'$, with sum as vertical composition. The whiskerings of homotopies  with chain maps are given by the obvious compositions of (degree 0 and degree 1) maps $\V \to \V$, rendering    two  distinct possible horizontal compositions of homotopies. In order that these coincide,  so that the interchange law for 2-categories holds, we must consider homotopies  $s\colon \V \to \V$ up to 2-fold homotopy.

From the chain complex of vector spaces $\V$  we also  derive a differential crossed module: $$\GL(\V)=(\beta\colon \gl^1(\V) \to \gl^0(\V), \t),$$ where $\gl^0(\V)$ is the Lie algebra of (degree 0) chain-maps, with the usual commutator defining the Lie algebra structure, and the underlying vector space of $\gl^1(\V)$ is the vector space of degree 1 maps $\V \to \V$ (homotopies) up to 2-fold homotopy, with a certain natural bracket. Of course $\beta(s)=s \partial+\partial s$, for $s \in \gl^1(\V)$, where $\dd$ is the boundary map in $\V$.

 A {(fake-flat)} local 2-connection  $(A,B)$ in the manifold $M$ is given by a $\gl^0(\V)$-valued 1-form $A$ in  $M$ and a $\gl^1(\V)$-valued 2-form $B$ in $M$, such that $\beta(B)=F_A=dA + \frac{1}{2} A \wedge A$, the curvature of $A$. A local 2-connection can be integrated to define a 2-dimensional holonomy $\H=(H^1,H^2)$. Explicitly, if $\gamma\colon [0,1] \to M$ is a {piecewise smooth} path we have a chain map $H^1(\gamma)\colon \V \to \V$; if $\Gamma\colon [0,1]^2 \to M$ is a {piecewise smooth} homotopy $\g \to \g'$ (relative to the end-points) connecting the paths  $\gamma$ and $\gamma'$,  called a 2-path, then we have a chain complex homotopy (up to 2-fold homotopy) $H^2(\Gamma)$ connecting the chain maps $H^1(\gamma)$ and $H^1(\gamma')$. These holonomy maps, particular cases of constructions in \cite{AC,BS2,SW2,FMP3}, preserve horizontal and vertical compositions of 2-paths, and of course the composition of paths. Moreover, if the pair $(A,B)$ is 2-flat, meaning that the 2-curvature 3-tensor ${\M}=d B+ A\wedge^\t B$ vanishes, the two dimensional holonomy $H^2(\Gamma)$ depends only on the homotopy class, relative to the boundary, of $\Gamma\colon [0,1]^2 \to M$.

Therefore, if we construct a 2-flat 2-connection $(A,B)$  in the configuration space $\C(n)/S_n$, the holonomy of it will define an invariant of braid-cobordisms $\S$, without branch points, by considering the two dimensional holonomy of the associated maps $\S'\colon  [0,1]^2 \to \C(n)/S_n$. {In particular, this renders a solution of the Zamolodchikov tetrahedron equation \cite{BC}, albeit in a weak (i.e. bicategorical) sense.}  An  important open problem is whether (for the example given in this article) this two dimensional holonomy is convergent, or for the least can be regularised, in the case when the braid cobordisms have branch points, in which case we need to compute the holonomy of a map {$\S'\colon [0,1]^2 \setminus \beta(\S)  \to \C(n)/S_n$, where $\beta(\S)$} is the (finite) set of branch points of $\S$, in which $\S'$ has a very particular kind of singularities. This would permit us to have a full fledged invariant of braid cobordisms. 

 Consider a  chain complex $\V$ and a positive integer  $n$.  Consider also a family of chain maps $t_{ab}\colon \V^{\btn n} \to \V^{\btn n}$, where $a,b$ are distinct elements of $\{1,\dots,n\}$, as well as chain homotopies (up to 2-fold homotopy) $K_{abc}$ of $\V^{\btn n}$, where $a,b,c$ are distinct elements of $\{1,\dots,n\}$. Here $\btn$ denotes the usual tensor product of chain complexes. In \cite{CFM11}, we  found necessary and sufficient conditions for a local 2-connection $(A,B)$ in $\C(n)$, of the form: 
\begin{equation}
 A =\sum_{1 \leq a <b \leq n} \w_{ab} t_{ab} \, , \qquad
  B = \sum_{1\leq a<b<c\leq n}  \w_{ba}\wedge\w_{ac}\,K_{bac}  + \w_{ab}\wedge\w_{bc}\, K_{abc} 
\end{equation}
to be flat and also so that its 2-dimensional holonomy descend to the quotient $\C(n)/S_n$ (which yields the obvious  $S_n$-covariance condition); see Theorem \ref{Main} below. An efficient way to construct data like this is to consider {infinitesimal 2-{Yang-Baxter operator}s (which we will also call infinitesimal 2-$\R$-matrices) in a differential crossed module $\lG=(\beta\colon \lh \to \lg,\t)$. These are pairs $(r,P)$, where $r \in \lg \tn \lg$ and $P$ lives in ${\overline{\lU}}^{(3)}$, satisfying further properties. Here $\overline{\lU}^{(3)}$ is the degree 1 component of the chain complex tensor product $(\beta\colon \lh \to \lg) \btn (\beta\colon \lh \to \lg)\btn (\beta\colon \lh \to \lg)$, modulo the images of the boundary map from the degree-2 component of it.} Therefore  ${\overline{\lU}}^{(3)}$ naturally maps to $\lg\tn \lg \tn \lg$, through a map $\b'$. Suppose we are given a categorical representation $\rho$ of $\lG$ in the chain complex $\V$, in other words a differential crossed module map $\rho\colon \lG \to \GL(\V)$, \cite{CFM11}. Put, see \eqref{imap}:
\begin{equation}
A = \sum_{1 \leq a<b \leq n} \w_{ab} \, \bar{\phi}_{ab}^\rho(r) \, , \qquad
B = \sum_{1 \leq a<b<c \leq n} \w_{ba}\wedge\w_{ac} \, \bar{\phi}_{bac}^\rho(P) + \w_{ab}\wedge \w_{bc} \, \bar{\phi}_{abc}^\rho(P) \, .
\end{equation}
In order that $(A,B)$ be a flat local  2-connection,  and that it is, moreover, $S_n$,   covariant, $(r,P)$ should satisfy (prior or after applying the categorical representation $\rho$) the following set of equations, categorifying the 4-term relation \eqref{4termR}:
\begin{equation}\label{irm}
\begin{split}
& r_{12}=r_{21} \\
&{\beta'}(P)=[r_{12}+r_{13},r_{23}]\\
& r_{14}\act (P_{213} + P_{234}) + (r_{12} + r_{23} + r_{24}) \act P_{314} - (r_{13}+r_{34})\act P_{214} = 0 \\
& r_{23}\act (P_{214} + P_{314}) -r_{14}\act (P_{423} +P_{123}) = 0 \\
& P_{123}+P_{231}+P_{312}=0 \\ 
& P_{123}=P_{132}
\end{split}
\end{equation}
In such a case, the pair $(A,B)$ will be called a  Knizhnik-Zamolodchikov 2-connection. {The above conditions, which we discuss in  detail here, define what we call an  infinitesimal 2-{Yang-Baxter operator}, or infinitesimal 2-$\R$-matrix, $(r,P)$ in $\lG$, free or eventually with respect to a categorical representation $\rho$ of $\lG$, see definitions \ref{irmatrix} and \ref{irmatrix2}.} {Relations \eqref{irm} where used in \cite{CFM11} to define the differential crossed module of horizontal 2-chord diagrams. In this paper we will construct a non-trivial representation of {these relations} by defining an infinitesimal 2-Yang-Baxter operator in the string Lie-2-algebra.} {For the time being, we note that given any Lie algebra $\lg$ and any symmetric tensor $r \in \lg \tn \lg$, if we put $P=[r_{12}+r_{13},r_{23}]$, then relations \eqref{irm} are satisfied by $(r,P)$, in $\lg^{\tn 4}$, where $\t$ is in this case the adjoint action of $\lg$ on $\lg$. This essential fact was proven in \cite{CFM11}. }

Let $\lG=(\partial\colon \lh \to \lg, \t)$ be a differential crossed module. Thus $\lg$ acts on $\lh$ by  derivations, and it is easy to see, by the crossed modules rules \eqref{peiffer}, that $\t$ descends to  a Lie algebra representation of $\pi_1(\lG) \doteq {\rm coker}(\partial)$ on $\pi_2(\lG) \doteq {\rm ker}(\lG)$. Similarly to the group crossed module case, treated in \cite{Br,ML}, we can in addition derive from $\lG$ a Lie algebra cohomology class, the $k$-invariant \cite{EML}, $k^3(\lG)\in H^3(\pi_1(\lG),\pi_2(\lG))$. Reciprocally \cite{ger66,W}, any Lie algebra cohomology class $\w \in \H^3(\lk,\la)$ can be induced from a differential crossed module $\lG$, with $\pi_1(\lG)\cong \lk$ and $\pi_2(\lG)=\la$, and we say that $\lG$ geometrically realises $\w$. Two crossed modules have the same $k$-invariant if, and only if, they are weak equivalent, in the model category of differential crossed modules, or, what is the same, if they are equivalent in the larger category of (not-necessarily strict) Lie 2-algebras \cite{BC}.

Let $\lg$ be a simple Lie algebra, acting in $\C$ trivially. The vector space $H^3(\lg,\C)$ is one dimensional and  generated by the cocyle $\w \colon \lg\wedge \lg \wedge \lg \to \C$, such that $\w(X,Y,Z)=\langle [X,Y], Z\rangle$, where $\langle, \rangle$ denotes the Cartan-Killing form. The string Lie 2-algebra is, by definition, a differential crossed module geometrically realising this cohomology class for the case of $\sl$; it is therefore defined only up to weak equivalence. 

Explicit realisations of the  string Lie 2-algebra appear in \cite{BSCS} and  \cite{wag06}. The realisation appearing in \cite{wag06} is completely algebraic, therefore fitting nicely in the framework of this article. Let us sketch this construction. The differential crossed module $\st$ associated to the string Lie 2-algebra (the string differential crossed module) has the form $ \st=(\dd\colon \F_0 \to \F_1 \rtimes_\alpha \sl,\t)$, where $\F_0$ is the $\sl$-module of formal power series in $\C$, and $\F_1$ is the $\sl$-module of formal differential forms in $\C$. Here $\sl$ acts on $\F_0$ and $\F_1$ through Lie derivatives, via a natural inclusion of $\sl$ into the Lie algebra of formal vector fields in $\C$. Both $\F_0$ and $\F_1$ are to be seen as being abelian Lie algebras. Finally  $\F_1 \rtimes_\alpha \sl $ denotes a semidirect product, twisted by a certain Lie algebra 2-cocycle $\alpha \colon \sl \wedge \sl \to \F_1$. The string differential crossed module can be embedded into the exact sequence:
\begin{equation}\label{es}
 \{0\} \to \C \ra{i} \F_0 \ra{\partial} \F_1 \rtimes_\alpha \sl \ra{\pi} \sl\to \{0\}  .
\end{equation} 
Here $\partial=(d,0)$, where $d$ denotes the formal de Rham differential.

Despite the fact that $\F_0$ is abelian, the differential crossed module associated to the underlying complex $\underline{\st}=(\F_0 \ra{\dd} \F_1 \rtimes_\alpha \sl)$ of $\st$ ( the relevant one at the level of 2-dimensional holonomy), namely: 
\begin{equation}\label{sl2a}
\GL(\F_0 \ra{\partial} \F_1 \rtimes_\alpha \sl)=\big(\beta\colon \gl^1(\F_0 \ra{\partial} \F_1 \rtimes_\alpha \sl) \to \gl^0(\F_0 \ra{\partial} \F_1 \rtimes_\alpha \sl), \t\big)
\end{equation}
is a differential crossed module of non-abelian Lie algebras. For instance, given two homotopies $s,t \colon \fsl \to \F_0$, both of which are elements of $\gl^1(\F_0 \ra{\partial} \F_1 \rtimes_\alpha \sl) $, their commutator is:
$$\{s,t\}=s \circ \dd \circ t-t\circ \dd \circ s, $$
which is clearly non-zero in general. 

Let $\lG=(\partial\colon \lh \to \lg,\t)$ be any differential crossed module. There exists a natural categorical representation $\rho=(\rho^1,\rho^0)\colon \lG \to \GL(\underline{\lG})$ of $\lG$ in its underlying complex $\underline{\lG}=(\partial\colon \lh \to \lg),$ called the adjoint {categorical} representation of $\lG$. For the case of Lie crossed modules this appeared in \cite{Wo}. Specifically, any element $X \in \lg$ is sent to the chain map $\rho^0_X\colon \underline{\lG} \to \underline{\lG}$, with:
$$
\lh \times \lg \ni (v,Y)  \stackrel{\rho^0_X}{\longmapsto}\big(X \t v, [X,Y]\big) \in \lh \times \lg,
$$ 
and each element $v \in \lh$ is sent to the chain homotopy $\rho^1_v$ of $ \underline{\lG}$, with:
$$ 
\lg \ni X \stackrel{\rho^1_v}{\longmapsto } -X \t v \in \lh. 
$$
In particular a $\lg$-invariant element $v\in \lh$ is always sent to the null homotopy.

Let us now go back to the string differential crossed module $\st=(\dd:\Fo\lra\fsl,\t)$. Looking at the exact sequence \eqref{es}, let us try to find an {infinitesimal 2-Yang-Baxter operator} $(\bar{r},P)$ in $\st$, such that $(\pi\otimes\pi)(\bar{r})=r$, where $r$ is the {infinitesimal Yang-Baxter operator} in $\sl$; see \eqref{rm}. Let $\overline{r}$ be the  obvious lift of $r$ to $(\F_1 \rtimes_\alpha \sl) \tn (\F_1 \rtimes_\alpha \sl)$. Explicitly:
$$
{\overline{r}= 2\, \zp{\ep} \tn \zp{\em} + 2 \, \zp{\em} \tn \zp{\ep} - 4 \, \zp{\eo} \tn \zp{\eo}= \sum_i \overline{s_i} \tn \overline{t_i} \in (\fsl)\tn (\fsl) \, .}
$$
Unlike $r$, the tensor $\overline{r}$ does not satisfy the 4-term relations \eqref{4termR}. However, since $r$ is an infinitesimal {Yang-Baxter operator}, the exactness of \eqref{es} implies, since $\pi([\bar r_{12}+\bar r_{13},\bar r_{23}])=[r_{12}+r_{13}.r_{23}]=0$, that $[\bar r_{12}+\bar r_{13},\bar r_{23}]$ is in the image of ${\dd'}\colon {\overline{\lU}}^{(3)} \to (\F_1 \rtimes_\alpha \sl)^{\tn 3}  $. Let $P^0$ be the most obvious lift of  $[\bar r_{12}+\bar r_{13},\bar r_{23}]$ to $\overline{\lU}^{(3)}$. 
Explicitly:
\begin{equation}
{P^0 = 16\,\big(\zp{\em}\tn\zp{\eo}\tn x + \zp{\em}\tn x\tn\zp{\eo} - \zp{\eo}\tn x\tn \zp{\em} - \zp{\eo}\tn\zp{\em}\tn x \,\big) \in \overline{\lU}^{(3)}.}
\end{equation} 
{Let also (where, if {$c\in \C$}, the constant polynomial with value $c$ is denoted by $c_{\Fo}$):
\begin{equation}
 C=\sum_i 2_{\Fo}  \otimes \overline{s_i} \tn \overline{t_i}  -\sum_i  \overline{s_i} \tn \overline{t_i} \tn 1_{\Fo}    -\sum_i  \overline{s_i} \tn 1_{\Fo}  \otimes  \overline{t_i} \in \overline{\lU}^{(3)}.
\end{equation}
The main result of this paper {(Theorem \ref{main2r})} states that relations \eqref{irm} are satisfied by $(\overline{r},P)=(\overline{r},P^0 + 4 C)$. Therefore $(\overline{r},P)$ is a free infinitesimal 2-Yang-Baxter operator in the string differential crossed module.
}

{It  is also remarkable, and essentially follows from the fact that, in $\st=(\dd:\Fo\lra\fsl,\t)$, the kernel $\ker(\dd)$ of $\dd$ is $(\fsl)$-invariant, that the pair $(\overline{r},P^0)$ is an infinitesimal 2-{Yang-Baxter} operator in the string differential crossed module, with respect to the adjoint {categorical} representation of it; {Theorem \ref{adj2R}}. However the same is also true for $(\overline{r},Q)$, where $Q$ is any inverse image, in ${\overline{\lU}}^{(3)}$, of the tensor $[\bar r_{12}+\bar r_{13},\bar r_{23}]$; {Remark \ref{taut}.}}

Since $(\pi \tn \pi)(\overline{r})=r,$ where $r\in \sl \tn \sl$ is the usual infinitesimal {Yang-Baxter operator} of $\sl$, given a positive integer $n$, we therefore have a lift of the $U_h(\sl)$ representation of the braid group in $\sl ^{ \tn n}$ to be an invariant of braid cobordisms, without branch points. The non-trivial part of the two dimensional holonomy will essentially live in the homology group: 
$$K^n=H_1\Big({\cal HOM}\left(\underline{\st}^{\btn n},\underline{\st}^{\btn n}\right)\Big).$$
Here  $\underline{\st}=(\partial\colon \F_0 \to \F_1 \rtimes_\a \sl)$ is the underlying complex of the string differential crossed module, $\btn$ denotes the tensor product of chain-complexes, and given chain complexes $\V$ and $\V'$, ${\cal HOM}(\V,\V')$ denotes the hom-chain complex. Clearly $K^1$ is the vector space $\hom(\sl,\C)$ of linear maps $\sl \to \C$. Since $\st$ is a complex of vector spaces with finite dimensional cohomology groups, \cite[10.24]{D}, we deduce 
$$ K^n \supset \hom(\sl,\C) \tn \hom(\sl, \sl) \tn \dots \tn \hom(\sl,\sl)\oplus\textrm{ cyclic permutations}.$$

To get a full fledged invariant of braid cobordisms from the holonomy of the Knizhnik-Zamolodchikov 2-connection derived from $(\overline{r},P)$, we need to investigate whether the integrals defining the two-dimensional holonomy of a general (possibly with branch points) braid cobordism $\S$ converge or not, or for the least find a regularisation process. Recall that braid cobordisms $\S$ yield in general, rather than a map $\S'\colon [0,1]^2 \to \C(n)$, a map $\S'\colon [0,1]^2 \setminus \beta(\S) \to \C(n)$, where  $\beta(\S) $ is the set of branch points of $\S$, which  are points where  $\S'$ has a very particular type of  singularities, \cite{CFM11}. We expect to address this very important issue in a future publication. {The case of loop braids, defined in \cite{BWC}, is a very promising one, since the singularities of the maps to the configuration space are of the simplest type in this case.} {Independent of this, we conjecture that a  braided monoidal bicategory  \cite{BNeu,Gu} can be derived from the framework developed in this paper and \cite{CFM11}, by making use of categorified Drinfeld associators, through a holonomy construction similar to \cite[chapters XIX and XX]{Ka}, \cite{Dr} and \cite{AF,LM}.}

{Let $\lg$ be a Lie algebra. Let $r \in \lg\tn \lg$ be a symmetric tensor. As mentioned before, the 4-term relation \eqref{4termR} is implied by the much stronger $\lg$-invariance condition \eqref{infBraiding}. In the latter case, the holonomy of the Knizhnik-Zamolodchikov connection $A$ of \eqref{kzi} is not only a braid-group representation, but also all holonomy maps are $\lg$-module intertwiners. For this reason, it is natural to address the categorification of the $\lg$-invariance condition  \eqref{infBraiding}, in the realm of a differential crossed module $\lG$ with $\pi_1(\lG)=\lg$. This can be accomplished by making use of the 2-category of weak categorical representations of the differential crossed module $\lG$, their  weak intertwiners, and 2-intertwiners connecting them; see \cite{E,BM} for the case of crossed modules of groups. In such a framework, the pair $(r,P)$ should induce a weak intertwiner $\underline{\lG} \btn \underline{\lG} \to \underline{\lG} \btn \underline{\lG}$, where $\underline{\lG}$ is the adjoint {categorical} representation of $\lG$ and $\btn$ denotes the tensor product of categorical representations, defined in  \cite{CFM11}. If an infinitesimal 2-{Yang-Baxter} operator $(r,P)$ in $\lG$ satisfies this {quasi-invariance} property, then the 1-dimensional and 2-dimensional holonomies of the associated Knizhnik-Zamolodchikov 2-connection will be weak 1- and 2-intertwiners for categorical representations of $\lG$. This framework will be fully developed in a future publication.}

\tableofcontents

\section{Preliminaries}

{\bf Note}: Given morphisms $f\colon A \to B$ and $g\colon B \to C$, of chain complexes or of vector spaces, their composition is written as $g \circ f$ or as  $gf$. We will often not distinguish a chain-complex of vector spaces from its underlying $\mathbb{Z}$-graded vector space. A chain map of chain complexes will always mean a degree 0 chain map.

\subsection{Differential and Lie crossed modules }
A differential crossed module $\lG=(\beta\colon \lh \to \lg,\t)$, see  \cite{B1,BC}, is given by a Lie algebra map $\beta \colon \lh \to \lg$ together with a left action $\t$ of $\lg$ on $\lh$ by Lie algebra derivations, such that the following conditions (called Peiffer relations) hold for each $X \in \lg$ and each $v,w \in \lh$: 
\begin{enumerate}
 \item $\b(X \t v)=[X, \b{(v)}]$
 \item $\b{(v)} \t w=[v,w]$
\end{enumerate}
 The fact that $\t$ is a Lie algebra action of $\lg$ on $\le$ by derivations means that the map 
$$\lg\times\lh \ni (X,v) \mapsto X \t v\in \lh$$ is bilinear, and, moreover, that for each $X,Y \in \lg$ and $u,v \in \lh$ :
\begin{enumerate}
 \item $X \t [u,v]=[X \t u, v]+[u, X \t v]$
 \item $[X,Y] \t u=X \t (Y \t u)-Y \t (X \t v)$
\end{enumerate}
A morphism $f=(\psi\colon \lh \to \lh', \phi\colon \lg \to \lg')$ between the differential crossed modules $\lG=(\beta\colon \lh \to \lg,\t)$ and $\lG'=(\beta'\colon \lh'\to \lg',\t')$ is given by Lie algebras maps $\psi\colon \lh \to \lh'$ and $ \phi\colon \lg \to \lg'$ making the following diagram commutative:
$$ \begin{CD}
    \lh @>\b> > \lg\\
 @V \psi VV  @VV \phi V  \\
\lh' @>\b' >> \lg'\\
   \end{CD}\quad ,
$$
and such that for each $v \in \lh$ and $X \in \lg$ we have:
$$\psi(X \t v)=\phi(X) \t' \psi{(v)}. $$
It is a well known fact that the category of differential crossed modules is equivalent to the category of strict Lie 2-algebras and strict maps. For details on this construction see \cite{B1,BC,BL}, and \cite{BHS} for the case of crossed modules of groups.

A Lie crossed module $\cG=(\beta\colon H \to G,\t)$ is given by a Lie group map $\beta\colon  H \to G$ and a smooth action of $G$ on $H$ by automorphisms, such that the following relations (called Peiffer relations) hold for each $g \in G$ and $h,h' \in H$:
\begin{enumerate}
 \item $\b(g \t h)=\b(g) h \b(g)^{-1}$
\item $\b(h) \t h'=hh'h^{-1}.$
\end{enumerate}
Morphisms of Lie crossed modules are defined in the obvious way.

Passing from Lie groups to Lie algebras yields a functor from the category of Lie crossed modules to the category of differential crossed modules. Standard Lie theory tells us that any differential crossed module of finite dimensional Lie algebras is the differential crossed module associated to a unique (up to isomorphism) Lie crossed module of simply connected groups. 

\subsection{Differential crossed modules from chain-complexes of vector spaces}\label{dcmcc}

This will be our most fundamental example of a differential crossed module. Let 
$$\V = \left\{ \dots \ra{\partial} V_i \ra{\partial} V_{i-1}\ra{\partial} \dots\right\}_{i \in \mathbb{Z}} $$
be a chain complex of vector spaces. We can define out of it a differential crossed module:
$$\GL(\V)=\left( \beta: \gl^1(\V) \to \gl^0(\V),\t \right) .$$
Let us explain the construction of this \cite{BC,KP,FMM,FMP3}. The Lie algebra  $(\gl^0(\V),\{\, ,\})$ is given by all chain maps $f\colon \V \to \V$ with Lie bracket $\{ \, ,\}$ given by the usual commutator of chain maps: 
$$\{f,g\}=[f,g]=fg-gf \, , \textrm{ for each } f,g \in \gl^0(\V).$$
Here a chain map $f \colon \V \to \V$ is given by a sequence of linear maps $f\colon V_i \to V_i$ satisfying the following compatibility condition with the boundary maps:
$$f_i \partial =\partial f_{i+1}\, , \quad \forall i \in \mathbb{Z}. $$
Note that we only consider  chain-maps of degree 0.

The underlying vector space of the Lie algebra $(\gl^1(\V),\{\, , \})$ is a quotient of the vector space ${\rm Hom}^1(\V)$ of degree 1 maps (homotopies) $s\colon \V \to \V$. In other words ${\rm Hom}^1(\V)$ is given by all sequences of linear maps $s=(s_i\colon V_i \to V_{i+1})_{i \in \mathbb{Z}}$, without any compatibility relation with the boundary maps $\partial$. There exists a linear map $\beta \colon {\rm Hom}^1(\V) \to \gl^0(\V)$, where $$\beta(s)=\partial s+s\partial.$$
Two chain maps $f$ and $f'$ are said to be homotopic if there exists a homotopy $s$ such that 
$$f'=f + \beta(s). $$
If two chain maps $f$ and $g$ are homotopic to the null chain map through homotopies $s$ and $t$, i.e. $\beta(s)=f$ and $\beta(t)=g$, then their commutator $\{f,g\}$ is homotopic to zero. In fact:
\begin{align*}
 \{f,g\}&=[s \partial + \partial s,t \partial+\partial t]\\
        &=s \partial t \partial +\partial st \partial +\partial s \partial t-t \partial s \partial -\partial ts \partial -\partial t \partial s\\
        &= \beta(s \partial t + st \partial-t \partial s - ts \partial) = \beta(s \partial t + \partial st-t \partial s - \partial ts )
\end{align*}
This calculation induces two Lie algebra structures in $\Hom^1(\V)$, both of them satisfying that the map $\b\colon \Hom^1(\V) \to \gl^0(\V)$ is a Lie algebra morphism. These are:
$$\{s,t\}_l= s \partial t + st \partial-t \partial s -ts \partial \, ,$$
$$\{s,t\}_r= s \partial t + \partial st-t \partial s - \partial ts \, .$$
The antisymmetry of these bilinear maps is immediate. The Jacobi identity follows from an explicit calculation, which we leave to the reader (and recommend the reader to perform).

There are actions by derivations of $\gl^0(\V)$ on both $(\Hom^1(V),\{,\}_l)$ and $(\Hom^1(V),\{,\}_r)$ . They both have the form:
$$ f \t s=f s - s f, \textrm{ where } s \in \Hom^1(\V) \textrm{ and } f \in \gl^0(\V).$$
That this is a Lie algebra representation is straightforward. That the map $s \mapsto f \t s$ is always a derivation for $\{,\}_l$ and $\{,\}_r$ follows from an explicit calculation, which we urge the reader to perform  by him/her-self. The fact that $f$ is a chain-map (that is: $f \partial=\partial f$) has a primary role here. 
\begin{Lemma}
 For each $f \in \gl^0(\V)$ and each $s,t \in \Hom^1(\V)$ we have:
\begin{enumerate}
 \item  $\beta(f \t s)=\{f,\beta(s)\}. $
 \item $\beta(t) \t s=\{t,s\}_l-ts \partial+\partial ts $
\item $\beta(t) \t s=\{t,s\}_r-st \partial+\partial st.$
\end{enumerate}
\end{Lemma}
\noindent Again the proof is straightforward. Therefore we have almost defined two differential crossed modules $(\beta\colon \Hom^1(\V) \to \gl^0(\V),\t)$, one for each Lie algebra structure $\{,\}_l$ and $\{,\}_r$ in $\Hom^1(\V)$, except that the second Peiffer identity only holds up to 2-fold homotopy. Indeed, let $\Hom^2(\V)$ be the space of degree 2 maps $\V \to \V$ and consider the usual map $\b\colon \Hom^2(\V) \to \Hom^1(\V)$ with 
\begin{equation}\label{bp}
\beta(a)=\partial a - a \partial , \textrm{ for each } a \in \Hom^2(\V).
\end{equation}
Notice $\beta^2=0$, thus $\beta\colon \Hom^1(\V) \to \gl^0(\V)$ descends to a map (also called $\beta$) from the vector space  quotient $$\gl^1(\V)=\frac{\Hom^1(\V)}{\beta(\Hom^2(\V))} $$ into $\gl^0(\V)$. 
\begin{Lemma}
 The vector subspace $\beta(\Hom^2(\V))\subset \Hom^1(\V)$ is a Lie algebra ideal both for $\{,\}_l$ and $\{,\}_r$. Moreover $\beta(\Hom^2(\V))$ is invariant under the action of $\gl^0(\V)$. In particular the Lie brackets $\{,\}_l$ and $\{,\}_r$ both descend to the quotient $\gl^1(\V)={\Hom^1(\V)}/{\beta(\Hom^2(\V))}$, defining the same Lie algebra structure in $\gl^1(\V)$  (whose Lie bracket we denote by $\{,\}$). Finally, the action $\t$ descends to an action of $\gl^0(\V)$ on $\gl^1(\V)$ by derivations. 
\end{Lemma}
\proof{
 We compute, for $a \in \Hom^2(\V)$ and $s\in\Hom^1(\V)$:
$$
\{\beta(a),s\}_l=\{\partial a -  a\partial,s\}_l =
\partial a \partial s + \partial a s \partial -a\partial s \partial =
\beta (a\partial s + a s \partial) 
$$
and
$$
\{\beta(a),s\}_r=\{\partial a -  a \partial,s\}_r = 
- \partial s \partial a + \partial s a \partial + s \partial a \partial = \beta (s a\partial - s\partial a) \, ,
$$
which proves the first assertion. The second assertion follows from the calculation (here $f \in \gl^0(\V)$ and $a \in \Hom^2(\V)$):
$$f \t \b(a) = [f,\partial a - a \partial]=  f\partial a - f  a \partial - \partial a f + a \partial f= \b(fa-af)\, .  $$
The only non-trivial instance of third assertion follows from the fact that 
$\{s,t\}_l=\{s,t\}_r+\beta(-st+ts) \, .$ \qedhere
} \\

\noindent
Given a chain complex $\V$ of vector spaces, we have thus constructed a differential crossed module
$$\GL(\V)=\left (\b\colon \gl^1(\V) \to \gl^0(\V),\t\right), $$
which will have an essential role in this article. Of course, this construction defines a functor from the category of chain-complexes of vector spaces and chain-maps to the category of differential crossed modules.

\subsection{Local 2-connections and their two-dimensional  holonomy}
Let $M$ be a manifold. Let also $\cG=(\beta\colon H \to G,\t)$ be a Lie crossed module. Let 
$\lG=(\beta\colon \lh \to \lg,\t)$ be the associated differential crossed module. A $\lG$-valued local 2-connection $(A,B)$ in $M$ is given by a 1-form $A\in \Omega^1(M;\lg)$ and a 2-form $B\in\Omega^2(M;\lh)$ such that $\beta(B)=F_A$, where 
$F_A=dA+\frac{1}{2}A\wedge^{[\, , \, ]}A$ is the curvature of $A$. Here $A\wedge^{[ \, , \, ]} A$ is  twice the antisymmetrisation of the tensor $X,Y\mapsto [A(X),A(Y)]$, for vector fields $X$ and $Y$ in $M$.

A local 2-connection determines a 2-dimensional holonomy $\H=(H^1,H^2)$ for paths and 2-paths in $M$. Let us explain this construction in the particular (and slightly simpler than the general) case when $\cG$ is the Lie crossed module associated to the differential crossed module $\GL(\V)=(\beta\colon \gl^1(\V) \to \gl^0(\V),\t)$, where $\V$ is a chain complex of vector spaces, \cite{FMM}. We also suppose that $\V=(V_i,\partial)_{i \in \mathbb{Z}}$ is such that all but a finite number of the $V_i$ are non trivial, and each of these is finite dimensional; in other words we suppose that $\V$ is a chain-complex of finite type.  In this case $\GL(\V)$ is a differential crossed module of finite dimensional Lie algebras.
\subsubsection{The $2$-category $\Aut(\V)$ for a chain complex $\V$}\label{autv}

{Define a 2-category $\Aut(\V)$ with a single object (in other words a monoidal category),  whose  morphisms are the  chain maps $\V \to \V$. The composition is done in the reverse order: 
$$ (\V \ra{f} \V  \ra{g} \V)= (\V \ra{fg} \V). $$
Given two chain maps $f,g\colon \V \to \V$, a 2-morphism $f\implies g$ is given by a pair $(f,s)$, where $s\in\gl^1(\V)=\Hom^1(\V)/\beta'(\Hom^2(\V))$, for which:}
$$f=g+ \beta(s)=g+\partial s+ s \partial.$$ The vertical and horizontal composition of 2-morphisms are, respectively:
$$\xymatrix{ &\ar@/^3pc/[rr]^{f+\beta(s)+\beta(t)}\ar[rr]|{f+\beta(s)} \ar@/_3pc/[rr]_{f} \ar @/^/ @{{}{ }{}} [rr]^{\Uparrow(f+\beta(s),t)} \ar @/_/ @{{}{ }{}} [rr]_{\Uparrow(f,s)}
 &  &&\quad =\quad &\ar@/^2pc/[rr]^{f+\beta(s)+\beta(t)} \ar @/^/ @{{}{ }{}} [rr]_{\Uparrow(f,s+t)} \ar@/_2pc/[rr]_{f}
 & & }$$

 $$\xymatrix{ &\ar@/^2pc/[rr]^{f_2}\ar@/_2pc/[rr]_{f_1} & \Uparrow(f_1,s) & \ar@/^2pc/[rr]^{g_2}\ar@/_2pc/[rr]_{g_1} & \Uparrow(g_1,t) & &\quad = & \ar@/^2pc/[rr]^{f_2g_2}\ar@/_2pc/[rr]_{f_1g_1} & \Uparrow(f_1g_1,s.t) & .}$$
Here:
\begin{align*}
 f_2g_2=f_1g_1+\beta(f_1 t+sg_2)=f_1g_1+\beta(sg_1+f_2 t)
\end{align*}
and
$$s.t= f_1 t+sg_2= sg_1+f_2 t \textrm{ in } \gl^1(\V).$$
Note:
\begin{align*}
 f_1 t+sg_2-(sg_1+f_2 t)=s(g_2-g_1)+(f_1-f_2)t=s \beta(t)-\beta(s)t=st \partial-\partial st=-\b(st).
\end{align*}
We leave it to the reader to verify that the interchange law and the remaining 2-category axioms are satisfied.

\subsubsection{The form of a 2-dimensional holonomy}
Parts of this appear in several places, most notably \cite{BS1,BS2,SW1,SW2,FMP1,FMP2,FMP3,GP}. However note that given that we are working in the 2-category $\Aut(\V)$, the formula for the two-dimensional holonomy  of a 2-path is considerably simpler than in those references, being analogous to the two-dimensional holonomy of a representation up to homotopy of a Lie algebroid \cite{AC}. 

Let $M$ be a manifold. A path $x \ra{\gamma} y$ is a piecewise smooth path $\gamma \colon [0,1] \to M$, where $x,y \in M$ are the initial and end points of $\gamma$. These paths can be concatenated in the obvious way, defining $\gamma_1 \gamma_2$, if the paths $\gamma_1$ and $\gamma_2$ are such that the end point of $\gamma_1$ coincides with the initial point of $\gamma_2$. If $x \ra{\gamma_1} y $ and  $x \ra{\gamma_2} y$ are paths, a 2-path $\Gamma\colon \gamma_1 \implies \gamma_2$ is a map $\Gamma \colon [0,1]^2 \to M$, such that:
\begin{enumerate}
 \item $\Gamma$ is piecewise smooth for some paving of the square $[0,1]^2$ by poligons.
 \item $\Gamma(s,0)=x$ and $\Gamma(s,1)=y$, for each $s \in [0,1]$.
 \item $\Gamma(0,t)=\gamma_1(t)$ and $\Gamma(1,t)=\gamma_2(t)$, for each $t \in [0,1]$.
\end{enumerate}
 Clearly 2-paths can be composed horizontally $\Gamma\Gamma'$ and vertically ${ \Gamma \atop \Gamma'}$ in the obvious way, as long as they coincide in the relevant sides of the square $[0,1]^2$ .
 
\begin{Definition}\label{2hol} 
Let $\V$ be a chain-complex of vector spaces and $M$ be a manifold. An $\Aut(\V)$-valued 2-dimensional holonomy \mbox{$\H=(H^1,H^2)$} is given by a map $H^1$ from the set of paths of $M$ into the set of chain-maps of $\V$ and a map $H^2$ from the set of 2-paths of $M$ into $\gl^1(\V)$ such that the following holds:
\begin{enumerate}
 \item {$H^1$ preserves the composition of paths, i.e.} $H^1(\gamma_1 \gamma_2)=H^1(\gamma_1) H^1(\gamma_2)$, given paths   $x \ra{\gamma_1} y  \ra{\gamma_2} z$.
 \item  (Globularity) If $x,y \in M$ and  $x \ra{\gamma_1} y $ and  $x \ra{\gamma_2} y$ are paths and $\gamma_1 \stackrel{\Gamma}{\implies} \gamma_2$ is a 2-path, then:
$$\beta(H^2(\Gamma))+H^1(\gamma_1)=H^1(\gamma_2).$$
 \item $H^2$ preserves the horizontal and vertical compositions of 2-paths and of homotopies (up to 2-fold homotopies).
\end{enumerate}
\end{Definition}
\begin{Remark}
 We can define a 2-dimensional holonomy with values in any 2-category in exactly the same fashion. We need however an additional  bit of data, which is a map $H^0$ from $M$ into the set of objects of the 2-category. The globularity condition should have an extra condition which is that if $x \ra{\gamma} y $  is a path in $M$ then $H^0(x) \ra{H^1(\gamma)} H^0(\gamma)$ has to be a 1-morphism in the 2-category. 
\end{Remark}
\begin{Remark}
 Continuing the previous remark, consider a manifold $M$ and a chain complex $\V$ of vector spaces. Let us define a 2-category $\Aut(\V,M)$, whose set of objects is $M$. Given $x,y \in M$, a  morphism $x \to y$ is a triple $(x,f,y)$, where $f\colon \V \to \V$ is a chain-map. The composition $( x \ra{(x,f,y)} y\ra{(y,f',z)} z)$ is $x \ra{(x,ff',z)} z$. The set of 2-morphisms $(x \ra{(x,f,y)} y) \implies (x \ra{(x,f',y)} y)$ is in one-to-one correspondence with the set of homotopies (up to 2-fold homotopy) $f \to f'$, with the obvious vertical and horizontal compositions. Clearly any $\Aut(\V)$-valued two dimensional holonomy induces a $\Aut(\V,M)$-valued 2-dimensional holonomy, where $H^0$ is the identity map. 
\end{Remark}
\noindent Below we will consider 2-dimensional holonomies taking values in a quotient of $\Aut(M,\V)$.
\begin{Remark}
 A coordinate free, and more general, definition of a 2-dimensional holonomy  can be stated in the framework of graded vector bundles $E$, provided with a fibrewise chain map $D\colon E \to E$, making each fibre into a chain-complex of vector spaces, \cite{AC}. Then the 2-category $\Aut(\V,M)$ can be substituted by the 2-category whose objects are the points of $M$, the morphisms $x \to y$ being the chain-maps $E_x \to E_y$, and the 2-morphisms  $(f \colon E_x \to E_y) \implies (f' \colon E_x \to E_y) $  being in one-to-one correspondence with chain complex homotopies $s$ (considered up to 2-fold homotopy) connecting $f$ and $f'$. We will not need nor use this generality.
\end{Remark}

\begin{Remark}
 {Considering appropriate thin-homotopy \cite{MP,FMP1,FMP2,FMP3,BS1} equivalence relations between paths and 2-paths, these will form a 2-category, in fact 2-groupoid, $\P_2(M)$, considering the compositions above. We could define a two dimensional holonomy with values in a 2-category ${\cal C}$ as being a (smooth) 2-functor $\P_2(M) \to {\cal C}$. }
\end{Remark}

\begin{Theorem}
Consider a chain complex $\V$ of vector spaces, which we suppose to be of finite type, and the differential crossed module (of finite dimensional Lie algebras)  
$$\GL(\V)=\left(\beta: \gl^1(\V) \to \gl^0(\V),\t \right) \, .$$
 Let $A$ be a 1-form in $M$ with values in $\gl^0(\V)$ and $B$ be a 2-form in $M$ with values in $\gl^1(\V)$ such that $\beta(B)=d A+\frac{1}{2} A \wedge^{[,]} A\doteq F_A$, the curvature of $A$. Then we can integrate the local 2-connection $(A,B)$ in order to obtain a 2-dimensional holonomy $\H=(H^1,H^2)$ with values in the 2-category $\Aut(\V,M)$, or equivalently, in this case, $\Aut(\V)$.
\end{Theorem}

\proof{Let us explain the construction of the 2-dimensional holonomy, as well as give the proof of this result. This is a particular case of \cite{BS1,BS2,SW1,SW2,FMP1,FMP2,FMP3}, very influenced by the exposition in \cite{AC}. In particular, on the contrary of the former references, and solely because we work in the 2-category $\Aut(\V)$, no differential equations are needed in order to express the 2-dimensional holonomy of a 2-path.

Given a path $\gamma$, then $H^1(\gamma)$ is simply the 1-dimensional holonomy of it, and lives in the Lie group ${\rm GL}^0(\V)$ of invertible chain maps $\V \to \V$. In other words $H(\gamma)=g_\g(1),$
where $t \mapsto g_\g(t)$ is the solution of the differential equation:
$$ \frac{d}{d t} g_\g(t)= g_\g(t) \,\,A\left( \frac{d}{d t}\g(t)\right), \textrm{ with } g_\g(0)=1.$$
This is a differential equation in the Lie group ${\rm GL}^0(\V)$. It can also be seen as a differential equation in the vector space $\gl^0(\V)$, which is the point of view we will take from now on. Note that if $\gamma(1)=\gamma'(0)$ then 
$$H(\gamma\gamma')=H(\gamma) H(\gamma'). $$
Given a 2-path $\g_0 \stackrel{\Gamma}{\implies} \g_1$, say $\Gamma(t,s)=\gamma_s(t)$ for $t,s \in [0,1]^2$, it is well known (for a proof see \cite{SW2,FMP1,FMP3}) that:
$$\frac{d}{ds} H^1(\gamma_s)=\int_0^1 g_{\g_s}(t) \,F_A\left( \frac{\partial}{\partial s}\Gamma(t,s),\frac{\partial }{\partial t} \Gamma(t,s) \right)\, g_{\g_s}(t)^{-1} dt\,\,H^1(\gamma_s) .  $$
Again, this can either  be seen  as a differential equation in the Lie group ${\rm GL}^0(\V)$, or as a differential equation in the vector space $\gl^0(\V)$ of chain maps $\V \to \V$. Looking at the latter picture, we therefore have:
$$H^1(\g_1)=H^1(\g_0) + \int_0^1 \int_0^1 g_{\g_s}(t)\, F_A\left( \frac{\partial}{\partial s}\Gamma(t,s),\frac{\partial t}{\partial t} \Gamma(t,s) \right)\, g_{\g_s}(t)^{-1}  \,\,H^1(\gamma_s)\, dt ds. $$
Therefore, given that $\beta(B)=F_A$, if we put:
$$H^2(\Gamma)= \int_0^1 \int_0^1 g_{\g_s}(t)\,B\left( \frac{\partial}{\partial s}\Gamma(t,s),\frac{\partial t}{\partial t} \Gamma(t,s) \right)\, g_{\g_s}(t)^{-1} \,H^1(\gamma_s)\, dt ds \in \gl^1(\V),$$
it thus follows that the globularity condition of Definition \ref{2hol} is satisfied. Compare with \cite[Proof of Prop. 3.13]{AC}. This is a particular case of the construction in \cite{BS2,SW1,FMP2}.

That $H^2$ preserves the  vertical compositions of 2-paths is trivial to check. Let us prove that $H^2$ preserves horizontal compositions. Consider points $x,y,z \in M$ and also paths $\g_0,\g_1 \colon x \to y$ and $\g_0',\g_1' \colon y \to z$, and finally 2-paths $\Gamma \colon \g_0 \implies \g_1$ and $\Gamma' \colon \g_0' \implies \g_1'$. Put $\gamma_s(t)=\Gamma(t,s)$ and $\gamma'_s(t)=\Gamma'(t,s)$, where $t,s \in [0,1]$. Define 2-paths for each $s\in[0,1]$ as $\Gamma_s(t,s')=\Gamma(t,1-ss')$  and  $\Gamma_s'(t,s')=\Gamma(t,ss')$ for $s',t \in [0,1]$.  Since $H^1(\g_s \g'_s)=H^1(\g_s) H^2(\g'_s)$ we have:
\allowdisplaybreaks{
\begin{align*}
H^2(\Gamma \Gamma')&=\int_0^1 \int_0^1 g_{\g_s}(t)\,B\left( \frac{\partial}{\partial s}\Gamma(t,s),\frac{\partial t}{\partial t} \Gamma(t,s) \right)\, g_{\g_s}(t)^{-1}\,H^1(\gamma_s) \,H^1(\g_s')\,dt ds \, + \\ 
&\quad \quad + \int_0^1 \int_0^1 H^1(\g_s)\,\,g_{\g'_s}(t)\,B\left( \frac{\partial}{\partial s}\Gamma'(t,s),\frac{\partial t}{\partial t} \Gamma'(t,s) \right)\, g_{\g'_s}(t)^{-1} H^1(\g_s') dt ds = \\
&=\int_0^1 \int_0^1 g_{\g_s}(t)\,B\left( \frac{\partial}{\partial s}\Gamma(t,s),\frac{\partial t}{\partial t} \Gamma(t,s) \right)\, g_{\g_s}(t)^{-1}\,H^1(\gamma_s) \,(H^1(\g_0')+\beta(H^2(\Gamma_s'))\,dt ds \, + \\ 
&\quad \quad + \int_0^1 \int_0^1 (H^1(\g_1)-\beta(H^2(\Gamma_{s}))\,\,g_{\g'_s}(t)\,B\left( \frac{\partial}{\partial s}\Gamma'(t,s),\frac{\partial t}{\partial t} \Gamma'(t,s) \right)\, g_{\g'_s}(t)^{-1} H^1(\g_s') dt ds = \\
&=H^2(\Gamma) H^1(\g'_0)+H^1(\g_1) H^2(\Gamma') \, + \\
&\quad \quad+\int_0^1 \int_0^1 g_{\g_s}(t)\,B\left( \frac{\partial}{\partial s}\Gamma(t,s),\frac{\partial t}{\partial t} \Gamma(t,s) \right)\, g_{\g_s}(t)^{-1}\,H^1(\gamma_s) \beta(H^2(\Gamma_s'))\,dt ds \, + \\ 
&\quad \quad - \int_0^1 \int_0^1 \beta(H^2(\Gamma_{s}))\,\,g_{\g'_s}(t)\,B\left( \frac{\partial}{\partial s}\Gamma'(t,s),\frac{\partial t}{\partial t} \Gamma'(t,s) \right)\, g_{\g'_s}(t)^{-1} H^1(\g_s') \,dt ds \,.
\end{align*}}
Given homotopies $s$ and $t$ then $\b(s) t=s \b(t)+\beta(st)$, equation \eqref{bp}, thus $s\beta(t)=\beta(s) t $ in $\gl^1(\V)$. Therefore:
\allowdisplaybreaks{
\begin{align*}
\int_0^1& \int_0^1 g_{\g_s}(t)\,B\left( \frac{\partial}{\partial s}\Gamma(t,s),\frac{\partial t}{\partial t} \Gamma(t,s) \right)\, g_{\g_s}(t)^{-1}\,H^1(\gamma_s) \beta(H^2(\Gamma_s'))\,dt ds = \\
&=\int_0^1\int_0^1 g_{\g_s}(t)\,F_A\left( \frac{\partial}{\partial s}\Gamma(t,s),\frac{\partial t}{\partial t} \Gamma(t,s) \right)\, g_{\g_s}(t)^{-1}\,H^1(\gamma_s) H^2(\Gamma_s')\,dt ds = \\
&=\int_0^1 \frac{d} {d s} \big(H^1(\g_s)\big)\,\, H^2(\Gamma_s') ds = \\
&=H^1(\g_1) H^2(\Gamma_1')- H^1(\g_0) H^2(\Gamma_0')-\int_0^1 H^1(\g_s) \frac{d} {d s} H^2(\Gamma_s') ds = \\
&=H^1(\g_1) H^2(\Gamma_1')-\int_0^1 H^1(\g_s) \frac{d} {d s} H^2(\Gamma_s')  ds\,,
\end{align*}}
where we have integrated by parts and use the fact that  $\Gamma'_0$ is a constant 2-path. We also have:
\allowdisplaybreaks{
\begin{align*}
\int_0^1 &\int_0^1 \beta(H^2(\Gamma_{s})\,\,g_{\g'_s}(t)\,B\left( \frac{\partial}{\partial s}\Gamma'(t,s),\frac{\partial t}{\partial t} \Gamma'(t,s) \right)\, g_{\g'_s}(t)^{-1} H^1(\g_s') dt ds = \\                                                                                                                                                                                         &=\int_0^1  (H^1(\g_1)-H^1(\g_{s}) )\int_0^1 g_{\g'_s}(t)\,B\left( \frac{\partial}{\partial s}\Gamma'(t,s),\frac{\partial t}{\partial t} \Gamma'(t,s) \right)\, g_{\g'_s}(t)^{-1} H^1(\g_s') dt ds = \\
&=\int_0^1  (H^1(\g_1)-H^1(\g_{s}) ) \frac{d} {d s} H^2(\Gamma_s')  ds=\int_0^1  H^1(\g_1) \frac{d} {d s} H^2(\Gamma_s')  ds-
\int_0^1  H^1(\g_{s})  \frac{d} {d s} H^2(\Gamma_s')  ds= \\
&= H^1(\g_1)  H^2(\Gamma_1') -  H^1(\g_1)  H^2(\Gamma_0')-
\int_0^1  H^1(\g_{s})  \frac{d} {d s} H^2(\Gamma_s')  ds=H^1(\g_1)  H^2(\Gamma_1') -\int_0^1  H^1(\g_{s})  \frac{d} {d s} H^2(\Gamma_s') 
\end{align*}}
since $\Gamma'_0$ is a constant 2-path. These calculations prove that $H^2$ preserves horizontal composites of 2-paths and of homotopies. 
\qedhere }

\begin{Remark}\label{cov1} 
Let $\V$ be a chain complex of vector spaces and $M$ be a manifold.
Let $(A,B)$ be a $\GL(\V)$-valued local 2-connection in $M$. Let $\stackrel{(A,B)}{\H}=(H^1,H^2)$ be its associated two-dimensional holonomy. Consider a smooth map $f\colon M \to M$. Let $f^*(A,B)=(f^*(A),f^*(B))$. Then $f^*(A,B)$ is a local 2-connection. Let $\stackrel{f^*(A,B)}{\H}$ be its holonomy. Then  $$\stackrel{(A,B)}{H^1}(f\circ \gamma)=\stackrel{f^*(A,B)}{H^1}(\gamma)$$ and  
$$\stackrel{(A,B)}{H^2}(f\circ \Gamma)=\stackrel{f^*(A,B)}{H^2}(\Gamma),$$
for each path $\g$ and a 2-path $\Gamma$.
\end{Remark}

\begin{Remark}\label{cov2}
In the notation of the previous remark, let $\phi\colon \V \to \V$ be an invertible chain-map. Then we have a 2-functor $\hat\phi\colon\Aut(\V) \to \Aut(\V)$, and also a differential crossed module map $\hat \phi \colon \GL(\V) \to \GL(\V)$, where any chain map and any homotopy (up to fold homotopy) are conjugated by $\f$, namely $\gl^0(\V) \ni g \mapsto \phi g\phi^{-1}\in\gl^0(\V) $ and $\gl^1(\V) \ni s \mapsto \phi s\phi^{-1}\in \gl^1(\V) .$ Then  $\hat\phi(A,B)=(\hat\phi(A),\hat\phi(B))$ is a local 2-connection. Moreover:
$$\stackrel{\hat\phi(A,B)}{\H}=\hat\phi\circ \stackrel{(A,B)}{\H}.$$
\end{Remark} 

\subsubsection{The 2-curvature 3-tensor}
\label{curv}
Let $\V$ be a chain complex of vector spaces and $M$ a manifold.
Let $(A,B)$ be a $\GL(\V)$-valued local 2-connection in $M$.
 Therefore $\beta(B)=F_A$, the curvature of $A$. The 2-curvature 3-tensor ${\M}$ of $(A,B)$ is defined as  $${\M}=d B+A \wedge^\t B,$$ where 
$ A \wedge^{\t} B$ is $3/2$ times the antisymmetrisation of the tensor $(X,Y,Z) \mapsto A(X) \t B(Y,Z)$, where $X,Y,Z$ are vector fields in $M$. In particular $$(A \wedge^\t B)(X,Y,Z)=A(X) \t B(Y,Z)+A(Y) \t B(Z,X)+A(Z) \t B(X,Y).$$
\begin{Theorem}
Suppose that $\Gamma_0$ and $\Gamma_1$ are 2-paths in $M$ which are homotopic relative to the boundary of $[0,1]^2$, through a piecewise smooth map $J\colon [0,1]^3 \to M$. Explicitly, suppose that $J(t,s,0)=\Gamma_0(t,s)$ and $J(t,s,1)=\Gamma_1(t,s)$. Also put $\gamma_{(x,s)}(t)=J(t,s,x)$, for $t,s,x \in  [0,1]$.
Then:
$$H^2(\Gamma_1)=H^2(\Gamma_0) +\int_0^1 \int_0^1\int_0^1g_{\g_{(x,s)}}(t) \,{\M}\left (\frac{\partial}{\partial x} J(t,s,x),\frac{\partial}{\partial s} J(t,s,x), \frac{\partial}{\partial t} J(t,s,x)\right)g_{\g_{(x,s)}}(t)^{-1} H^1({\g_{(x,s)}})\, dtdsdx. $$
\end{Theorem}
\proof{
Let $\Gamma_x(t,s)=J(t,s,x)$ for  $t,s,x \in  [0,1]$. Given $g \in {\rm GL}^0(\V)$, an invertible chain map, and $s \in \gl^1(\V)$, we put $g \t s=gsg^{-1}$. It is easy to see that  $\gl^1(\V) \ni s \mapsto g \t s \in \gl^1(\V)$ is a Lie algebra map.  We have:
$$H^2(\Gamma_x)= \int_0^1 \int_0^1 g_{\g_{(x,s)}}(t)\t \,B\left( \frac{\partial}{\partial s}J(t,s,x),\frac{\partial t}{\partial t} J(t,s,x) \right) \,H^1(\gamma_{(x,s)})\, dt ds \in \gl^1(\V).$$
The following is well known (however for a proof see \cite{FMP3}):
\begin{align*}
\frac{\partial}{\partial s} g_{\g_{(x,s)}}(t) & = \int_0^t g_{\g_{(x,s)}}(t') \,F_A\left( \frac{\partial}{\partial s}\Gamma_x(t',s),\frac{\partial }{\partial t'} \Gamma_x(t',s) \right)\, g_{\g_{(x,s)}}(t')^{-1}\, g_{\g_{(x,s)}}(t) \,dt'+g_{\g_{(x,s)}}(t)\,A\left( \frac{\partial}{\partial s} J(t,x,s)\right) \, ,\\
\frac{\partial}{\partial x} g_{\g_{(x,s)}}(t) & =\int_0^t g_{\g_{(x,s)}}(t') \,F_A\left( \frac{\partial}{\partial x}\Gamma_x(t',s),\frac{\partial }{\partial t'} \Gamma_x(t',s) \right)\, g_{\g_{(x,s)}}(t')^{-1}\, g_{\g_{(x,s)}}(t) \,dt'+g_{\g_{(x,s)}}(t)\,A\left( \frac{\partial}{\partial x} J(t,x,s)\right) \, .
\end{align*}
Also, by definition:
$$\frac{\partial}{\partial t} g_{\g_{(x,s)}}(t)=g_{\g_{(x,s)}}(t)\,A\left( \frac{\partial}{\partial t} J(t,x,s)\right),$$
and finally:
\begin{align*}
 \frac{\partial}{\partial x}  B\left( \frac{\partial}{\partial s}J(t,s,x),\frac{\partial t}{\partial t} J(t,s,x) \right) & = \frac{\partial}{\partial x} J^*(B) \left (\frac{\partial}{\partial s},\frac{\partial}{\partial t}\right) \\
&=J^*(d B) \left (\frac{\partial}{\partial x},\frac{\partial}{\partial s},\frac{\partial}{\partial t} \right)-\frac{\partial}{\partial s} J^*(B) \left (\frac{\partial}{\partial t},\frac{\partial}{\partial x}\right)-\frac{\partial}{\partial t} J^*(B) \left (\frac{\partial}{\partial x},\frac{\partial}{\partial s}\right) \, .
\end{align*}
Then, integrating by parts, and using the fact that the homotopy $J$ is relative to the boundary of $[0,1]^2$:
{\allowdisplaybreaks
\begin{equation}\label{ref}
\begin{split}
& \frac{d}{dx} H^2(\Gamma_x)=\iint\limits_{[0,1]^2} g_{\g_{(x,s)}}(t)  {\M}\left (\frac{\partial}{\partial x} J(t,s,x),\frac{\partial}{\partial s} J(t,s,x), \frac{\partial}{\partial t} J(t,s,x)\right)g_{\g_{(x,s)}}(t)^{-1} H^1({\g_{(x,s)}}) \,dtds \, + \\
& + \iint\limits_{[0,1]^2} \int_0^t\left ( g_{\g_{(x,s)}}(t') \,F_A\left( \frac{\partial}{\partial s}\Gamma_x(t',s),\frac{\partial }{\partial t'} \Gamma_x(t',s) \right)\, g_{\g_{(x,s)}}(t')^{-1}\, g_{\g_{(x,s)}}(t) \,\right)  \t \,B\left( \frac{\partial}{\partial t}J(t,s,x),\frac{\partial }{\partial x} J(t,s,x) \right) \,H^1(\gamma_{(x,s)})\,dt' dt ds \, +\\
& + \iint\limits_{[0,1]^2} \int_0^t \left(g_{\g_{(x,s)}}(t') \,F_A\left( \frac{\partial}{\partial x}\Gamma_x(t',s),\frac{\partial }{\partial t'} \Gamma_x(t',s) \right)\, g_{\g_{(x,s)}}(t')^{-1}\, g_{\g_{(x,s)}}(t) \right)\, \t \,B\left( \frac{\partial}{\partial s}J(t,s,x),\frac{\partial t}{\partial t} J(t,s,x) \right) \,H^1(\gamma_{(x,s)})\,dt' dt ds \, + \\
& + \iiint\limits_{[0,1]^3} g_{\g_{(x,s)}}(t)\t \,B\left( \frac{\partial}{\partial s}J(t,s,x),\frac{\partial t}{\partial t} J(t,s,x) \right)g_{\g_{(x,s)}}(t') \,F_A\left( \frac{\partial}{\partial x}\Gamma_x(t',s),\frac{\partial }{\partial t'} \Gamma_x(t',s) \right)\, g_{\g_{(x,s)}}(t')^{-1}\, H^1(\gamma_{(x,s)}) \,dt'dtds \, + \\
& + \iiint\limits_{[0,1]^3}  g_{\g_{(x,s)}}(t)\t \,B\left( \frac{\partial}{\partial t}J(t,s,x),\frac{\partial }{\partial x} J(t,s,x) \right)g_{\g_{(x,s)}}(t') \,F_A\left( \frac{\partial}{\partial s}\Gamma_x(t',s),\frac{\partial }{\partial t'} \Gamma_x(t',s) \right)\, g_{\g_{(x,s)}}(t')^{-1}\, H^1(\gamma_{(x,s)}) \,dtdt'ds \, .
\end{split}
\end{equation}}
Recall that if  $s,t \in \gl^1(V)$ we have $\beta(s)t=s \beta(t)$. Since $\beta(B)=F_A$ the last two terms of \eqref{ref} can be written as:
{\allowdisplaybreaks
\begin{equation}\label{ref2}
\begin{split}
& \iiint\limits_{[0,1]^3} g_{\g_{(x,s)}}(t) \,F_A \left( \frac{\partial}{\partial s}J(t,s,x),\frac{\partial t}{\partial t} J(t,s,x) \right)g_{\g_{(x,s)}}(t)^{-1} g_{\g_{(x,s)}}(t')  \t \,B\left( \frac{\partial}{\partial x}\Gamma_x(t',s),\frac{\partial }{\partial t'} \Gamma_x(t',s) \right)\,  H^1(\gamma_{(x,s)}) \,dt'dtds \, + \\
& + \iiint\limits_{[0,1]^3} g_{\g_{(x,s)}}(t) \,F_A\left( \frac{\partial}{\partial t}J(t,s,x),\frac{\partial }{\partial x} J(t,s,x) \right) g_{\g_{(x,s)}}(t)^{-1} g_{\g_{(x,s)}}(t') \t \,B\left( \frac{\partial}{\partial s}\Gamma_x(t',s),\frac{\partial }{\partial t'} \Gamma_x(t',s) \right) H^1(\gamma_{(x,s)}) \,dt'dtds = \\
& = \iint\limits_{[0,1]^2} \int_0^{t'}  g_{\g_{(x,s)}}(t) \,F_A \left( \frac{\partial}{\partial s}J(t,s,x),\frac{\partial t}{\partial t} J(t,s,x) \right)g_{\g_{(x,s)}}(t)^{-1} g_{\g_{(x,s)}}(t')  \t \,B\left( \frac{\partial}{\partial x}\Gamma_x(t',s),\frac{\partial }{\partial t'} \Gamma_x(t',s) \right)\,  H^1(\gamma_{(x,s)}) \,dtdt'ds \, + \\
& + \iint\limits_{[0,1]^2} \int_0^{t}  g_{\g_{(x,s)}}(t) \,F_A \left( \frac{\partial}{\partial s}J(t,s,x),\frac{\partial t}{\partial t} J(t,s,x) \right)g_{\g_{(x,s)}}(t)^{-1} g_{\g_{(x,s)}}(t')  \t \,B\left( \frac{\partial}{\partial x}\Gamma_x(t',s),\frac{\partial }{\partial t'} \Gamma_x(t',s) \right)\,  H^1(\gamma_{(x,s)}) \,dt'dtds \, +\\
& + \iint\limits_{[0,1]^2} \int_0^{t}  g_{\g_{(x,s)}}(t) \,F_A \left( \frac{\partial}{\partial t}J(t,s,x),\frac{\partial x}{\partial x} J(t,s,x) \right)g_{\g_{(x,s)}}(t)^{-1} g_{\g_{(x,s)}}(t')  \t \,B\left( \frac{\partial}{\partial s}\Gamma_x(t',s),\frac{\partial }{\partial t'} \Gamma_x(t',s) \right)\,  H^1(\gamma_{(x,s)}) \,dt'dtds\, + \\
& + \iint\limits_{[0,1]^2} \int_0^{t'}  g_{\g_{(x,s)}}(t) \,F_A\left( \frac{\partial}{\partial t}J(t,s,x),\frac{\partial }{\partial x} J(t,s,x) \right) g_{\g_{(x,s)}}(t)^{-1} g_{\g_{(x,s)}}(t') \t \,B\left( \frac{\partial}{\partial s}\Gamma_x(t',s),\frac{\partial }{\partial t'} \Gamma_x(t',s) \right) H^1(\gamma_{(x,s)}) \,dtdt'ds
\end{split}
\end{equation}}
where we have split an integral over $[0,1]^2$ in a sum of integrals over two triangles.
Using the expression of the action of ${\mathfrak gl}^0(\V)$ on $\gl^1(\V)$, namely $b \t s=bs-sb$, for a chain map $b \in \gl^1(\V)$ and homotopy $s$, the second and third terms of \eqref{ref} can be written as:
{\allowdisplaybreaks
\begin{equation}
\label{ref3}
\begin{split}
& \iint\limits_{[0,1]^2} \int_0^t\left ( g_{\g_{(x,s)}}(t') \,F_A\left( \frac{\partial}{\partial s}\Gamma_x(t',s),\frac{\partial }{\partial t'} \Gamma_x(t',s) \right)\, g_{\g_{(x,s)}}(t')^{-1}\, \right) g_{\g_{(x,s)}}(t) \, \t \,B\left( \frac{\partial}{\partial t}J(t,s,x),\frac{\partial }{\partial x} J(t,s,x) \right) \,H^1(\gamma_{(x,s)})\,dt' dt ds \, +\\
& + \iint\limits_{[0,1]^2} \int_0^t \left(g_{\g_{(x,s)}}(t') \,F_A\left( \frac{\partial}{\partial x}\Gamma_x(t',s),\frac{\partial }{\partial t'} \Gamma_x(t',s) \right)\, g_{\g_{(x,s)}}(t')^{-1} \right) \, g_{\g_{(x,s)}}(t)\, \t \,B\left( \frac{\partial}{\partial s}J(t,s,x),\frac{\partial t}{\partial t} J(t,s,x) \right) \,H^1(\gamma_{(x,s)})\,dt' dt ds \, +\\
& - \iint\limits_{[0,1]^2} \int_0^t g_{\g_{(x,s)}}(t) \, \t \,B\left( \frac{\partial}{\partial t}J(t,s,x),\frac{\partial }{\partial x} J(t,s,x) \right) \left ( g_{\g_{(x,s)}}(t') \,F_A\left( \frac{\partial}{\partial s} \Gamma_x(t',s),\frac{\partial }{\partial t'} \Gamma_x(t',s) \right)\, g_{\g_{(x,s)}}(t')^{-1}\, \right)\,H^1(\gamma_{(x,s)})\,dt' dt ds \, + \\
& - \iint\limits_{[0,1]^2} \int_0^t  \, g_{\g_{(x,s)}}(t)\, \t \,B\left( \frac{\partial}{\partial s}J(t,s,x),\frac{\partial t}{\partial t} J(t,s,x) \right) \left(g_{\g_{(x,s)}}(t') \,F_A\left( \frac{\partial}{\partial x}\Gamma_x(t',s),\frac{\partial }{\partial t'} \Gamma_x(t',s) \right)g_{\g_{(x,s)}}(t')^{-1} \right)\, \,H^1(\gamma_{(x,s)})\,dt' dt ds \, .
\end{split}
\end{equation}}
We can see that the sum of \eqref{ref2} and \eqref{ref3} is zero, by using
the fact that if $s,t \in \gl^1(\V)$ then $ \partial(t)s=t \partial(s)$.
Therefore \eqref{ref} simplifies to
$$\frac{d}{dx} H^2(\Gamma_x)=\int_0^1 \int_0^1g_{\g_{(x,s)}}(t)  {\M}\left (\frac{\partial}{\partial x} J(t,s,x),\frac{\partial}{\partial s} J(t,s,x), \frac{\partial}{\partial t} J(t,s,x)\right)g_{\g_{(x,s)}}(t)^{-1} H^1({\g_{(x,s)}}) \, dtds,$$
which leads at once to the statement of the theorem. \qedhere }\\

A local 2-connection  $(A,B)$ is said to be 2-flat if the 2-curvature 3-form ${\M}=d B+ A \wedge^\t B$ vanishes. An immediate corollary of the previous Theorem is:

\begin{Corollary}
 If $(A,B)$ is 2-flat and if $\Gamma_0$ and $\Gamma_1$ are homotopic relative to the boundary, then 
$$H^2(\Gamma_0)=H^2(\Gamma_1) \, .$$
\end{Corollary}

\subsubsection{Manifolds with a group action}\label{mga}

Let $M$ be a manifold. Let $\V=\{V_n, \partial\colon V_n \to V_{n-1}\}_{n \in \mathbb{Z}}$ be a chain complex of vector spaces. Suppose that $\V$ is of finite type. Consider a local 2-connection  $(A,B)$ with values in $\GL(\V)=(\beta\colon \gl^1(\V) \to \gl^0(\V),\t)$.  
We therefore have, see Definition \ref{2hol}, an $\Aut(\V)$-valued two dimensional holonomy $ \H=(H^1,H^2)$.

Suppose that we have a left action  of a (discrete) group $G$ on $M$ by diffeomorphisms, free (stabilisers are trivial) and properly discontinuous. {We denote the action as $g \mapsto L_g$ and $L_g(x)=g(x)$, for $x \in M$.} Consider the quotient manifold $M/G$. Denote the  projection map as  $M \ni x \mapsto [x] \in M/G$.  By standard covering space theory, applied to the covering $M \to M/G$, any path $[x] \ra{[\gamma]} [y]$ in $M/G$ can be lifted to a path $x\ra{\gamma} y$ in  $M$, and any two such lifts are related by the action of a unique element of $G$. Similarly, if $([x] \ra{[\gamma_0]} [y])\stackrel{[\Gamma]}{\implies} ( [x] \ra{[\gamma_1]} [y])$ is a 2-path in $M/G$ then  it can be lifted to a 2-path $\Gamma$ in $M$, connecting lifts  $\gamma_0$ and $\gamma_1$ of  $[\gamma_0]$ and $[\gamma_1]$, with $\gamma_0(0)=\gamma_1(0)$ and $\gamma_0(1)=\gamma_1(1)$, and any two lifts like these are related by the action of a unique element of $G$.

Let us consider also a left action of $G$ on $\V$ by chain-complex maps, $G \ni g \mapsto \tau_g \in {\rm GL}^0(\V)$. We thus have a right action $R$ of $G$ on the 2-category $\Aut(\V)$, by 2-functors. This has the form: \begin{equation}\label{actV}
\begin{split}
R_g(f) & =  \tau_{g^{-1}}\circ f \circ \tau_g\,\,, \textrm{ for each } f\in \gl^0(\V) \textrm{ and each } g \in G \, , \\
R_g(s) & =  \tau_{g^{-1}}\circ s \circ \tau_g\,\,, \textrm{ for each } s \in \gl^1(\V) \textrm{ and each } g \in G \, .
\end{split}
\end{equation}

Let us define a 2-category $\Aut(\V,M,G)$. The objects are the points of the quotient manifold $M/G$. A morphism $[x] \to [y]$ is an equivalence class of triples of the form $(x',f,y')$, where $x' \in [x]$, $y' \in [y]$ and $f$ is a chain map $\V \to \V$. Two of these are said to be equivalent if they are related by the transformation:
$$(x,f,y)\mapsto (g(x),\tau_g f \tau_{g^{-1}},g(y))\doteq L_g(x,f,y).$$
The composition of $(x,f,y)\colon [x] \to [y]$ and $(y',f',z)\colon [y] \to [z]$ is: 
$$(x,f\tau_{g^{-1}} f'\tau_{{g}},g^{-1}(z)) \, ,$$
where $y'=g(y)$. This is independent of the representatives chosen. Indeed if (here $a,b \in G$):
$$(x',f_1,y'')= (a(x),\tau_a f \tau_{a^{-1}} ,a(y))$$
and 
$$(y''',f_1',z')\mapsto (b(y'),\tau_b f' \tau_{b^{-1}},b(z))$$
then $y'''=b(y')=b(g(y))=(bga^{-1})(y'')$, and the composition of these two morphisms is \begin{align*}
\left( a(x),\tau_a f \tau_{a^{-1}} \tau_{(bga^{-1})^{-1}}  \tau_b f' \tau_{b^{-1}}\tau_{(bga^{-1})},(bga^{-1})^{-1}(b(z))\right)&= \left( a(x),\tau_a f \tau_{g^{-1}} f' \tau_{g} \tau_{a^{-1}},ag^{-1}(z)\right)\\
&= L_a\left((x,f\tau_{g^{-1}} f'\tau_{{g}},g^{-1}(z))\right) \, .
\end{align*}
It is easy to check that the composition of 1-morphisms is associative.

Analogously, the set of 2-morphisms $([x] \ra{(x',f',y')} [y]) \implies ([x] \ra{(x'',f'',y'')} [y]) $ is the set of all matrices
$$\begin{pmatrix}
(x',f'',y')\\
   s\\
(x'',f',y'')
  \end{pmatrix}
 $$
where $s\in \gl^1(\V)$ is such that $\beta(s)+f'=\tau_{g^{-1}} f'' \tau_g$, where $g(x')=x''$, considered up to the equivalence relation:
$$\begin{pmatrix}
(x',f'',y')\\
   s\\
(x'',f',y'')
  \end{pmatrix}=\begin{pmatrix}
L_a(x',f'',y')\\
   \tau_a s \tau_{a^{-1}}\\
L_a(x'',f',y'')
  \end{pmatrix}
 $$
for arbitrary $a \in G$. These compose horizontally and vertically, analogously to 1-morphisms.

\begin{Theorem}\label{hcov}
Let $M$ be a manifold and $\V$ be a chain complex of vector spaces. Let $(A,B)$ be a {$\GL(\V)$-valued} local 2-connection {in $M$}. Suppose that there is a free and properly discontinuous left action $G \ni g \mapsto L_g \in \diff(M)$ of a discrete group $G$ on $M$. Suppose also that we have a {left} action of $G$ on $\V$ by chain-complex maps,  made into a right action of $G$ on $\Aut(\V)$ by 2-functors, by using \eqref{actV}. If these actions satisfy:
\begin{equation}\label{quo}
(L_{g^{-1}})^*(A)=R_g(A) \quad \mbox{ and } \quad
(L_{g^{-1}})^*(B)=R_g(B) \, ,
\end{equation}
for each $g \in G$, then the 2-dimensional holonomy of $(A,B)$ with values in $\Aut(\V,M)$ descends to a 2-dimensional holonomy in the manifold $M/G$  with values in the 2-category $\Aut(\V,M,G)$.
\end{Theorem}
{\proof 
This essentially follows from Remarks \ref{cov1} and \ref{cov2} and standard covering space theory. If we have a path $[x]\ra{[\gamma]}  [y]$ in $M/G$, we lift it to a path $x'\ra{\gamma} y'$. Put $$H^1([\gamma])=[x]\ra{(x',\stackrel{(A,B)}{H^1}(\gamma),y')} [y] \, .$$
If we choose another lift $L_g(\gamma)$ of $[\gamma]$ then
\begin{align*}
\stackrel{(A,B)}{H^1}(L_g(\gamma))=\stackrel{L_g^*(A,B)}{H^1}(\gamma)=\stackrel{R_{g^{-1}}(A,B)}{H^1}(\gamma)=\tau_{{g}} \stackrel{(A,B)}{H^1}(\gamma) \tau_{{g}^{-1}}\,\,.
\end{align*}
Thus
\begin{align*}
[x]\ra{(L_g(x'),H^1(L_g(\gamma)),L_g(y'))} [y]=[x]\ra{(L_g(x'), \tau_{{g}} {H^1}(\gamma) \tau_{{g}^{-1}} ,L_g(y'))} [y]=[x] \ra{(x',H^1(\gamma),y')} [y]\,,
\end{align*}
and this shows that $H^1(\gamma)$ is well defined. The same analysis applies to 2-paths. 
\qedhere}

\begin{Definition}
A local 2-connection  satisfying \eqref{quo} will be called a $G$-covariant local 2-connection.
\end{Definition}

\noindent
Equations $\eqref{quo}$ need only to be checked on  group generators of $G$. Indeed if \eqref{quo} holds  for $g$ and $h$ in $G$, then:
\begin{align*}
 &(L_{(gh)^{-1}})^*(A)=(L_{(h^{-1}g^{-1})})^*(A)=(L_{h^{-1}}L_{g^{-1}})^*(A)=
(L_{g^{-1}})^*\big((L_{h^{-1}})^*(A)\big)=(L_{g^{-1}})^*\big(R_h(A)\big)=R_h\big(R_g(A)\big)=R_{gh}(A)\,,
\end{align*}
and the same for $B$. 

\subsection{Categorical representations of differential crossed modules on chain-complexes of vector spaces}

\subsubsection{Definition of a categorical representation}
Similar constructions appear for example in \cite{BM,E,Wo}.
A categorical representation of a differential crossed module $\lG=(\beta\colon  \lh \to \lg, \t)$ on a chain-complex of vector spaces $\V=(V_i,\partial)_{i\in\Z}$ is by definition a morphism of differential crossed modules $\rho=(\rho^1,\rho^0)\colon\lG \to \GL(\V)$, see Section \ref{dcmcc}.
We thus have Lie algebra maps 
\begin{alignat*}{2}
\rho^0 & : \lg \to \gl^0(\V) \, ,  \quad &\lg \ni X &\mapsto \rho^0_X \in \gl^0(\V) \\
\rho^1 & : \lh \to \gl^1(\V) \, ,  \quad &\lh \ni v &\mapsto \rho^1_v \in \gl^1(\V) \, .
\end{alignat*}
These are to satisfy, for each $X,Y \in \lg$ and $u,v \in \lh$:
\begin{itemize}
\item[]
\begin{inparaenum}[i)]
\item $\rho^0_{[X,Y]}=\{\rho^0_X,\rho^0_Y\} \, , \qquad $
\item $\rho^1_{[u,v]}=\{\rho^1_u,\rho^1_v\} \, , \qquad $
\item $\beta(\rho^1_u)=\rho^0_{\beta(u)}\, , \qquad $
\item $\rho^1_{X \t u}=\rho^0_X \t \rho^1_u\, . \qquad $
\end{inparaenum}
\end{itemize}
We will frequently drop the indices on $\rho^0$ and $\rho^1$ when they are obvious from the context. 
\subsubsection{The adjoint categorical representation of a differential crossed module}
A very simple and natural example, which will play a major role in this paper,  of a categorical representation is the adjoint categorical representation of a differential crossed module $\lG=(\partial\colon  \lh \to \lg, \t)$ on its underlying (short) complex of vector spaces $\underline{\lG}=(\partial\colon \lh \to \lg)$. In the case of Lie crossed modules this appeared in \cite{Wo}. 

The adjoint categorical representation $\rho=(\rho^1,\rho^0)$ on generic elements $X,Y \in \lg$ and $v\in \lh$ has the  form: 
\begin{align}
\label{rho0ad}
\rho^0_X(v,Y) & = ( X \t v, [X,Y]) \\
\label{rho1ad} 
\rho^1_v(X) & = -X \t v \, ,
\end{align}
where we characterize the chain map of short complexes $\rho^0_X$ by its underlying pair of maps, and $\rho^1_v:\lg\to\lh$ as it should.  Clearly relation i) above is true. Also for each $u,v \in \lh$ and $X \in \lg$:
\begin{align*}
\rho^1_{[v,u]}(X)=-X\t[v,u]
\end{align*}
and, since the underlying complex $(\partial\colon  \lh \to \lg)$  of $\lG$ has length two, thus $\rho^1_u \rho^1_v=\rho^1_v \rho^1_u=0$:
\begin{align*}
\{\rho^1_v,\rho^1_u\}(X)&=(\rho^1_v\partial\rho^1_u-\rho^1_u\partial\rho^1_v)(X)=-\rho_v^1(\partial(X \t u))+\rho_u^1(\partial(X \t v)) \\
& = \partial(X \t u) \t v- \partial(X \t v) \t u=[X,\partial(u)] \t v-[X,\partial{(v)}] \t u\\
& = X \t \partial(u) \t v- \partial(u) \t X \t v-X \t \partial{(v)} \t u+ \partial{(v)} \t X \t u\\
& = X \t [u,v]-[u,X \t v]-X \t [v,u]+[v,X \t u].
\end{align*}
Thus, by using $X\t[v,u]=[X \t v,u]+[v,X \t u]$, we see that $\rho^1_{[v,u]}=\{\rho^1_v,\rho^1_u\}$, which proves ii). To prove iii) note that for each $u,v \in \lh$ and $X \in \lg$:
$$ \rho^0_{\partial(u)}(v,X)=(\partial(u) \t v, [\partial(u), X])=([u,v], [\partial(u),X]),$$
whereas:
$$ \beta(\rho^1_u)(v,X)=\big(\rho^1_u(\partial{(v)}),\partial(\rho^1_u(X))\big)=(-\partial{(v)} \t u,-\partial(X \t u)\big)=-([v,u],[X,\partial(u)]) \, .$$
Finally we have for $X,Y \in \lg$ and $u \in \lh$: $$\rho^1_{X \t u}(Y)=-Y \t X \t u.$$
On the other hand: 
$$
(\rho^0_X \t \rho^1_u)(Y) = (\rho^0_X  \rho^1_u)(Y)-( \rho^1_u\rho^0_X )(Y) = 
-X \t Y \t u+[X,Y] \t u=-Y \t X \t u \, .
$$
by definition of a Lie algebra representation, and this proves iv).

\subsection{Tensor product of complexes}
\subsubsection{The symmetric monoidal category of chain complexes}\label{smccc}

Most of the  framework presented here is quite classical. See for example \cite{D} for the tensor product of complexes and also \cite{Sch} for graded tensor calculus.

Let $\V^j=\{V_i^j, \partial_i\colon V_i^j \to V_{i-1}^j\}_{i \in \mathbb{Z}}$ be a family of  complexes of vector spaces, indexed by a $j \in \{1, \dots,n\}$, for a positive integer $n$. Let us recall the  definition of the chain complex tensor product $$\btn_{j=1}^n \V^j=\{W_k, \partial \colon W_k \to W_{k-1}\}_{k \in \mathbb{Z}}.$$
This complex is such that the vector space of elements of degree $k$ is:
$$W_k=\bigoplus_{i_1+i_2+\dots +i_n=k} V_{i_1}\tn V_{i_2} \tn \dots \tn V_{i_n}   \,\,, $$
with boundary, for $x_i^j \in V_i^j$, where $i \in \mathbb{Z}$ and $j \in  \{1,\dots,n\}$:
\begin{multline}
\label{tpc}
\partial\big(x_{i_1}^1 \tn x_{i_2}^2 \tn x_{i_3}^3 \tn \dots \tn x_{i_n}^n\big)=
\partial(x_{i_1}^1) \tn x_{i_2}^2 \tn x_{i_3}^3 \tn \dots \tn x_{i_n}^n+(-1)^{i_1} x_{i_1}^1 \tn \partial(x_{i_2}^2) \tn x_{i_3}^3 \tn \dots \tn x_{i_n}^n + \\ 
+(-1)^{i_1+i_2}x_{i_1}^1 \tn x_{i_2}^2 \tn \partial(x_{i_3}^3) \tn \dots \tn x_{i_n}^n+ \dots +(-1)^{i_1+i_2+i_3 + \dots +i_{n-1}}x_{i_1}^1 \tn x_{i_2}^2 \tn x_{i_3}^3 \tn \dots \tn \partial(x_{i_n}^n).
\end{multline}
The tensor product of chain complexes is associative up to (the obvious) isomorphism of complexes. Given chain complexes $\U=\{U_i, \partial\colon U_i \to U_{i-1}\}_{i\in\Z}$ and  $\V=\{V_i, \partial\colon V_i \to V_{i-1}\}_{i\in\Z}$ there is an isomorphism $\U \btn \V \to \V \btn \U$; it has the form, for $u_i \in U_i$ and $v_j \in V_j$:
$$u_i \tn v_j \mapsto (-1)^{ij} v_j \tn v_i . $$
This gives the category of chain complexes  and (degree zero) chain-maps  the structure of a symmetric monoidal category. More generally, for each $k \in \{1,\dots, n-1\}$, we have an isomorphism $\tau_{X_k}$, of chain complexes, for each transposition, $X_k$ exchanging $k$ and $k+1$. Here 
$$\V^1 \btn \V^2 \btn \dots \btn \V^k \btn \V^{k+1} \btn \dots\btn \V^n \ra{\tau_{X_k}} \V^1 \btn \V^2 \btn \dots \btn \V^{k+1} \btn \V^k  \btn \dots \btn \V^n, $$
where if $x_i^j \in V^j_i$ we have $$x_{i_1}^1 \tn x_{i_2}^2 \tn x_{i_3}^3 \tn \dots \tn x_{i_k}^\tn k x_{i_{k+1}}^{k+1}  \tn \dots \tn  x_{i_n}^n\stackrel{\tau_{X_k}}{\longmapsto} (-1)^{i_k i_{k+1}} x_{i_1}^1 \tn x_{i_2}^2 \tn x_{i_3}^3 \tn \dots \tn  x_{i_{k+1}}^{k+1}\tn x_{i_k}^k   \tn \dots \tn  x_{i_n}^n .$$
These satisfy the well known symmetric group $S_n$ relations, namely: $$\tau_{X_k} \tau_{X_{k+1}} \tau_{X_k}=\tau_{X_{k+1}} \tau_{X_k}\tau_{X_{k+1}}.$$
In particular,  given a permutation $\s$ of the symmetric group $S_n$ we can define a chain isomorphism $\tau_\s\colon \V^{\btn n} \to \V^{\btn n} $ (where $\V^{\btn n}$ denotes the tensor product of $\V$ with itself $n$ times)   as:
\begin{equation}\label{actionSn}
 x_{i_1}^1 \tn x_{i_2}^2 \tn x_{i_3}^3 \tn \dots \tn x_{i_n}^n \mapsto \epsilon(\s,i_1,\dots i_n)\,\,x_{i_{\s^{-1}(1)}}^{\s^{-1}(1)} \tn x_{i_{\s^{-1}(2)}}^{\s^{-1}(2)} \tn x_{i_{\s^{-1}(3)}}^{\s^{-1}(3)} \tn \dots \tn x_{i_{\s^{-1}(n)}}^{\s^{-1}(n)}, 
\end{equation}
where
$$\epsilon(\s,i_1,\dots i_n) =\prod_{\big\{k,l\in \{1,\dots,n\} \textrm{ such that } k<l \textrm{ and } \sigma(k) > \sigma(l)\big\}}(-1)^{i_ki_l}.$$
Clearly if $\sigma,\sigma'$ are elements of the symmetric group $S_n$, it holds
$\tau_{\s\s'}=\tau_{\s} \tau_{\s'}$. We thus have a representation $\tau$ of $S_n$ on $\V^{\btn n}$ by degree zero chain maps, which will have a primary role later.

Given complexes $\U=\{U_i, \partial\colon U_i \to U_{i-1}\}_{i \in \mathbb{Z}}$ and  $\V=\{V_i, \partial\colon V_i \to V_{i-1}\}_{i \in \mathbb{Z}}$ we define $\Hom^n(\U,\V)$, the space of maps of degree $n$, as being the space of sequences $a=(a_i)$ of linear maps $a_i\colon U_i \to V_{i+n}$, called $n$-fold homotopies. There exists a chain complex of vector spaces: $$\cHom(\U,\V)=\{\Hom^n(\U,\V), \beta \colon \Hom^n(\U,\V) \to \Hom^{n-1}(\U,\V)\}_{n \in \mathbb{Z}} ,$$ 
where if $a\colon \U \to \V$ is a map of degree $n$ we define:
\begin{equation}\label{hc}\beta(a)=\partial a-(-1)^n a \partial.
\end{equation}

\noindent
Given complexes $\V^i$ and $\cW^i$, where $i=1, \dots, n$, there exists a degree zero chain  map:
$$ \overline{\bigotimes}_{i=1}^{\; n} \cHom (\V^i,\cW^i) \stackrel{\kappa}{\longrightarrow} \cHom (\btn_{i=1}^n \V^i,\btn_{i=1}^n \cW^i ), $$
sending $f^1 \tn f^2 \tn \dots \tn f^n$ to $f^1 \btn f^2 \btn \dots \btn f^n,$ 
for each sequence of maps $f^k \colon \V^k \to \cW^k$ of degree $m_k$ $(k =1,\dots, n)$.  Here, if $x^k$ are degree $i_k$ elements of $\V^k$, we put:
\begin{equation}
\label{signchi}
(f^1 \btn f^2 \btn \dots \btn f^n)(x^1 \otimes x^2 \otimes \dots \otimes x^n) = (-1)^{\chi(\{m_k\},\{i_k\})} f^1(x^1) \otimes f^2(x^2) \otimes \dots \otimes f^n(x^n)
\end{equation}
where 
$$ \chi(\{m_k\},\{i_k\}) =i_1(m_2+\dots +m_n) +i_2(m_3 + \dots + m_n) +\dots + i_{n-1} m_n \, . $$
The following result will be crucial later.

\begin{Lemma}
\label{cov4}
For any permutation $\s \in S_n$ it holds that
\begin{equation}\label{cov5}
\tau_{\sigma} \,\circ\, \big( \kappa (f^1\tn \ldots \tn f^n)\big) \,\circ \, \tau_{\sigma}^{-1} = \kappa\left(\tau_\sigma(f^1 \tn \ldots \tn f^n)\right) \, ,
\end{equation}
as degree $m_1+m_2+\dots+m_n$ maps $\btn_{i=1}^n \V^i \to\btn_{i=1}^n \cW^i$.
\end{Lemma}

\noindent 
A word on notation: $\tau_\s$ is used here to denote the morphisms associated to the permutation $\s$, however in three different complexes, namely, by order of appearance in the previous formula: $\btn_{i=1}^n \cW^i$, $\btn_{i=1}^n \V^i$ and $\overline{\otimes}_{i=1}^{\; n} \cHom (\V^i,\cW^i) $.

\proof{It is enough to prove the result for the transpositions $X_k$ exchanging $k$ and $k+1$. Put $\tau_k=\tau_{X_k}$, thus:
$$\tau_k(x^1\ot \ldots \ot x^k \ot x^{k+1} \ot \ldots \ot x^n ) = (-1)^{i_ki_{k+1}} \, (x^1\ot\ldots \ot x^{k+1}\ot x^k\ot \ldots \ot x^n ) \, .$$
We then compute the left-hand-side of \eqref{cov5} on $x^1\ot \ldots \ot x^k \ot x^{k+1} \ot \ldots \ot x^n$ 
\begin{equation*}
\begin{split}
& \tau_k \, (f^1\btn \ldots \btn f^n) \, \tau_k^{-1}(x^1\ot \ldots \ot x^k \ot x^{k+1} \ot \ldots \ot x^n)  = \\
& = (-1)^{i_ki_{k+1}} \, \tau_k \, (f^1\btn \ldots \btn f^n) (x^1\ot\ldots \ot x^{k+1}\ot x^k\ot \ldots \ot x^n ) \\
& = (-1)^{i_k i_{k+1}} \, (-1)^{\chi(\{m_k\},\{i_k\})} \, (-1)^{m_{k+1}i_{k+1} + m_{k+1}i_k} \, \tau_k \, \left( f^1(x^1) \tn \ldots \tn f^k(x^{k+1}) \tn f^{k+1}(x^k) \tn \ldots \tn f^n(x^n) \right) \\
& = (-1)^{i_k i_{k+1}} \, (-1)^{\chi(\{m_k\},\{i_k\})} \, (-1)^{m_{k+1}i_{k+1} + m_{k+1}i_k} \, (-1)^{(m_k+i_{k+1})(m_{k+1}+i_k)} \left( f^1(x^1) \tn \ldots \tn f^{k+1}(x^k)\tn f^k(x^{k+1}) \tn  \ldots \tn f^n(x^n)\right )  \\
& = (-1)^{\chi(\{m_k\},\{i_k\})} \, (-1)^{m_ki_k + m_{k+1}i_k + m_k m_{k+1}} \, \big( f^1(x^1) \tn \ldots \tn f^{k+1}(x^k)\tn f^k(x^{k+1}) \tn  \ldots \tn f^n(x^n)\big ) \, .
\end{split}
\end{equation*}
A shorter computation for the right hand side gives:
\begin{equation*}
\begin{split}
\kappa \big(\tau_\sigma(f^1 \tn \ldots & \tn f^n)\big) (x^1\ot \ldots \ot x^k \ot x^{k+1} \ot \ldots \ot x^n) = \\ 
 & = (-1)^{m_k m_{k+1}} \, (f^1\btn\ldots\btn f^{k+1}\btn f^k\btn\ldots\btn f^n)(x^1\tn \ldots \tn x^n) \\
 & = (-1)^{m_k m_{k+1}} \, (-1)^{\chi(\{m_k\},\{i_k\})} \, (-1)^{m_{k+1} i_k + m_k i_k} \big( f^1(x^1) \tn \ldots \tn f^{k+1}(x^k)\tn f^k(x^{k+1}) \tn  \ldots \tn f^n(x^n) \big) \, . \qedhere
\end{split}
\end{equation*}
}
\subsubsection{Insertion maps}
\label{im}
Let $\V=(\dots \ra{\partial } V_i \ra{\partial} V_{i-1} \ra{\partial } \dots)$ be a chain complex of vector spaces. Then $\GL(\V^{\btn n})$ is a differential crossed module. We can also perform the  tensor product of the underlying chain complex of $\GL(\V)$ with it self $k$-times. The latter is not a differential crossed module, however for every $n\geq k$ it maps to $\GL(\V^{\btn n})$ naturally. This paragraph is devoted to carefully describe this map, which will have a major role in our construction.

We first introduce some notation. Given $\V$, we denote $(\V)_2\doteq \V_2=(V_1 \ra{\partial} V_0)$, which is a chain complex of length two. Let also $\overline{\V_2}$ be the chain-complex $\overline{\V_2}=(V_1/\partial(V_2) \ra{\partial} V_0)$. We clearly have functors $(\, . \,)_2$ and  $\overline{(\, . \, )_2}$ from the category of chain complexes to the category of length two chain complexes (of vector spaces). Given another chain complex of vector spaces $\V'$, we will normally abbreviate $\overline{f_2}\colon \overline{\V_2}\to \overline{\V'_2} $ as $\overline{f}$ for a chain map $f\colon \V \to \V'$. For every differential crossed module $\lG= (\beta\colon \lh \to \lg, \t)$, let  $\underline{\lG}=(\beta\colon \lh \to \lg)$ be its underlying chain complex of vector spaces. 

Let $k$ and $n$ be positive integers with $k \leq n$.   Let $\psi \colon \{1,\dots,k\} \to \{1,\dots, n\}$ be an injective map. The goal of this section is to define and study the properties of the chain map of length two chain complexes:
$$
F_\psi\colon \overline{(\underline{\GL(\V)}^{\btn k })_2 }\to  \underline{\GL(\V^{\btn n})}\, . 
$$
In each of these complexes, the single boundary map will be denoted by $\beta$. We start by the case when $\psi$ is increasing, $\psi(i+1)> \psi{(i)}$ for every $i$. In this case we define $F_\psi\colon (\underline{\GL(\V)}^{\btn k })_2 \to  \underline{\GL(\V^{\btn n})} $ as:
$$F_\psi (f_1 \tn \dots \tn f_k)= \id\btn\ldots \btn f_1\btn\ldots \btn f_k\btn\ldots\btn\id \doteq  \kappa(\id\tn\ldots \tn f_1\tn\ldots \tn f_k\tn\ldots\tn\id),$$
where we have inserted $f_i$ in the $\psi{(i)}^{\rm th}$ position of the $n$-fold tensor product $\V^{\btn n}$, inserting identities in the remaining positions. Each $f_i$ is to be either a chain-map  $\V \to \V$ or a homotopy of $\V$, up to 2-fold homotopy. Only one homotopy up to 2-fold homotopy can appear in the list $f_1, \dots ,f_k$ (since we are only concerned with what happens in degree 0 and 1).

Some details need checking. First, let us show that $F_{\Psi}$ is well defined. Let $i \in  \{1,\dots,k\}$. Suppose $f_j, j \neq i$ are chain maps $\V \to \V$ and that $s_i=s_i'+\beta(a_i)$, where $s_i,s_i' \in \Hom^1(\V)$ and $a_i \in \Hom^2(\V)$. In this case we have:
\begin{align*}
F_\psi (f_1 \tn \dots \tn s_i \tn \dots \tn f_k)=F_\psi (f_1 \tn \dots \tn s_i' \tn \dots \tn f_k)+ \beta(\id\btn\ldots \btn f_1\btn\ldots \btn a_i \btn \dots \btn f_k\btn\dots \btn\id)
\end{align*}
since, given that all maps $f_j$ are chain maps, thus $\beta(f_j)=0$, by \eqref{tpc}:
\begin{align*}
 \beta(\id\btn\ldots \btn f_1\btn\ldots \btn a_i \btn \dots \btn f_k\btn\dots \btn\id)&=\beta\big(\kappa(\id\tn\ldots \tn f_1\tn\ldots \tn a_i \tn \dots \tn f_k\tn\dots \tn\id)\big)\\&=\kappa\big(\beta(\id\tn\ldots \tn f_1\tn\ldots \tn a_i \tn \dots \tn f_k\tn\dots \tn\id)\big)\\
&=\kappa\big(\id\tn\ldots \tn f_1\tn\ldots \tn \beta(a_i) \tn \dots \tn f_k\tn\dots \tn\id\big)\\&=
\kappa\big(\id\tn\ldots \tn f_1\tn\ldots \tn s_i-s_i' \tn \dots \tn f_k\tn\dots \tn\id\big)\\
&=
\id\btn\ldots \btn f_1\btn\ldots \btn s_i-s_i' \btn \dots \btn f_k\btn\dots \btn\id,
\end{align*}
and we conclude. Next, $F_\psi$ is a chain-map since
\begin{align*}
\beta\big(F_\psi (f_1 \tn \dots \tn s_i \tn \dots \tn f_k)\big)&=\beta \big(\kappa(\id\tn\ldots \tn f_1\tn\ldots \tn s_i \tn \dots \tn f_k\tn\dots \tn\id)\big)\\
&=\kappa \big(\beta(\id\tn\ldots \tn f_1\tn\ldots \tn s_i \tn \dots \tn f_k\tn\dots \tn\id)\big)\\
&=\kappa(\id\tn\ldots \tn f_1\tn\ldots \tn \beta(s_i) \tn \dots \tn f_k\tn\dots \tn\id)\\
&=F_\psi(f_1 \tn \dots \tn \beta(s_i) \tn \dots \tn f_n).
\end{align*}
It is also true that $F_\psi$ descends to a map of length two chain complexes of vector spaces:
$$F_\psi\colon (\overline{\underline{\GL(\V)}^{\btn k })_2} \to  \underline{\GL(\V^{\btn n})}. $$
Indeed, consider a general element of the degree 2 component of $ {\underline{\GL(\V)}^{\btn k }}$. It is a linear combination of elements of the form:  $f_1 \tn \dots \tn s_i \tn \dots \tn s_j\tn \dots \tn f_k$, where only two of the terms in this tensor product are homotopies (up to 2-fold homotopy)  namely $s_i$ and $s_j$, the remaining ones being chain-maps.
Then $$F_\psi\big( \beta( f_1 \tn \dots \tn s_i \tn \dots \tn s_j\tn \dots \tn f_k)\big)=0.$$
 In other words, by \eqref{tpc}:
$$F_\psi(  f_1 \tn \dots \tn \beta(s_i)\tn \dots \tn s_j \tn \dots \tn f_k)=F_\psi(  f_1 \tn \dots \tn s_i\tn \dots \tn \beta(s_j)\tn \dots \tn f_k).$$
This is because, by definition and \eqref{tpc}, and inserting $f_l$ in the $\psi(l)^{\rm th}$ position of the tensor product we have:
\begin{align*} \beta &\big(\kappa( \dots \tn f_1 \tn \dots \tn s_i\tn \dots \tn s_j \tn \dots\tn f_k \dots )\big ) = \kappa \big(\beta( \dots \tn f_1 \tn \dots \tn s_i\tn \dots \tn s_j \tn \dots\tn f_k \dots )\big ) = \\
&= \kappa \big( \dots \tn f_1 \tn \dots \tn \beta(s_i)\tn \dots \tn s_j \tn \dots\tn f_k \dots \big )- \kappa \big( \dots \tn f_1 \tn \dots \tn s_i\tn \dots \tn \beta(s_j) \tn \dots\tn f_k \dots \big )= \\
&=F_\psi(  f_1 \tn \dots \tn \beta(s_i)\tn \dots \tn s_j \tn \dots \tn f_k)-F_\psi(  f_1 \tn \dots \tn s_i\tn \dots \tn \beta(s_j)\tn \dots \tn f_k) \, .
\end{align*}

To define $F_{\Psi}$ for a generic injective map $\psi\colon  \{1,\dots,k\} \to \{1,\dots,n\}$, not necessarily increasing, we observe that $\Psi$ {factors} uniquely as $\psi'\circ \s$, where $\psi'$ is increasing and $\s \in S_k$, the symmetric group of degree $k$. We then put
\begin{equation}\label{cfd2}
F_\psi\doteq F_{\psi'} \circ \overline{\tau_\s} .
\end{equation}
Here $\tau_\s\colon \underline{\GL(\V)}^{\btn k } \to \underline{\GL(\V)}^{\btn k }$ is the chain map associated to $\sigma$, see \eqref{actionSn}, and $\overline{\tau_\s}$ is  its projection onto a map:  
$$
\overline{\tau_\s}\colon  \overline{(\underline{\GL(\V)}^{\btn k })_2} \to\overline{(\underline{\GL(\V)}^{\btn k })_2} \, .
$$

\begin{Lemma}
\label{refC}
Let $k,k' \leq n$. Given  $\psi\colon\{1,\dots, k\} \to \{1, \dots, n\}$ and
   $\psi'\colon\{1,\dots, k'\} \to \{1, \dots, n\}$, if $$\psi(\{1,\dots, k	\}) \cap \psi'(\{1,\dots, k'\})=\emptyset,$$ then in the differential crossed module $\GL(\V^{\btn n})$ we have:
\begin{align*}
\{ F_\psi(f_1 \tn \dots \tn f_k),F_{\psi'} (g_1\tn \dots \tn g_k') \}&=0,\\
 F_\psi(f_1 \tn \dots \tn f_k)\t F_{\psi'} (g_1\tn \dots \tn t_j \tn\dots \tn g_k') &=0,\\
\{ F_\psi(f_1 \tn \dots \tn s_i\tn \dots \tn f_k),F_{\psi'} (g_1\tn \dots \tn t_j \tn\dots \tn g_k') \}&=0.
\end{align*}
\end{Lemma}
\proof The first two assertions follow from the definition of $\GL(\V^{\btn n})$, given that in this case  $F_\psi(f_1 \tn \dots \tn f_k)$ and $F_{\psi'} (g_1\tn \dots \tn g_k') $ commute as degree 0 maps, and so do $F_\psi(f_1 \tn \dots \tn f_k)$ and $ F_{\psi'} (g_1\tn \dots \tn t_j \tn\dots \tn g_k') $. Also 
$\beta(F_\psi(f_1 \tn \dots \tn s_i\tn \dots \tn f_k))=F_\psi(f_1 \tn \dots \tn \beta( s_i)\tn \dots \tn f_k) $ and $F_{\psi'} (g_1\tn \dots \tn t_j \tn\dots \tn g_k')$ do commute as maps $\V^{\btn n} \to \V^{\btn n}$. Therefore, by the differential crossed module rules:
\begin{align*}
 \{ F_\psi(f_1 \tn \dots \tn s_i\tn \dots \tn f_k),F_{\psi'}& (g_1\tn \dots \tn t_j \tn\dots \tn g_k') \}\\&=\beta\big( F_\psi(f_1 \tn \dots \tn s_i\tn \dots \tn f_k)\big) \t F_{\psi'} (g_1\tn \dots \tn t_j \tn\dots \tn g_k') =0 \, . \qedhere
\end{align*}
\noindent
Let now $\psi\colon \{1,\dots,k\} \to \{1,\dots,n\}$ be an injective map, and $\psi=\psi'\circ \s$ the factorization as before, $\s \in S_k$ and $\psi'$ increasing. Suppose that $\s'\in S_n$. Then $\s'\circ \psi'=\psi'' \circ \s''$, uniquely, where $\psi''$ is increasing and $\s'' \in S_k$. Note that $\psi'' \s'' \s=\s' \psi' \s=\s' \psi$. This notation is used in the proof of the following lemma.
\begin{Lemma}
\label{ref4}
As chain maps $(\overline{\underline{\GL(\V)}^{\btn k })_2} \to  \underline{\GL(\V^{\btn n})}$, we have:
$$\tau_{\s'}\circ F_\psi \circ \tau_{\s'^{-1}}=F_{\sigma' \psi}.$$
\end{Lemma}
\proof{We start by supposing that $\psi'$ is increasing. Let us see that
$\tau_{\s'} \circ F_{\psi'} \circ \tau_{\s'^{-1}}=F_{\sigma' \psi'}$ in this case.
Let us be given $f_1 \tn \dots \tn f_k$ in $\overline{(\underline{\GL(\V)}^{\btn k })}_2$, where $f_i$ is either a chain map or a homotopy, up to 2-fold homotopy. Then, by definition and Lemma \ref{cov4}:
\begin{align*}
\tau_{\s'}\circ  F_{\psi'}(f_1 \tn \dots \tn f_k) \circ \tau_{\s'^{-1}}&\doteq\tau_{\s'}\circ\left (\id\btn\ldots \btn f_1\btn\ldots \btn f_k\btn\ldots\btn\id\right)\circ\tau_{\s'^{-1}}\\
&=\kappa\big(\tau_{	\s'}\left(\id\tn\ldots \tn f_1\tn\ldots \tn f_k\tn\ldots\tn\id\right)\big)\\
&=F_{\psi''} \tau_{\s''}(f_1\tn \dots\tn f_k)\doteq F_{\psi''\s''}(f_1\tn \dots\tn f_k)\\
&=F_{\s'\psi'}(f_1\tn \dots\tn f_k)\,,
\end{align*}
where in the third identity we used \eqref{actionSn}. Now, for not necessarily increasing $\psi$, we have:
\begin{align*}
  \tau_{\s'}\circ F_\psi (f_1 \tn \dots  \tn  f_n) \circ \tau_{\s'^{-1}}&\doteq\tau_{\s'} \circ \big(F_{\psi'} \left( \tau_{\s} (f_1 \tn \dots \tn f_n)\right)\big) \circ \tau_{\s'^{-1}}\\
&=F_{\psi''} \left( \tau_{\s''}\tau_{\s} (f_1 \tn \dots \tn f_n)\right)\big) \\
&=F_{\psi'' \s'' \s} \left( f_1 \tn \dots \tn f_n \right)\\
&=F_{\s'\psi} \left( f_1 \tn \dots \tn f_n \right) \, . \qedhere 
\end{align*}
}

\section{Categorified Knizhnik-Zamolodchikov Connections}

\subsection{Knizhnik-Zamolodchikov 2-connections in  configuration spaces}
We now recall some results from \cite{CFM11}.
Let us consider and fix a positive integer $n$. Let $\cW$ be a chain complex of vector spaces. Let us suppose that we have a left action $\s \mapsto \tau_\s$ of the symmetric group $S_n$ on $\cW$ by chain-complex isomorphisms. The motivation for this is the case when $\cW=\V^{\btn n}$ is the tensor product of a chain-complex $\V$ with it self $n$ times, together with the standard action of $S_n$, see \eqref{actionSn}.

Let $\C(n)$ be the configuration space of $n$ distinct particles in the complex plane:
$$\C(n)=\{(z_1,\dots,z_n) \in \C^n\colon z_i \neq z_j \textrm{ if } i \neq j\}.$$
This has an obvious left action $S_n \ni \sigma \mapsto L_\s \in {\rm diff}(\C(n)) $ of the symmetric group $S_n$. The configuration space of $n$ indistinguishable particles in $\C$ is defined as $\C(n)/S_n$.

Recalling the definition of the crossed module $\GL(\cW)=(\beta \colon \gl^1(\cW) \to \gl^0(\cW),\t)$, see Section \ref{dcmcc}, and also Section \ref{mga}, we thus have a right  action $\sigma \mapsto R_\s$ of $S_n$ on $\GL(\cW)$ by differential crossed module maps. Here
\begin{align*}
R_\s(f) & \doteq \tau_{\s^{-1}} \, f \, \tau_{\s} \, , \quad \textrm{ for each } f \in \gl^0(\cW) \textrm{ and } \s \in S_n  \, ,\\
R_\s(s) & \doteq \tau_{\s^{-1}} \, s \, \tau_{\s} \, , \quad \, \textrm{ for each } s \in \gl^1(\cW) \textrm{ and } \s \in S_n \, .
\end{align*}
We will write the Lie algebra brackets on $\gl^i(\cW)$ as $\{,\}$, for $i=0,1$.

Define closed 2-forms $\w_{ij}$ in the configuration space $\C(n)$, for $1\leq i,j \leq n$ and $i \neq j$:
$$\w_{ij}=\frac{dz_i - dz_j}{z_i-z_j} \, . $$
Clearly for each $\s\in S_n$: 
\begin{equation}\label{cov3}
L_\s^*(\w_{ij})=\w_{\s^{-1}{(i)}\s^{-1}(j)}.
\end{equation}
An easy calculation proves the well known Arnold's relation \cite{Ar}, for each distinct indices $i,j,k \in \{1,\dots,n\}$:
\begin{equation}\label{ai}
\w_{ij} \wedge \w_{jk}+\w_{jk} \wedge \w_{ki}+\w_{ki} \wedge \w_{ij}=0 .
\end{equation}
Also, for each distinct indices $i,j \in \{1,\dots,n\}$, we have $\w_{ij}=\w_{ji}$.

For $i < j$, with $i,j \in \{1,\dots,n\}$, we consider arbitrary degree zero chain maps $t_{ij} \colon \cW \to \cW$ and suppose that
\begin{equation}
\label{c1}
 \{t_{ij},t_{i'j'}\} = 0 \, , \quad \textrm{ if } \{i,j\}\cap \{i',j'\}=\emptyset \, .
\end{equation}
Define the following $\gl^0(\cW)$-valued 1-form:
\begin{equation}\label{defA}
A=\sum_{1 \leq i < j \leq n} \w_{ij} t_{ij} 
\end{equation}
Its curvature $F_A=d A+\frac{1}{2}A\wedge^{\{,\}} A=\frac{1}{2}A\wedge^{\{,\}} A$ is:
$$F_A=\sum_{a<b<c} R_{bac} \,\w_{ba} \wedge \w_{ac} + R_{abc} \,\w_{ab} \wedge \w_{bc} ,
 $$
where if $a<b<c$ we put:
$$R_{abc} = \{t_{ab} + t_{ac} , t_{bc}\} \textrm{ and } R_{bac}=\{t_{ab}+t_{bc},t_{ac}\}.$$
This  calculation appears for example in \cite{CFM11}, and previously in several other places \cite{BN,Ko,Ka,Kh1}.

\begin{Theorem}
\label{mainprevious}
In the conditions of \eqref{c1}, the connection form $A=\sum_{1 \leq i < j \leq n} \w_{ij} t_{ij}$ of \eqref{defA} is flat if and only if the following relation, called the 4-term relation, is satisfied:
\begin{equation}\label{4t}\{t_{ab} + t_{ac} , t_{bc}\}=0=  \{t_{ab} + t_{bc} , t_{ac}\} \, , \quad \textrm{ for } 1 \leq a<b<c \leq n.
\end{equation}
In this case the connection form $A$ of \eqref{defA} will be called a Knizhnik-Zamolodchikov connection.
\end{Theorem}
\noindent {Relations \eqref{c1} and \eqref{4t} are sometimes called infinitesimal braid group relations.}

In order to define a local 2-connection $(A,B)$ on $\C(n)$ with values in $\GL(\cW)=(\beta\colon \gl^1(\cW) \to \gl^0(\cW),\t)$, we need a $\gl^1(\cW)$-valued 2-form $B$ such that $\beta(B)=F_A$. We  also want to impose that $(A,B)$  be a flat 2-connection, in other words that the $2$-curvature $3$-form ${\M}=d B+A \wedge^\t B$ vanishes, see Section \ref{curv}.

To match the condition $\beta(B)=F_A$ we define a $\gl^1(\cW)$-valued  2-form $B$ in the configuration space $\C(n)$ as:
\begin{equation}
\label{defB}
B = \sum_{a<b<c} K_{bac} \, \w_{ba}\wedge\w_{ac} + K_{abc} \, \w_{ab}\wedge\w_{bc},
\end{equation}
for some $K_{abc}, K_{bac} \in\gl^1(\cW)$ such that:
\begin{equation}\label{c2} 
\begin{split}
\b(K_{abc})&=R_{abc}, \\ \b(K_{bac})&=R_{bac},\\
t_{ab} \act K_{ijk}&=0 \quad \textrm{ if } \{a,b\} \cap \{i,j,k\}=\emptyset,
\end{split} 
\end{equation}
where we have  $1\leq a<b<c \leq n$ and either $1 \leq i<j <k \leq n$ or $1\leq j<i<k \leq n$.
The following crucial Theorem is proved in \cite{CFM11}.

\begin{Theorem}
\label{iffflat}
Given a $\gl(\cW)$-valued $2$-connection $(A,B)$ on $\C(n)$ with $A$ as in \eqref{defA}, where the relations in  \eqref{c1} hold, and $B$ as in \eqref{defB}, such that  \eqref{c2} holds, then the $2$-curvature $3$-form ${\M}=d B + A \wedge^\t B$ vanishes if and only if the following conditions are satisfied:
\begin{equation}\label{relfl}
\begin{split}
& t_{ad}\act (K_{bac} + K_{bcd}) + (t_{ab} + t_{bc} + t_{bd}) \act K_{cad} - (t_{ac}+t_{cd})\act K_{bad} = 0 \\
& t_{bd}\act (K_{abc} + K_{acd}) + (t_{ab} + t_{ad} + t_{ac}) \act K_{cbd} - (t_{bc}+t_{cd})\act K_{abd} = 0 \\
& t_{bc}\act (K_{bad} + K_{cad}) + t_{ad}\act (K_{cbd} + K_{bcd} -K_{abc}) = 0 \\
& t_{ac}\act (K_{abd} + K_{cbd}) + t_{bd}\act (K_{cad} + K_{acd} -K_{bac}) = 0 \\
& t_{cd}\act (K_{bac} + K_{bad}) + (t_{ab} + t_{bc} + t_{bd}) \act K_{acd} - (t_{ac}+t_{ad})\act K_{bcd} = 0 \\
& t_{cd}\act (K_{abc} + K_{abd}) + (t_{ab} + t_{ac} + t_{ad}) \act K_{bcd} - (t_{bd}+t_{bc})\act K_{acd} = 0 \\
\end{split}
\end{equation}
with $a<b<c<d \in \{1,\ldots ,n\}$. 
\end{Theorem}
\noindent {Relations \eqref{c1}, \eqref{c2} and  \eqref{relfl}  can be interpreted as being infinitesimal  relations for braid cobordisms.}
\begin{Remark}\label{red}
A useful observation for later: note that if we exchange $a$ and $b$ in the first, third and fifth condition, and putting $t_{ba}=t_{ab}$, we get, respectively, the second, fourth and sixth conditions. Exchanging $ a$ and $ c$ in the first equation also yields  the fifth, provided that we have $K_{bca}=K_{bac}$.
\end{Remark}

\begin{Remark}
\label{exbianchi}
 The relations \eqref{relfl}  are satisfied in the differential crossed module $(\id: \gl^0(\cW) \to \gl^0(\cW),\t^{ad})$, where $\t^{ad}$ is the adjoint action of $\gl^0(\cW)$ on itself, if we put $K_{abc}=R_{abc}\in \gl^0(\cW)$. Indeed in this case \eqref{relfl} are the components of the Bianchi identity $d F_A+A \wedge {F}_A=0$, always satisfied by the curvature form $F_A$ of $A$.
\end{Remark}
\begin{Remark}\label{gn}
 {Even though we stated  Theorem \ref{iffflat}  in the context of the differential crossed module  $\GL(\cW)=(\beta \colon \gl^1(\cW) \to \gl^0(\cW),\t)$, for a chain complex $\cW$, this result  remains true, with the obvious modifications, for any differential crossed module $\lG=(\beta\colon \lh \to \lg,\t)$; see \cite[Corollary 23]{CFM11}.}
\end{Remark}

In the conditions of Theorem \ref{iffflat}, for the case when \eqref{relfl} holds, in order that the holonomy of the local 2-connection  $(A,B)$ descend to the quotient manifold $\C(n)/S_n$,  we must impose conditions \eqref{quo}. For instance, as far as the 1-form $A$ is concerned, these say  that given $\s \in S_n$ we must have
\begin{align*}
L_{\s}^* \big(\sum_{1 \leq i < j \leq n} \w_{ij} t_{ij} \big)=R_{\sigma^{-1}}\big(\sum_{1 \leq i < j \leq n} \w_{ij} t_{ij} \big),
 \end{align*}
or:
\begin{align*}
\sum_{1 \leq i < j \leq n} \w_{\sigma^{-1}{(i)}\s^{-1}(j)} t_{ij} =\sum_{1 \leq i < j \leq n} \w_{ij} (\tau_{\s}t_{ij}\tau_{\s^{-1}}).
 \end{align*}
Thus we must have:
$$ \tau_{\s}t_{ij}\tau_{\s^{-1}} =\left \{ \begin{CD} t_{\s{(i)} \s(j)}\textrm{ if } \s{(i)} < \s(j) \\
                                              t_{\s(j) \s{(i)}}\textrm{ if } \s(j) < \s{(i)}                                                   
                                             \end{CD} \right. \, .
$$ 
Therefore defining
$$t_{ab}=t_{ba}, \textrm{ if } a> b,$$
the condition 
\begin{align*}
L_{\s}^* \big(A\big)=R_{\sigma^{-1}}\big(A \big)
 \end{align*}
holds for each permutation $\s \in S_n$ if, and only if, for each $s \in S_n$ we have 
\begin{equation}
 \tau_{\s}t_{ij}\tau_{\s^{-1}}=t_{\s{(i)} \s(j)}.
\end{equation}

We now put  for any distinct $i,j,k \in \{1,\dots n\}$:
\begin{equation*}
R_{ijk}=\{t_{ij}+t_{jk},t_{jk}\}.
\end{equation*}
This is compatible with the previous formulae, given that $t_{ij}=t_{ji}$. Also:
$$R_{ijk}+R_{jki}+R_{kij}=0 \textrm{ and } R_{ijk}=R_{ikj}. $$

The same kind of argument that gives conditions for the covariance of the 1-form $A$ under the action of the symmetric group permits us to find out the conditions that make the relation \eqref{quo} true for the case of the $B$ form. This was sketched in \cite{CFM11}. The stated result was:
\begin{Theorem}\label{Main}
Choose a positive integer $n$.
 Let $\cW$ be a chain-complex, provided with an action $\s \mapsto \tau_\s$ of $S_n$ on it by chain-complex isomorphism. 
Choose chain maps 
$$t_{ab} \colon \cW \to \cW, \textrm { for distinct } 1\leq a,b \leq n.$$ 
These should satisfy, for all distinct $i,j$ and distinct $i',j'$:
\begin{equation}\label{cf1}
 \begin{split}
  t_{ij}&=t_{ji},\\
  \{t_{ij},t_{i'j'}\}&=0, \textrm{ if } \{i,j\}\cap \{i',j'\}=\emptyset.
\end{split}
\end{equation}
Define, for each distinct $a,b,c \in \{1,\dots, n\}$:
\begin{equation}\label{c4t}
R_{abc} = \{t_{ab} + t_{ac} , t_{bc}\}.
\end{equation}
Also choose homotopies (up to 2-fold homotopy) $K_{abc} \in \gl^1(\cW)$, for each distinct $a,b,c \in \{1,\dots, n\}$, satisfying:
\begin{equation}\label{fc2} 
\begin{split}
\b(K_{abc})&=R_{abc}, \\
t_{ab} \act K_{ijk}&=0 \textrm{ if } \{a,b\} \cap \{i,j,k\}=\emptyset,
\end{split}
\end{equation}
 where $a,b,c$ are distinct indices and $i,j,k$ are distinct indices, in both cases in $\{1,\dots,n\}$.
Then the local 2-connection  $(A,B)$ in \eqref{defA} and \eqref{defB} is covariant under the action of the symmetric group \eqref{quo}, if, and only if, for any distinct indices $a,b,c\in  \{1,\dots,n\}$ we have that:
\begin{equation}\label{ja}
K_{abc}+K_{bca}+K_{cab}=0, \quad \quad \quad K_{bca}=K_{bac} \, ,
\end{equation}
and that for each permutation $\sigma \in S_n$:
\begin{align}\label{cov}t_{\sigma(a) \sigma(b)}=  \tau_{\sigma} t_{ab} \tau_{\sigma^{-1}}\textrm{ and } K_{\s(a)\s(b) \s(c)}=\tau_{\s} K_{abc} \tau_{\sigma^{-1}}.
 \end{align}
Morever, in this case the pair $(A,B)$ is 2-flat if and only if, for any distinct indices $a,b,c,d \in  \{1,\dots,n\}$ we have:
\begin{equation}
\begin{split}\label{referir}
& t_{ad}\act (K_{bac} + K_{bcd}) + (t_{ab} + t_{bc} + t_{bd}) \act K_{cad} - (t_{ac}+t_{cd})\act K_{bad} = 0 \\
& t_{bc}\act (K_{bad} + K_{cad}) -t_{ad}\act (K_{dbc} +K_{abc}) = 0.
\end{split}
\end{equation}
\end{Theorem}
\begin{Remark}
{As we did for Theorem \ref{iffflat}, we stated Theorem \ref{Main}  in the context of the differential crossed module  $\GL(\cW)=(\beta \colon \gl^1(\cW) \to \gl^0(\cW),\t)$, derived from a  chain complex $\cW$. However,  Theorem \ref{Main}   remains true, with the obvious modifications, for any differential crossed module $\lG=(\beta\colon \lh \to \lg,\t)$; see \cite[Corollary 23]{CFM11}.}
\end{Remark}

\noindent
{Let us give full details  of the proof of the most important part of Theorem \ref{Main}}, which is that if equations \eqref{cf1} to \eqref{referir} hold then condition \eqref{relfl} of Theorem \ref{mainprevious} and equation \eqref{quo} are satisfied.
{\proof
Concerning the $S_n$ covariance, the first equation of \eqref{quo} was already discussed. As for the second equation of \eqref{quo}, note that we can rewrite the 2-form $B$ of \eqref{defB} as, by using \eqref{ja} and Arnold identity \eqref{ai}:
\begin{align}
B &= \sum_{a<b<c} K_{bac} \, \w_{ba}\wedge\w_{ac} + K_{abc} \, \w_{ab}\wedge\w_{bc}\label{a}\\
 & = \sum_{a<b<c} -(K_{cba}+K_{acb}) \, \w_{ba}\wedge\w_{ac}- K_{acb} (\, \w_{bc}\wedge\w_{ca}+\w_{ca}\wedge\w_{ab})\\
&= \sum_{a<b<c} -K_{cba} \, \w_{ba}\wedge\w_{ac}- K_{acb} \, \w_{bc}\wedge\w_{ca}\\
& = \sum_{a<b<c} K_{acb} \, \w_{ac}\wedge\w_{cb} + K_{cab} \, \w_{ca}\wedge\w_{ab}\label{b}.
\end{align}
Analogously:
\begin{align}\label{c}
B& = \sum_{a<b<c} K_{cba} \, \w_{cb}\wedge\w_{ba} + K_{bca} \, \w_{bc}\wedge\w_{ca}.
\end{align}
Thus, summing \eqref{a},\eqref{b} and \eqref{c} we have:
\begin{align*}
 B&= \frac{1}{3}\sum_{1\leq a_1 < a_2  < a_3\leq n} \quad  \sum_{\s \in S_3} K_{a_{\s(1)} a_{\s(2)} a_{\s(3)} } \,\, \w_{a_{\s(1)} a_{\s(2)}}\, \wedge\, \w_{a_{\s(2)} a_{\s(3)}}\\
  &= \frac{1}{3\times 3!}\sum_{ a_1 \neq a_2, a_1 \neq a_2,   a_1 \neq a_3} \quad  \sum_{\s \in S_3} K_{a_{\s(1)} a_{\s(2)} a_{\s(3)} } \,\, \w_{a_{\s(1)} a_{\s(2)}}\, \wedge\, \w_{a_{\s(2)} a_{\s(3)}}
\end{align*}
Therefore for any permutation $\s' \in S_n$ we have, by \eqref{cov3}:
$$L_{{\s'}^{-1}}^*(B)=\frac{1}{3\times 3!}\sum_{ a_1 \neq a_2, a_1 \neq a_2,   a_1 \neq a_3} \quad  \sum_{\s \in S_3} K_{a_{\s(1)} a_{\s(2)} a_{\s(3)} } \,\, \w_{\s'(a_{\s(1)}) \s'(a_{\s(2)})}\, \wedge\, \w_{\s'(a_{\s(2)})\s'( a_{\s(3)})}.$$
Whereas, by \eqref{cov}:
$$R_{{\s'}}(B)=\tau_{{\s'}^{-1}} B\tau_{{\s'}}=  \frac{1}{3\times 3!}\sum_{ a_1 \neq a_2, a_1 \neq a_2,   a_1 \neq a_3}\quad  \sum_{\s \in S_3} K_{\s'^{-1}(a_{\s(1)}) \s'^{-1}( a_{\s(2)} ) \s'^{-1}(a_{\s(3)} )} \,\, \w_{a_{\s(1)} a_{\s(2)}}\, \wedge\, \w_{a_{\s(2)} a_{\s(3)}}.$$
These last two clearly coincide. As for the 2-flatness, all equations of \eqref{relfl}  hold because of \eqref{referir} and the $S_n$-covariance, see Remark \ref{red}.\qedhere
}\\

Theorem \ref{Main} implies that, provided that the covariance properties \eqref{cov} hold, to define the pair $(A,B)$ in  \eqref{defA} and \eqref{defB}, with zero 2-curvature, one only needs to specify the chain map $t_{12}=t \in \gl^0(\cW)$ and the chain-homotopy (up to 2-fold homotopy) $K_{123}=K \in \gl^1(\cW)$. Then obligatory we must have:
\begin{equation}\label{a1}
\begin{split}
t_{\sigma(1) \sigma(2)}&=  \tau_{\sigma}\, t \, \tau_{\sigma^{-1}}\doteq R_{\s^{-1}}(t),\\
 K_{\s(1)\s(2) \s(3)}&=\tau_{\s}\, K\, \tau_{\sigma^{-1}}	\doteq R_{\s^{-1}}(K);
\end{split}
\end{equation}
for any permutation $\s \in S_n$. Therefore $t$ and $K$ must satisfy, for any $\s,\s' \in S_n$:
\begin{equation}\label{a2}
\begin{split}
\tau_{\sigma}\, t \,\tau_{\sigma^{-1}}&=\tau_{\sigma'} t \tau_{\sigma'^{-1}}, \textrm{ if }\s(1)=\s'(1) \textrm{ and } \s(2)=\s'(2);\\ 
\tau_{\sigma} \,K\, \tau_{\sigma^{-1}}&=\tau_{\sigma'} K \tau_{\sigma'^{-1}} 
\textrm{ if }\s(1)=\s'(1), \s(2)=\s'(2) \textrm{ and } \s(3)=\s'(3);
\end{split}
\end{equation}
and also, in the crossed module $\GL(\cW)$:
\begin{equation}\label{a3}\beta(K) = \{t_{12} + t_{13} , t_{23}\}.\end{equation}
To ensure the 2-flatness and the $S_n$-equivariance of $(A,B)$, the following conditions are to be satisfied:
\begin{equation}\label{a4}
\begin{split}
 t_{14}\act (K_{213} + K_{234}) + (t_{12} + t_{23} + t_{24}) \act K_{314} - (t_{13}+t_{34})\act K_{214} &= 0 ,\\
 t_{23}\act (K_{214} + K_{314}) -t_{14}\act (K_{423} +K_{123}) &= 0,
\end{split}
\end{equation}
plus:
\begin{equation}\label{a5}
  t_{12} = t_{21} \, , \qquad 
  K_{123}+K_{231}+K_{312} = 0 \, , \qquad 
  K_{231} = K_{213} \\
\end{equation}
and in addition,  for any pair / triple of distinct indices   $a,b$ and $i,j,k$:
\begin{equation}\label{a6}
\begin{split}
 {t_{ij},t_{ab}}&=0, \textrm{ if } \{i,j\}\cap \{a,b\}=\emptyset,\\
t_{ab} \act K_{ijk}&=0 \textrm{ if } \{a,b\} \cap \{i,j,k\}=\emptyset.  
\end{split}         
\end{equation}
The following lemma simplifying the conditions of Theorem \ref{Main} will be very useful later.
\begin{Lemma}\label{Main3}
Let $\cW$ be a chain complex of vector spaces. Let $n$ be a positive integer. Suppose we have an action of $S_n$ on $\cW$ by chain-complex isomorphisms. Consider a chain map $t=t_{12}\colon \cW \to \cW$ and a chain homotopy $K=K_{123} \in \gl^1(\cW)$. Suppose \eqref{a2} holds, and define $t_{ab}$ and $K_{abc}$ by using \eqref{a1}.
 Then the pair $(A,B)$  as in \eqref{defA} and \eqref{defB} is a $\GL(\V)$-valued local 2-connection  in the configuration space $\C(n)$, with zero 2-curvature 3-tensor, and covariant with respect to the action of $S_n$, see equations \eqref{quo}, if, and only if, equations \eqref{a3}, \eqref{a4}, \eqref{a5} and \eqref{a6} are satisfied.
\end{Lemma}
{\proof The claim follows  almost trivially from our discussion. Let us look at the conditions of Theorem \ref{Main}.
The second equation of \eqref{cf1}, the second part of \eqref{fc2} and \eqref{cov} are exactly equations \eqref{a6} and \eqref{a1}. The remaining conditions follow from equations \eqref{a1}, \eqref{a2}, \eqref{a3}, \eqref{a4} and \eqref{a5}, by applying the right action $\s \mapsto R_\s$ of $S_n$ on $\gl(\cW)$ by differential crossed module morphisms to each side of \eqref{a2}, \eqref{a3}, \eqref{a4} and \eqref{a5}, and considering the yielded set of equations for each $\s$ in $S_n$.
\qedhere }

\subsection{Infinitesimal {2-Yang-Baxter operators} in a differential crossed module}\label{i2rm}

In this subsection we actively use the results and notation of Sections \ref{smccc} and  \ref{im}. 

\subsubsection{Insertion maps for a categorical representation}
\label{Sinsmap}

Let $\lG=(\beta\colon \lh \to \lg, \t)$ be a differential crossed module. Consider its underlying (short) complex of vector spaces $\underline{\lG}=(\beta\colon \lh \to \lg).$ Given a positive integer $k$, define $\lU^{(k)}$ as being the degree-1 part of the tensor product of $\underline{\lG}$  with it self $k$-times. Define also $\overline{\lU}^{(k)}$ as being $\lU^{(k)}$, modulo the boundary of the degree two part of $\underline{\lG}^{\btn n}$.
Explicitly:
$$\lU^{(k)}=\lh \otimes\lg \otimes \lg \otimes\dots \otimes \lg + \textrm{ cyclic permulations},$$
and
$ {\overline{\lU}}^{(k)}$ equal is to $\lU^{(k)}$ modulo the relations:
\begin{multline}
\label{irel}
X_1 \otimes \dots \otimes X_m \otimes \beta{(v)} \otimes X_{m+1} \otimes \dots \otimes X_{m'-1} \otimes w \otimes X_{m'+1} \otimes \dots \otimes X_n = \\
X_1 \otimes \dots \otimes X_m \otimes v \otimes X_{m+1} \otimes \dots \otimes X_{m'-1} \otimes \beta(w) \otimes X_{m'+1} \otimes \dots \otimes X_k
\end{multline}
where all $X_i$ live in $\lg$, and $v,w \in \lh$, furthermore $1\leq m<  m'\leq k$.
There exists an obvious map ${\beta'}\colon \lU^{(k)} \to \lg^{\btn k}$, which descends to the quotient ${\overline{\lU}}^{(k)}$, defined as being:
$${\beta'} =\beta \tn 1 \tn \dots \tn 1 +\textrm{ cyclic permulations}. $$
Clearly $\b'$ is the first non-trivial boundary map of the tensor product of $(\beta \colon \lh \to \lg)$ with it self $k$ times. In the notation of Section \ref{im} we thus have that
$$\overline{(\underline{\lG}^{\btn k })_2}=(\beta' \colon {\overline{\lU}}^{(k)} \to \lg^{\tn k}). $$

Let $(\V,\partial)$ be a chain-complex of vector spaces. Let $\GL(\V)$ be the differential crossed module constructed in Section \ref{dcmcc}. Consider its underlying short complex of vector spaces $\underline{\GL(\V)}=\left(\beta\colon \gl^1(\V) \to \gl^0(\V)\right)$.
Let $\rho\colon \lG\to \GL(\V)$ be a categorical representation of $\lG$. Suppose that we are given $k$ distinct indices $a_i$ in $\{1, \dots ,n\}$, describing therefore an injective map $\psi\colon \{1,\dots,k\} \to \{1,\dots, n\}$, with $i \mapsto a_i$. We define:
$$\phi_{a_1\ldots a_k}^{{(1)}} :\overline{\mathfrak{U}}^{(k)} \rightarrow \gl^1(\V^{\btn n})$$ 
as being:
\begin{equation}
\label{phiUk}
\phi_{a_1\ldots a_k}^{{(1)}} \left( \sum_{i=1}^k u_{i}^1\tn \ldots \tn u_{i}^k\right)= \sum_{i=1}^k \, \id\btn\ldots \btn \rho(u_{i}^1)\btn\ldots \btn\rho(u_{i}^k)\btn\ldots\btn\id
\end{equation}
for each
$$ \sum_{i=1}^k \, u_{i}^1\tn \ldots \tn u_{i}^k \in \mathfrak{U}^{(k)}.$$ 
Within \eqref{phiUk}, in every summand we have inserted the $\rho$ image of $u_{i}^r$ in the $a_r^{\rm th}$ factor of the tensor product as the only non trivial entries (the exact order in which the $\rho(u_i^r)$ appear in the tensor product may not be the one indicated in \eqref{phiUk}). The definition of
$$ \phi_{a_1\ldots a_k}^{{(0)}} : \lg^{\tn k} \rightarrow \gl^0(\V^{\btn n}) $$
is entirely similar (we will frequently drop the indices in $\phi^{(0)}$ and $\phi^1{(1)}$ when they are obvious from the context).  Then the pair $(\phi_{a_1\ldots a_k}^{{(1)}}, \phi_{a_1\ldots a_k}^{{(0)}})\doteq F^\rho_\psi$ defines a chain-map $F^\rho_\psi\colon \overline{(\underline{\lG}^{\btn k })_2} \to \underline{\GL(\V^{\btn n})}$, in the notation of Section \ref{im}. This map is the following composition, which proves that it is well defined:
\begin{equation}
\label{cfd}
\overline{(\underline{\lG}^{\btn k })_2} \ra{\overline{\rho^{\btn k}}}  \overline{(\underline{\GL(\V)}^{\btn k })_2} \ra{F_\psi} \underline{\GL(\V^{\btn n})}.
\end{equation}
We thus have the following commutative diagram of vector space maps:
\begin{equation}
\label{dphicd}
\xymatrix{
\bar{\mathfrak{U}}^{(k)}\ar[d]_-{\beta'}\ar[rr]^-{\phi^{(1)}_{a_1\ldots a_k}} & & \gl^1(\V^{\btn n})\ar[d]^-{\b} \\
\lg^{\tn k}\ar[rr]^-{\phi^{(0)}_{a_1\ldots a_k}} & & \gl^0(\V^{\btn n})
}\,\,\, ,
\end{equation}
covariant with respect to the actions of the symmetric groups $S_k$ and $S_n$, see equations \eqref{refA} and \eqref{refB} below.

\subsubsection{Infinitesimal {2-Yang-Baxter operators}}

Let $\V$ be a chain complex. If {$\sigma \in S_k$} and if $L$ is an homogeneous element  in $\overline{(\V^{\btn k})_2}$  we denote
$$L_{\s(1)\dots \s(k)}=\overline{\tau_{\s}}(L) \, , $$
where $\overline{\tau}$ is the projection of the standard action of $S_k$ on $\V^{\btn k }$ to  $\overline{(\V^{\btn k})_2}$. If $L=\sum_{i} X_i \tn Y_i \tn Z_i \in \overline{(\V^{\btn k})_2} $ then: 
\begin{align*}
L_{123}&=L\,,
&L_{132}&= \sum_{i} X_i \tn Z_i \tn Y_i\,,
&L_{231}&= \sum_{i} Z_i \tn X_i \tn Y_i.
\end{align*}
Note that no minus signs need to be inserted here since all but possibly one of the elements $X_i,Y_i$ and $Z_i$ have degree 0, for each $i$.

\begin{Definition}[Free infinitesimal {2-Yang-Baxter operator}]\label{irmatrix}
Let $\lG =(\beta\colon \lh \to \lg)$ be a differential crossed module. A  free infinitesimal {2-Yang-Baxter operator (or free infinitesimal 2-$\R$-matrix)} $(r,P)$ in $\lG$ is given by:
\begin{enumerate}
   \item A tensor $r \in \lg \tn \lg$,
   \item A tensor $P \in {\overline{\lU}}^{(3)},$
\end{enumerate}
such that:
 \begin{enumerate}
\item $r_{12}=r_{21},$
\item $P_{123} + P_{231} + P_{312} = 0,$
\item $P_{123} = P_{132},$
\item ${\beta'}(P)=[r_{12}+r_{13},r_{23}],$
\item $r_{14}\t (P_{213} + P_{234}) +(r_{12} + r_{23} + r_{24})\t P_{314} - (r_{13}+r_{34})\t P_{214} = 0,$
\item $r_{23}\t (P_{214} + P_{314}) - r_{14}\t (P_{423}+P_{123}) = 0 .$
\end{enumerate}
The last two identities are to hold in $\overline{\lU}^{(4)}$.
\end{Definition}
A word on notation: here if $r=\sum_{i}x_i\tn y_i\in \lg \tn \lg $ and $P=\sum_{i} U_i \tn V_i \tn W_i\in \overline{\lU}^{(3)}$ then, for example:
\begin{equation}\label{spec}
\begin{split}
[r_{12}+r_{13},r_{23}]&=\sum_{i,j} x_i \tn [y_i,x_j]\tn y_j+x_i \tn x_j \tn [y_i,y_j]\, ,\\
r_{14}\t P_{213}&=\sum_{i,j} (x_i \t V_j) \tn  U_j \tn W_j \tn y_i \, , 
\end{split}
\end{equation}
 where in the second equation $\t$ may either mean the action of $\lg$ on $\lh$ or the adjoint action of $\lg$ on $\lh$.  The second equation of \eqref{spec} does descend to the quotient $\bar{\mathfrak{U}}^{(4)}$. Indeed given $u, v \in \lh$ we have $\partial(u) \tn v=u \tn \partial{(v)}$, and, on the other hand, if $X\in \lg$, we have:
\begin{align*}
(X \t \partial(u)) \tn v =\partial( X \t u) \tn v=(X \t u) \tn \partial{(v)} \, .
\end{align*}

\begin{Example}
\label{exid}
 {Let $\lg$ be a Lie algebra. Let $\lG=(\id\colon \lg \to \lg,\t^{\mathrm{ad}})$, where $\t^{\mathrm{ad}}$ is the adjoint action; this is a differential crossed module. In this case ${\overline{\lU}}^{(n)}\cong\lg^{\tn n }$. If $r\in \lg \tn \lg$ is symmetric, and if we put $P=[r_{12}+r_{13},r_{23}]\in \lg \tn \lg \tn \lg$, then the pair $(r,P)$ is a free infinitesimal 2-Yang-Baxter operator in $\lG$. For a proof see \cite{CFM11}. }
\end{Example}
\noindent {Below we will construct an infinitesimal 2-Yang-Baxter operator in the differential crossed module associated to  the string Lie-2-algebra.}

It is also useful to deal with representations of the conditions which characterize an infinitesimal {2-Yang-Baxter operator.} For this reason we introduced insertion maps. Indeed, given a categorical representation $\rho\colon \lG\to \GL(\V)$ we have, in $\gl^0(\V \tn \V \tn \V)$:
\begin{equation}
\begin{split}
 \overline{\rho^{\btn 3}}([r_{12}+r_{13},r_{23}]) &=(\rho \btn \rho \btn \rho)[r_{12}+r_{13},r_{23}]\\&=\sum_{i,j} \rho(x_i) \tn \rho([y_i,x_j])\tn \rho( y_j) +\rho(x_i) \tn \rho( x_j)  \tn \rho( [y_i,y_j])\\
&=\sum_{i,j} \rho(x_i) \tn \{\rho(y_i),\rho(x_j)\}\tn \rho( y_j) +\rho(x_i) \tn \rho( x_j)  \tn  \{\rho(y_i),\rho(y_j)\}\\
&=\{\phi_{12}{(r)}+\phi_{13}(r),\phi_{23}(r)\} \, .
\end{split}
\end{equation}
Also, in $\GL(\V \tn \V \tn \V)$:
\begin{equation}\label{4t1}
\begin{split}
\beta(\phi_{123}(K))&=\phi_{123}( \beta'(K))=\phi_{123}([r_{12}+r_{13},r_{23}])\\
&=\overline{\rho^{\btn 3}}([r_{12}+r_{13},r_{23}])=\{\phi_{12}{(r)}+\phi_{13}(r),\phi_{12}(r)\} \, .
\end{split}
\end{equation}
We can also see that, since $\rho$ is a categorical representation, and where $\t$ is either the adjoint action of $\gl^0(\V)$ on $\gl^0(\V)$ or the already defined action of $\gl^0(\V)$ on $\gl^1(\V)$:
\begin{equation}\label{ref5}
\begin{split}
\overline{\rho^{\btn 4}}( r_{14}\t P_{213})&=\sum_{i,j} \overline{\rho^{\btn 4}} \big((x_i \t V_j) \tn  U_j \tn W_j \tn y_i\big) \\
&=\sum_{i,j}  \overline{\rho}\big(x_i \t V_j\big) \btn  \overline{\rho}( U_j) \btn  \overline{\rho}(W_j) \btn \overline{\rho}( y_i)\\
&=\sum_{i,j} \left ( \overline{\rho}(x_i) \t \overline{\rho}( V_j)\right) \btn  \overline{\rho}( U_j) \btn  \overline{\rho}(W_j) \btn \overline{\rho}( y_i)\\
&=\phi_{14}(r) \t \phi_{231}(P) \, .
\end{split} 
\end{equation}

\begin{Definition}[{Infinitesimal {2-Yang-Baxter operator} with respect to a categorical representation}]\label{irmatrix2}
 {In the conditions of Definition \ref{irmatrix}, let $\rho\colon \lG \to \gl(\V)$ be a categorical representation of $\lG$. Then $(r,P)$ is said to be an infinitesimal 2-{Yang-Baxter operator} (or infinitesimal 2-$\R$-matrix)} in $\lG$, with respect to $\rho$, if
conditions 1. to {6}. of Definition \ref{irmatrix} hold after applying $\overline{\rho^{\btn k}}$, where $k$ can be, depending in the condition, 2, 3 or 4.
\end{Definition}

Recall now Theorem \ref{Main} and Lemma \ref{Main3}. We now show that (representations of) infinitesimal {2-Yang-Baxter operators} provide natural examples of chain maps and homotopies satisfying the conditions of Theorem \ref{Main}, hence yielding 2-flat connections on $\C(n)$, covariant with respect to the action of $S_n$.  Let $\rho\colon \lG\to \GL(\V)$ be a categorical representation of a differential crossed module. By \eqref{cfd} and \eqref{cfd2} we have, for each injective map $\psi\colon \{1,\dots,k\} \to \{1, \dots ,n\}$ and each $\s \in S_k$:
\begin{equation}\label{refA}
F^\rho_\psi(\tau_\s(L))=F^\rho_{\psi \circ \tau}(L)
\end{equation}
since by \eqref{actionSn}, and given that  
$\overline{\rho^{\btn k}}\colon \overline{(\underline{\lG}^{\tn k })_2} \to  \overline{(\underline{\GL(\V)}^{\tn k })_2} $  is a chain-map of degree 0, it is
$$\overline{\rho^{\btn k} } \circ \tau_{\s}=\tau_{\s} \circ \overline{\rho^{\btn k}} \, . $$
Also, by Lemma \ref{ref4} we have, for each $\s \in S_n$ and each $L$,
\begin{equation}\label{refB}
\tau_{\s} F_\psi^\rho(L) \tau_{\s^{-1}}= F_{\sigma \psi}^\rho(L). 
\end{equation}
\begin{Lemma}
Fix a positive integer $n$. Let $\V$ be a chain-complex of vector spaces. Let $S_n$ act on $\V^{\btn n}$ in the usual way, equation \eqref{actionSn}. Let $\lG=(\beta \colon \lh \to \lg,\t)$ be a differential crossed module and $\rho\colon \lG \to \GL(\V)$ be a categorical representation of $\lG$. Let $(r,P)$ be an infinitesimal {2-Yang-Baxter operator} in $\lG$, with respect to $\rho$. If we define:
\begin{align*}
t=t_{12} & =\phi_{12}(r)\in \gl^0(\V^{\btn n}) \\
K=K_{123} & =\phi_{123}(P)\in \gl^1(\V^{\btn n})
\end{align*}
then all conditions of Theorem {\ref{Main}}  are satisfied, in the formulation of Lemma \ref{Main3}.
\end{Lemma}
\proof{
 This essentially follows immediately from \eqref{refA}, \eqref{refB} and Lemma \ref{refC}. Explicitly, conditions \eqref{a2} holds since if $\s \in S_n$ is such that $\s(1)=\s'(1),$ $\s(2)=\s'(2)$ and $\s(3)=\s'(3)$ then, putting (from now on until the end of the proof) $\psi(1)=1, \psi(2)=2, \psi(3)=3$:
\begin{equation*}
 \tau_{\s} K \tau_{\s^{-1}} = \tau_{\s} F_\psi^\rho(P) \tau_{\s^{-1}} = F^\rho_{\s \psi}(P) = F^\rho_{\s' \psi}(P) = \tau_{\s'} F_\psi^\rho(P) \tau_{\s'^{-1}} = \tau_{\s'} K \tau_{\s'^{-1}},
\end{equation*}
and analogously if  $\s(1)=\s'(1)$ and $\s(2)=\s'(2)$:
$$\tau_{\s} \, t\, \tau_{\s^{-1}}= \tau_{\s'}\, t\, \tau_{\s'^{-1}}.$$
From \eqref{a1} and \eqref{refB} we also have that, for distinct $a,b,c \in \{1,\dots,n\}$, $$K_{abc}=\phi_{abc}(P) \textrm{ and } t_{ab}=\phi_{ab}(r).$$ 
From \eqref{4t1} and \eqref{ref5} we thus obtain \eqref{a3} and \eqref{a4}.
Conditions \eqref{a6} follow from Lemma \ref{refC}. 

All equations of \eqref{a5} are proved in exactly the same way. For example, considering the permutations in $S_n$ and $S_3$ defined as $\s_0=\id$, $\s_1=(231)$ and $\s_2=(312)$, we have:
\begin{align*}
 K_{123}+K_{231}+K_{312}&\doteq \tau_{\s_0}K\tau_{\s_0^{-1}}+\tau_{\s_1}K\tau_{\s_1^{-1}}+\tau_{\s_2}K\tau_{\s_2^{-1}}\\
&\doteq \tau_{\s_0}F^\rho_{\psi}(P)\tau_{\s_0^{-1}}+\tau_{\s_1}F^\rho_{\psi}(P) \tau_{\s_1^{-1}}+\tau_{\s_2}F^\rho_{\psi}(P)\tau_{\s_2^{-1}}\\
&= F^\rho_{\s_0\psi}(P)+F^\rho_{\s_1\psi}(P) +F^\rho_{\s_2\psi}(P)\\
&= F^\rho_{\psi\s_0}(P)+F^\rho_{\psi\s_1}(P) +F^\rho_{\psi\s_2}(P)\\
&= F^\rho_{\psi}(\tau_{\s_0} (P))+F^\rho_{\psi}(\tau_{\s_1}(P)) +F^\rho_{\psi}(\tau_{\s_2}(P))\\
&= F^\rho_{\psi}(P_{123}+P_{231}+P_{312})=0.
\end{align*}
\qedhere}

As an immediate  corollary we have  one of the main results of this paper, also stated in \cite{CFM11} for  free infinitesimal 2-$\R$-matrices.

\begin{Theorem}\label{qws}
Consider  a positive integer $n$.  Let $\lG=(\beta \colon \lh \to \lg,\t)$ be a differential crossed module. Let $\rho\colon \lG \to \GL(\V)$ be a categorical representation of $\lG$, where $\V$ is a chain-complex of vector spaces. Let $(r,P)$ be an infinitesimal {2-Yang-Baxter operator} in $\lG$ with respect to $\rho$. Then putting, for each distinct $a,b,c \in \{1,\dots,n\}$:
$$t_{ab}=\phi_{ab}(r) \quad \mbox{ and } \quad K_{abc}=\phi_{abc}(P), $$
the condition of Theorem \eqref{Main} are satisfied. Therefore the pair $(A,B)$:
\begin{equation}\label{def2Knizhnik-Zamolodchikov}
\begin{split}
A&=\sum_{1 \leq a < b\leq n} \w_{ab} \,\phi_{ab}(r) \\
B&=\sum_{1 \leq a < b<c\leq n} \w_{ab} \wedge\w_{bc} \,\phi_{abc}(P) +\w_{ba} \wedge\w_{ac} \,\phi_{bac}(P) 
\end{split}
\end{equation}
defines a 2-flat $\GL(\V)$-valued local 2-connection in the configuration space $\C(n)$, covariant under the action of $S_n$. 
\end{Theorem}
\noindent These local 2-connections $(A,B)$ are our proposal for a  Knizhnik-Zamolodchikov 2-connection.

\section{{An infinitesimal 2-Yang-Baxter operator in the string Lie-2-algebra}}

\subsection{Lie algebra cohomology, abelian extensions and differential crossed modules}

We recall a general construction of differential crossed modules that will be used in this section to provide an explicit presentation (due to Wagemann \cite{wag06}) of the string Lie 2-algebra (discussed in the Introduction), or more precisely of its associated differential crossed module $\st$.

Recall that given a Lie algebra $\lg$ with an action $\t$ on a vector space $V$, we have a cochain complex $C^\bullet(\lg,V)=\big(C^n(\lg, V), \delta^V\big)$, where $C^n(\lg,V)=\hom(\wedge^n(\lg), V) $. For $\w\colon \wedge^n(\lg)\to V$ the differential reads:
$$
\delta^V(\w)(X_0,\dots, X_{n})=\sum_{0 \leq i \leq n}(-1)^i X_i\t \left( \w \, \big(X_0,\dots, \hat{X_i}, \dots, X_n\big) \right) + \sum_{1 \leq i < j \leq n} (-1)^{i+j} \w\, \big ([X_i,X_j], X_0,  \dots, \hat{X_i}, \dots, \hat{X_j}, \dots , X_n\big),
$$
where as usual $\hat{X_i}$ means omitting $X_i$. 

An exact sequence of Lie algebras:
\begin{equation}
\label{abex}
0 \longrightarrow V_3 \stackrel{\nu}{\longrightarrow} \mathfrak{e} \stackrel{\pi}{\longrightarrow} \gg \longrightarrow 0
\end{equation} 
is said to be an abelian Lie algebra extension of $\gg$ by $V_3$ if $\nu(V_3)$ is abelian. Note that $V_3$ has a natural $\gg$-module structure, described explicitly on $\nu(V_3)$ by $x\t \nu{(u)}:= [\pi^{-1}(x),\nu{(u)}]$ for $x\in\gg,u\in V_3$. It is very easy to check that such expression is well defined and that it does not depend on the choice of the pre-image of $x$. It is well known that abelian extensions are classified by the Lie algebra cohomology group $H^2(\gg,V_3)$. We sketch the correspondence; in one direction, take $\alpha\in H^2(\gg,V_3)$ and define $V_3\rtimes_{\alpha} \gg$ as the vector space $V_3\oplus\gg$ endowed with the Lie bracket
\begin{equation} 
\label{lbr}
\big[(v,x),(w,y)\big] := \big(x\t w- y\t v + \alpha(x,y),[x,y]\big) \, ,
\textrm{ where } v,w\in V_3 , \, x,y\in\gg \, .
\end{equation}
Then set $\mathfrak{e}= V_3\rtimes_{\alpha} \gg$ in \eqref{abex}. In the opposite direction, take any section of $\pi$, i.e. a linear map $a:\gg\rightarrow\mathfrak{e}$ such that $\pi\circ a = \id_{\gg}$, and denote by $\alpha:\gg\wedge\gg\to\mathfrak{e}$ the failure of $a$ to be a Lie algebra morphism, namely, for $x,y \in \lg$:
$$ \alpha(x,y) := a([x,y]) - [a(x),a(y)] \,.$$
Since $\pi\circ\alpha=0$ we know that $\alpha$ takes value in $\textrm{im }(\nu)\subset \mathfrak{e}$. This means there exists a unique $\alpha':\gg\wedge\gg\to V_3$ such that $\nu\circ\alpha'=\alpha$. Such $\alpha'$ is a $2$-cochain on $\gg$ with values in $V_3$, i.e. $\alpha'\in C^2(\gg,V_3)$. It is a standard result, which we nevertheless prove for completeness, that $\alpha'$ is a $2$-cocycle. 
\begin{Lemma}
The two cochain $\alpha'\in C^2(\gg,V_3)$ satisfies $\delta^{V_3}\alpha'=0$, namely $\alpha'$ is a $2$-cocycle on $\gg$ with values in $V_3$. 
\end{Lemma}
\proof{
Note that the $\gg$-module structure on $V_3$ (see discussion above) is defined on $\nu(V_3)\subset\mathfrak{e}$ by an arbitrary section of $\sigma$, which we now take to be $a$. Almost by definition $\nu:V_3\to \textrm{im} (\nu)$ is an injective $\gg$-module map. We show that $\nu (\delta^{V_3}\alpha')=0$: 
\begin{equation*}
\begin{split}
\nu & (\delta^{V_3}\alpha')(x_1,x_2,x_3) =  \\
& = - \, \nu\alpha'([x_1,x_2],x_3) + \nu\alpha'([x_1,x_3],x_2) - \nu\alpha'([x_2,x_3],x_1) +  \nu \, (x_1\t\alpha'(x_2,x_3) - x_2\t\alpha'(x_1,x_3)+x_3\t\alpha'(x_2,x_3) \, )  \\
& = - \, \alpha([x_1,x_2],x_3) + \alpha([x_1,x_3],x_2) - \alpha([x_2,x_3],x_1) + x_1\t\alpha(x_2,x_3) -x_2\t\alpha(x_1,x_3)+x_3\t\alpha(x_2,x_3) \\
& = - \, [a([x_1,x_2]),a(x_3)] + a([x_1,x_2],x_3) + [a([x_1,x_3]),a(x_2)] - a([x_1,x_3],x_2) + [a([x_2,x_3]),a(x_1)] + a([x_2,x_3],x_1) \, + \\
& \; \quad + [a(x_1),[a(x_2),a(x_3)]] - a([x_2,x_3]) ] - [a(x_2),[a(x_1),a(x_3)]] - a([x_1,x_3]) ] + [a(x_3),[a(x_1),a(x_2)]] - a([x_1,x_2]) ] \\
& = - \, [a([x_1,x_2]),a(x_3)] + [a([x_1,x_3]),a(x_2)] - [a([x_2,x_3]),a(x_1)]  a([x_1,x_2],x_3) - a([x_1,x_3],x_2) + a([x_2,x_3],x_1) \, + \\
& \; \quad + [a(x_1),[a(x_2),a(x_3)]] - [a(x_2),[a(x_1),a(x_3)]] + [a(x_3),[a(x_1),a(x_2)]] - [a(x_1),a([x_2,x_3])] + [a(x_2),a([x_1,x_3])] \, + \\ 
& \; \quad - [a(x_3),a([x_1,x_2])] = 0 
\end{split}
\end{equation*}
where in the last step we used twice the Jacobi identity. \qedhere
}\\

\noindent Note that there is no dependence on the choice of the section $a\colon \lg \to \mathfrak{e}$ (i.e. different sections produce cohomologous cocycles). 

Next, consider a sequence of $\gg$-modules (when necessary thought as abelian Lie algebras):
\begin{equation}
\label{seqmod}
0 \longrightarrow V_1 \stackrel{i}{\longrightarrow} V_2 \stackrel{p}{\longrightarrow} V_3 \longrightarrow 0 \, .
\end{equation}
This yields a short exact sequence of cochain complexes:
\begin{equation}\label{seqmod1}
0 \longrightarrow C^\bullet(\lg,V_1) \stackrel{i}{\longrightarrow} C^\bullet(\lg,V_2) \stackrel{p}{\longrightarrow} C^\bullet(\lg,V_3) \longrightarrow 0 \, ,
\end{equation}
whose cohomology long exact sequence will have an important role below. 

Consider a Lie algebra 2-cocycle $\alpha$. Putting together the two sequences \eqref{abex} and \eqref{seqmod}, where the former is derived from $\alpha$, thus $\mathfrak{e}= V_3\rtimes_{\alpha} \gg$, we get an exact sequence of $\e$-modules:
\begin{equation}
\label{yonpr}
\xymatrix{
 & & 0 \ar[dr] & & 0 & & \\  
 & & & V_3 \ar[dr]^-{\nu}\ar[ur] & & & \\
0 \ar[r] & V_1 \ar[r]^-{i} & V_2 \ar[ur]^-{p} \ar@.[rr]^-{\mu:=\nu p} & & \mathfrak{e}\ar[r]^-{\pi} & \gg\ar[r] & 0
}
\end{equation}
where every $\gg$-module has been trivially extended to an $\mathfrak{e}$-module, by using the morphism $\pi \colon \mathfrak{e} \to \gg$. The following theorem shows that we have obtained a differential crossed module $\mu:V_2\to\mathfrak{e}$. It may be seen as a constructive example of Gerstenhaber correspondence \cite{ger66} between weak equivalence classes of differential crossed modules $(\partial\colon \le \to \lg,\t)$  and $H^3(\coker{(\dd)},\ker{(\dd)})$. 

\begin{Theorem}[Wagemann]
\label{thmwag}
In the above setting, \eqref{yonpr} is a differential crossed module $\mu:V_2\rightarrow\mathfrak{e}$. Its associated cohomology class in $H^3(\gg,V_1)$ is the image of the $2$-cocycle defining the abelian Lie algebra extension \eqref{abex} under the connecting homomorphism in the long exact cohomology sequence associated to \eqref{seqmod1}.
\end{Theorem}

\proof{We give the most important steps of the proof, for more details see \cite[Thm 3]{wag06} and references therein.
That $\mu:V_2\to\mathfrak{e}$ satisfies the axioms of a differential crossed modules it is a standard exercise that we leave to the reader. More interesting is the second claim about cohomology classes. 

To associate a cocycle $\gamma\in H^3(\gg,V_1)$ to the differential crossed module $\mu:V_2\to\mathfrak{e}$ we can proceed as follows. 
We already know how to associate to the abelian extension $0\to V_3\to\mathfrak{e}\to\gg\to 0$  a $2$-cocycle $\alpha'\in H^2(\gg,V_3)$. We go on by considering $b:\textrm{im } (p)\to V_2$ a section of $p$, i.e. $p\circ b=\id_{\textrm{im} (p)}$; we then set $\beta(x_1,x_2):= b(\alpha'(x_1,x_2))$. Hence $p(\beta(x_1,x_2))=\alpha'(x_1,x_2)$ and $\beta \in C^2(\gg,V_2)$. We now show that $p \, \big((\delta^{V_2}\beta)(x_1,x_2,x_3)\big)=0$. Indeed:
\begin{equation*}
\begin{split}
p \, \big( & (\delta^{V_2} \beta)(x_1,x_2,x_3) \big) = \\
& = p \, \big( - \beta([x_1,x_2],x_3) + \beta([x_1,x_3],x_2) - \beta([x_2,x_3],x_1) + x_1\t\beta(x_2,x_3) - x_2\t\beta(x_1,x_3) + x_3\t\beta(x_1,x_2) \big) \\
& = - \alpha'([x_1,x_2],x_3) + \alpha'([x_1,x_3],x_2) - \alpha'([x_2,x_3],x_1) + x_1\t\alpha'(x_2,x_3) - x_2\t\alpha'(x_1,x_3) + x_3\t\alpha'(x_1,x_2) \\
& = (d^{V_3}\alpha')(x_1,x_2,x_3) = 0 \,,
\end{split}
\end{equation*}
where we used that $p$ is a $\gg$-module map. This shows that $\delta^{V_2}\beta$ takes value in $\textrm{im} {(i)}$. We can further lift by considering $c:i(V_1)\subset V_2 \to V_1$ a section of $i$, and introducing $\gamma:=c(d^{V_2}\beta)$. So we have $\gamma\in C^3(\gg,V_1)$, and $\delta^{V_1}\gamma=0$ from
$$ i \, \delta^{V_1}\gamma = \delta^{V_2} (i \, \gamma) = \delta^{V_2} (i \, c \, d^{V_2}\beta ) = \delta^{V_2} (\delta^{V_2}\beta) = 0 $$ 
where we denoted by the same symbol $i$ the injective chain map $i:C^{\bullet}(\gg,V_1)\to C^{\bullet}(\gg,V_2)$ induced from the sequence \eqref{seqmod}. We have therefore determined the cohomology class $\gamma\in H^3(\gg,V_1)$ associated to the differential crossed module. Of course one has further to show that the choice of different sections $a'$, $b'$ or $c'$ would lead to a $\gamma'$ such that $[\gamma']=[\gamma]$ in $H^3(\gg,V_1)$. (The process we used to define a three dimensional cohomology class from $\mu\colon V_2 \to \mathfrak{e}$ generalises readily to any differential crossed module.)

On the other hand, the image of $\alpha'$ under the connecting homomorphism $\delta:H^2(\gg,V_3)\to H^3(\gg,V_1)$ is computed exactly in the same way we obtained $\gamma$:
\begin{equation*}
\xymatrix{
 0\ar[r] & C^2(\gg,V_1)\ar[r]^i \ar[d]^-{\delta^{V_1}} & \stackrel{\beta=b(\alpha')}{C^2(\gg,V_2)} \ar[r]^{p} \ar[d]^-{\delta^{V_2}} & \stackrel{\alpha'}{C^2(\gg,V_3)} \ar[r] \ar[d]^-{\delta^{V_3}} \ar @/_{15pt}/[l]_b & 0 \\
 0 \ar[r] & \stackrel{\gamma = c(\delta^{V_2}\beta)}{C^3(\gg,V_1)} \ar[r]^i & \stackrel{\delta^{V_2}\beta}{C^3(\gg,V_2)} \ar[r]^{p} \ar @/_{15pt}/[l]_c & C^3(\gg,V_3) \ar[r] & 0
}
\end{equation*}
so that indeed $\delta \alpha' = c(\delta^{V_2}(b(\alpha'))) = \gamma \in H^3(\gg,V_1)$.
\qedhere}

\subsection{{The differential crossed module $\st$ associated to the string Lie 2-algebra}}\label{tsl2a} In this section we apply Theorem \ref{thmwag} to a particular abelian extension of $\sl$, recovering the differential crossed module {$\st=(\dd:\Fo\lra\fsl,\t)$}, associated to the  string Lie 2-algebra as a particular instance of \eqref{yonpr}. As mentioned above, this construction is due to Wagemann \cite{wag06}.

We start by fixing the notation. Let $W_1$ be the Lie algebra of vector fields in one variable $x$, with Lie bracket given by the commutator of vector fields: 
$$
\big[ f(x) \, \frac{\d}{\d x} \, , \, g(x) \, \frac{\d}{\d x} \big] =  \big(f \, \frac{\d g}{\d x} - \frac{\d f}{\d x} \, g \big)(x) \, \frac{\d}{\d x} = \big(fg' - f'g\big)(x) \, \frac{\d}{\d x} \, , \qquad \forall \, f(x)\,\frac{\d}{\d x} \, , \, g(x)\,\frac{\d}{\d x} \in W_1 \, , $$
where $f'$ denotes the derivative of $f$. 
Identify $\sl\subset W_1$ as the Lie subalgebra generated by:
\begin{equation}\label{incl}
\em = \frac{\d}{\d x} \, , \qquad\qquad \eo = x \, \frac{\d}{\d x} \, , \qquad\qquad \ep = x^2 \, \frac{\d}{\d x},
\end{equation}
so that the commutation relations read:
\begin{equation}\label{com}
 [\eo , \em ] = - \, \em \, , \qquad\qquad [\em , \ep] = 2\eo \, , \qquad\qquad [\eo , \ep] = \ep \, .
\end{equation}
The Cartan-killing form $\langle \, , \, \rangle:\sl\ot\sl\to\mathbb{C}$ is taken with the normalisation: $\langle x,y \rangle = -(1/2) \, \mathrm{Tr}\,(\mathrm{ad}_x\circ \mathrm{ad}_y)$, in terms of the adjoint representation $\mathrm{ad}_x(y)=[x,y]$ for $x,y \in \sl$. An orthonormal basis is given by the matrices:
\begin{equation}
\label{ortbas}
s_1 = \left( \begin{array}{cc} 0 & \ii \\ \ii & 0 \end{array} \right) \, , \qquad
s_2 = \left( \begin{array}{cc} 0 & 1   \\ -1  & 0 \end{array} \right) \, , \qquad  
s_3 = \left( \begin{array}{cc} \ii & 0 \\ 0& -\ii \end{array} \right)
\end{equation}
satisfying $\langle s_i , s_j \rangle = \delta_{ij}$, $i,j=1,2,3$. The relation between the two basis of $\sl$ is: 
\begin{equation}
\label{basis}
s_1 = \ii \, (\ep - \em) \, , \qquad s_2 = \ep + \em \, , \qquad s_3 = 2\ii \, \eo \, .
\end{equation}
\noindent
Let $\Fo$ be the space of {complex} polynomials in the variable $x$, and $\Fu$ the space of formal one-forms  $f(x)\d x$, {over the space of real numbers}, where $f(x)$ is {a complex}  polynomial. We consider $\Fo$ and $\Fu$ to be abelian Lie algebras, i.e. with the trivial Lie bracket. They are both  $W_1$-modules via the Lie derivative:
\begin{equation}
\label{w1act}
f(x)\frac{\d}{\d x} \, \t \, g(x) = (fg')(x) \, , \quad f(x)\frac{\d}{\d x} \, \t \, g(x)\d x = (fg' + f'g)(x) \, \d x \, , \qquad \forall \, f(x) \, \frac{\d}{\d x} \in W_1 \, , g(x) \in \Fo \, , g(x)\, \d x \in \Fu 
\end{equation}
hence they are $\sl$-modules as well, by restriction of the module structure. \\

Consider the $2$-cochain on $\sl$ with values in $\Fu$, $\alpha\in C^2(\sl,\Fu)$, defined as, in the basis $\{e_{-1},e_{0},e_{1}\}$ of $\sl$: 
\begin{equation}
\label{defalfa}
\alpha(\eo , \ep)= - \alpha(\ep,\eo)= 2\d x \, , \quad  \textrm{ and zero otherwise } .\end{equation} It is a direct computation to verify that $\delta^{\Fu}\alpha=0$ and that $[\alpha]\in H^2(\sl,\Fu)$ is a non trivial cohomology class. We denote $\fsl$ the abelian extension of $\sl$ by $\Fu$ associated to $\alpha$. We recall from \eqref{lbr} that $\fsl$ is the vector space $\Fu\oplus\sl$ endowed with the Lie bracket:
\begin{equation}
\label{fslbr}
\big[(a,y),(b,z)\big] := \big( y\t b - z\t a + \alpha(y,z) \, , [y,z] \, \big) \, , \quad \forall \, a,b \in \Fu \, , \, y,z \in \sl \, .
\end{equation}
Note that every $\sl$-module $F$ extends trivially to a  $\fsl$-module by setting:
\begin{equation}
\label{fslmod}
(a,y)\t f := y\t f \, , \qquad (a,y)\in\fsl , f \in F. 
\end{equation}
{In particular, we will be interested in the case $F=\Fo\,$. We now apply the general construction of the previous section to the abelian extension:}
\begin{equation}
\label{absl}
0\longrightarrow \Fu \stackrel{\nu}{\longrightarrow} \fsl \stackrel{\pi}{\longrightarrow} \sl \longrightarrow 0
\end{equation} 
together with the sequence of $\sl$-modules (restricting the $W_1$-module structure)
\begin{equation}
\label{seqF}
0 \longrightarrow \mathbb{C} \stackrel{i}{\longrightarrow} \Fo \stackrel{p}{\longrightarrow} \Fu \longrightarrow 0
\end{equation}
{where $p=d^{DR}$ is the formal de Rham differential. The differential crossed module \eqref{yonpr} associated to  this example is denoted by $\st=(\dd:\Fo\lra\fsl,\t)$, and it fits inside the  four-terms exact sequence:}
\begin{equation}
\label{mainseq} 
0 \lra \mathbb{C} \stackrel{i}{\lra} \Fo \stackrel{\dd}{\lra} \fsl \stackrel{\pi}{\lra} \sl \lra 0 \, . 
\end{equation}  

For the reader's convenience we repeat the relevant structures {in $\st=(\dd:\Fo\lra\fsl,\t)$}: the Lie bracket in $\fsl$ is given in \eqref{fslbr}, considering \eqref{defalfa}, while in $\Fo$ is trivial. The $(\fsl)$-action on $\Fo$  is of the type \eqref{fslmod}, trivially extending the one of $\sl$. The maps $\dd$ and $\pi$ are explicitly $\dd(f)=(\nu\circ \d^{DR})(f)=(\d f,0)$ and $\pi(\omega,y)= y$, for $f\in\Fo \, , \, (\omega,y)\in\fsl$; hence $\mathrm{ker} (\dd) = \mathbb{C}$ and $\mathrm{coker} (\dd)=\sl$.

From Theorem \ref{thmwag} we know that $\dd: \Fo \longrightarrow \fsl$ geometrically realises the $3$-cocycle $\delta(\alpha)\in H^3(\sl,\mathbb{C})$, where $\delta$ is the connecting homomorphism in the long exact cohomology sequence associated to \eqref{seqF}. We conclude that it is an explicit realisation of the string Lie 2-algebra by the following lemma of \cite{wag06}.

\begin{Lemma}[Wagemann]
The cocycle $\delta{\alpha}$ is proportional to the Cartan-Killing $3$-cocycle $\Xi\in H^3(\sl,\mathbb{C})$, explicitly given by 
$\Xi(x,y,z) = \langle [x,y] , z \rangle \, , \forall \, 
x,y,x\in\sl \, .$
\end{Lemma}

\proof{
We compute both cocycles. Recall from the proof of Theorem \ref{thmwag} that $\delta\alpha$ is obtained as $\delta \alpha = c({\d}^{\Fo}\beta)$, with $\beta(\eo,\ep)=2x$ (and zero otherwise) being the simplest lift of $\alpha'$ to $\Fo$. We evaluate the $3$-cocycle ${\d}^{\Fo}\beta$ on $(\em,\eo,\ep)$ and the only non-zero contribution comes from $\em\t\beta(\ep,\eo) = \em\t 2x= 2$. Hence $\delta\alpha'(\em,\eo,\ep)=2$. 

On the other hand $\Xi(\em,\eo,\ep) = \langle [\em,\eo],\ep \rangle = \langle \em , \ep \rangle$. Passing to the orthonormal basis \eqref{ortbas} via \eqref{basis} we get $ (1/4) \langle \ii \,  s_1 + s_2 , \ii \, s_1 - s_2 \rangle = - \, (1/2)$. 
 \qedhere
}

\begin{Remark}\label{crux}
{Looking at $\st=(\dd:\Fo\lra\fsl,\t)$, by using \eqref{incl}, \eqref{w1act} and \eqref{fslmod}, we can see that $(\fsl)$  acts trivially on the set of constant functions of $x$ in $\Fo$. Note that this set of constant polynomials in $x$ is exactly the kernel of the map $\dd\colon \Fo \to \fsl$, thus $(\fsl) \t \ker({\dd})=\{0\}$. This fact will be crucial later.}
\end{Remark}

\subsection{An infinitesimal {2-Yang-Baxter operator} in the string Lie-2-algebra} 
In this section we use the notation and nomenclature of Section \ref{i2rm}.
Let $\lg$ be a Lie algebra provided with a non-degenerate, symmetric bilinear form $\langle \, , \rangle$. Let $\{x_i\}$ be a basis of $\lg$ and $\{y^i\}$ be the dual basis of $\lg^*$ identified with $\lg$ by using $\langle \, , \rangle$. Then $r=\sum_i x_i \tn {y_i} \in \lg\tn \lg$ is symmetric and satisfies the 4-term relation $[r_{12}+r_{13},r_{23}]=0$, {in $(\mathcal{U}(\gg))^{\tn 3}$, where $\mathcal{U}(\gg)$ is the enveloping algebra of $\gg$,} and it is therefore an infinitesimal {Yang-Baxter operator in $\lg$}; see the Introduction.

We want to determine elements {$ \overline{r} \in (\fsl)^{ \otimes 3}$ and $P \in \overline{\lU}^{(3)}$, in such a way that $(\overline{r},P)$ is an infinitesimal {2-Yang-Baxter operator} in the  differential crossed module} $\st=(\dd:\Fo\lra\fsl,\t)$,  associated to the  string Lie 2-algebra. Our guiding principle in doing so is to use a lift of the infinitesimal Yang-Baxter operator $r$ in $\sl$, by the obvious section of  $\pi:\fsl\lra\sl$. In the notation of  Section \ref{tsl2a}, the element $r\in \sl\tn\sl$ associated to the Cartan-Killing form on $\sl$ is:
\begin{equation}
\label{r}
r = 2 \, \em\tn\ep + 2 \, \ep\tn\em - 4 \, \eo\tn\eo \, .
\end{equation}
Consider the section $\sigma: \sl \rra \fsl$ defined as $\sigma(e_i):=(0,e_i)$. Denote by $\overline{r}$ the lift of  $r \in \sl\tn \sl$ via $\sigma$:
\begin{equation}
\label{rbar}
\overline{r}=(\sigma\otimes\sigma)(r) = 2\, \zp{\ep} \tn \zp{\em} + 2 \, \zp{\em} \tn \zp{\ep} - 4 \, \zp{\eo} \tn \zp{\eo} \, \in \fsl\tn\fsl \, . 
\end{equation}
{Since $\sigma$ is not a Lie algebra map, $\overline{r}$ does not satisfy the 4-term relation $[\overline{r}_{12} +\overline{r}_{13} ,\overline{r}_{23} ]$ in  \eqref{4termR}, unlike $r$. However, by Definition \ref{irmatrix}, the tensor $P$ must be a pre-image,  under $\dd'\colon \overline{\lU}^{(3)} \to (\fsl)^{3 \otimes}$, of the tensor $[\overline{r}_{12} +\overline{r}_{13} ,\overline{r}_{23} ]\in(\fsl)\tn(\fsl) \tn (\fsl) $.}
\begin{Proposition}\label{invi}
The tensor $[\rb_{12} + \rb_{13},\rb_{23}]$ explicitly reads:
\begin{multline}
\frac{1}{16} \, [\rb_{12} + \rb_{13},\rb_{23}]  = 
\zp{\em}\tn\zp{\eo}\tn (\d x,0) + \zp{\em}\tn (\d x,0) \tn \zp{\eo} \, + \\ 
- \zp{\eo}\tn (\d x,0) \tn\zp{\em} - \zp{\eo}\tn \zp{\em}\tn (\d x,0) \, .
\end{multline}
{Every  element $P    \in \overline{\mathfrak{U}}^{(3)}$ such that ${\dd'}(P)= [\rb_{12} + \rb_{13},\rb_{23}]$ therefore has the form:}
\begin{equation}
\label{P}
\begin{split}
P &= P^0 + \lambda C \\
   &\doteq 16 \big(\zp{\em}\tn\zp{\eo}\tn x + \zp{\em}\tn x\tn\zp{\eo} - \zp{\eo}\tn x\tn \zp{\em} - \zp{\eo}\tn\zp{\em}\tn x \big) + \lambda C \,,
\end{split}
\end{equation} 
{where  $C \in \mathrm{ker}\big( \dd'\colon \overline{\lU}^{(3)} \to (\fsl)^{3 \otimes}\big)$  and $\lambda \in \C$.}
\end{Proposition}
{\proof
By direct computation:{\allowdisplaybreaks
\begin{equation*}
\begin{split}
[\rb_{12} + \rb_{13},\rb_{23}]  &= \, \big[ \, 2 \, \eep \tn \eem \tn 1 + 2 \, \eem \tn \eep \tn 1 - 4 \, \eeo \tn \eeo \tn 1 \, + \\
&\quad + 2 \, \eep \tn 1 \tn \eem + 2 \, \eem \tn 1 \tn \eep - 4 \, \eeo \tn 1 \tn \eeo \, , \\
& \qquad 2 \cdot 1 \tn \eep \tn \eem + 2 \cdot 1 \tn \eem \tn \eep - 4 \cdot 1 \tn \eeo \tn \eeo \big] = \\
& = \, 4 \, \eep \tn ( \al{\em}{\ep},2\eo) \tn \eem - 8 \, \eep \tn (\al{\em}{\eo},\em) \tn \eeo \, + \\
&\quad + 4 \, \eem \tn (\al{\ep}{\em},-2\eo) \tn \eep - 8 \, \eem \tn (\al{\ep}{\eo},-\ep) \tn \eeo \, + \\
&\quad - 8 \, \eeo \tn (\al{\eo}{\ep},\ep) \tn \eem - 8 \, \eeo \tn (\al{\eo}{\em},-\em) \tn \eep \, + \\
& \quad+ 4 \, \eep \tn \eem \tn (\al{\em}{\ep},2\eo) -8 \, \eep \tn \eeo \tn (\al{\em}{\eo},\em) \, + \\
& \quad+ 4 \, \eem \tn \eep \tn (\al{\ep}{\em},-2\eo) \tn - 8 \, \eem \tn \eeo \tn (\al{\ep}{\eo},-\ep) \, + \\
& \quad- 8 \, \eeo \tn \eep \tn (\al{\eo}{\em},-\em) - 8 \, \eeo \tn \eem \tn (\al{\eo}{\ep},\ep) \, .  
\end{split}
\end{equation*}
}
Denoting with $l.i$ the $i^{th}$ term in the $l^{th}$ line, the following cancellations occur: $1.1$ with $4.2$, $1.2$ with $4.1$. We are left with
\begin{equation*}
8 \, \eem \tn \eeo \tn (2\d x, 0) + 8 \, \eem \tn (2\d x,0) \tn \eeo 
- 8 \, \eeo \tn (2\d x,0) \tn \eem - 8 \, \eeo \tn \eem \tn (2\d x,0) \, ,
\end{equation*}
explicitly $(2.1+5.2)+(2.2+5.1)+(3.1+5.1)+(3.2+6.2)$.
The second claim is easily verified.
\qedhere }\\

{
Given that $\beta'(P)=[\overline r_{12}+\overline r_{13},\overline r_{23}]$, the pair  $(\overline{r},P)=(\overline{r},P^0+\lambda C)$ {will be an} infinitesimal 2-Yang-Baxter operator in $\st$, if, and only if, the following relations hold, freely or evaluated in some  categorical representation of the string differential crossed module, definitions \ref{irmatrix} and \ref{irmatrix2}:
\begin{enumerate}[(i)]
\item \label{1i} $\overline{r}_{12}=\overline{r}_{21}$,
\item \label{2i}  $P_{123} + P_{231} + P_{312} = 0$,
\item \label{3i} $P_{123} = P_{132}$,
\item \label{4i} $\rb_{14}\t (P_{213} + P_{234}) +(\rb_{12} + \rb_{23} + \rb_{24})\t P_{314} - (\rb_{13}+\rb_{34})\t P_{214} = 0$,
\item \label{5i} $\rb_{23}\t (P_{214} + P_{314}) - \rb_{14}\t (P_{423}+P_{123}) = 0.$
\end{enumerate}}\noindent {Our main result (Theorem \ref{main2r}) states that there exist $C \in \mathrm{ker} {(\dd')}$ and $\lambda \in \C$, so that $(\overline{r},P^0+\lambda C)$ is a free infinitesimal 2-Yang-Baxter operator in  $\st$. We now start discussing the proof of  this result.} 

{Noting that  $\overline{r}_{12}=\overline{r}_{21}$, thus relation \eqref{1i} is satisfied, let us now evaluate the left-hand-side of relations {\eqref{2i}} to \eqref{5i} on $P^0$. We immediately see that (concerning conditions {\eqref{2i}} and {\eqref{3i}}):
\begin{equation}
 \begin{split}
P^0_{123} + P^0_{231} + P^0_{312} &= 0,\\
P^0_{123} &= P^0_{132}.	
\end{split}
\end{equation}}
{Concerning conditions {\eqref{4i}} and {\eqref{5i}}, we  compute, in a lengthy but straightforward way,  $\rb_{14}\t (P^0_{213} + P^0_{234}) +(\rb_{12} + \rb_{23} + \rb_{24})\t P^0_{314} - (\rb_{13}+\rb_{34})\t P^0_{214}$ and 
 $\rb_{23}\t (P^0_{214} + P^0_{314}) - \rb_{14}\t (P^0_{423}+P^0_{123})$, in \ref{reliii} and \ref{reliv}; see \eqref{12t} and \eqref{canc}. While relation {\eqref{5i}} holds for $P^0$, the left-hand-side of {\eqref{4i}} produces non-zero terms when applied to $P^0$.  In section \ref{corrC}, we discuss how it is possible to choose  $C \in \mathrm{ker}\big( \dd'\colon \overline{\lU}^{(3)} \to (\fsl)^{3 \otimes}\big)$ and $\lambda \in \C$, in {order to} cancel these terms.  Nevertheless, by the calculations in \ref{reliii}, we will be able to see that the left-hand-side of relation {\eqref{4i}} applied to $P^0$ acts trivially in the adjoint {categorical} representation of $\st$, so that $(\bar r,P^0)$ is found to be an infinitesimal $2$-Yang-Baxter operator, with respect to the adjoint {categorical} representation. The latter result can however be obtained in a more general way:  }

\begin{Theorem}
\label{adj2R}
The {following elements $\overline{r}\in (\fsl)\tn(\fsl)$ and $P^0\in \overline{\mathfrak{U}}^{(3)}$:}
\begin{align*}
\overline{r} & = 2\, \zp{\ep} \tn \zp{\em} + 2 \, \zp{\em} \tn \zp{\ep} - 4 \, \zp{\eo} \tn \zp{\eo} \,,\\
P^0  & = 16 \, \big(\zp{\em}\tn\zp{\eo}\tn x + \zp{\em}\tn x\tn\zp{\eo} - \zp{\eo}\tn x\tn \zp{\em} - \zp{\eo}\tn\zp{\em}\tn x \big) \, ,
\end{align*}
define an infinitesimal {2-Yang-Baxter operator}, with respect to  the adjoint categorical representation of $\st$. 
\end{Theorem}

\proof{
{The crucial fact here is that, in   $\st=(\partial \colon \Fo \to \fsl ,\t)$,  the kernel of ${\dd'}\colon {\overline{\lU}}^{(4)} \to (\fsl)^{\otimes 4}$ acts trivially in the chain complex $\underline{\st}^{ \otimes 4}$, via the insertion map \ref{Sinsmap}. This is a consequence of the fact that the kernel of $\partial \colon \Fo \to  \fsl$ (which is the set of constant functions in $\Fo$) is $(\fsl)$-invariant, Remark \ref{crux}.}

{Let us give full details. Since we are considering chain complexes of length 2, of vector spaces, K\"unneth Theorem tells us that:}
\begin{align*}
\ker &\big({\dd'}\colon {\overline{\lU}}^{(4)} \to (\fsl)^{\otimes 4}\big)\\&=\big(\ker (\dd \colon \Fo \to \fsl)\big) \otimes (\{0\} \rtimes \sl) \otimes  (\{0\} \rtimes \sl) \otimes (\{0\} \rtimes \sl)\oplus  \textrm{ cyclic permutations }.\end{align*}
{From the explicit expression of the insertion maps for a categorical representation (see \ref{Sinsmap} and in particular \eqref{phiUk}) and of the map   $\rho=\rho^1$ in the adjoint categorical representation, see \eqref{rho1ad}, we conclude, since   $\ker(\partial \colon \Fo \to  \fsl)$ is $(\fsl)$-invariant, that the kernel of ${\dd'\colon {\overline{\lU}}^{(4)} \to (\fsl)^{\otimes 4}}$ acts trivially in the fourth tensor power of chain complex $\underline{\st}$, associated to $\st$.}

{By Example \ref{exid},  if we apply ${\dd'}\colon {\overline{\lU}}^{(4)} \to (\fsl)^{\otimes 4}$ to the left-hand-side of relations {\eqref{4i}} and {\eqref{5i}}, above,  on $P^0$, we get zero,  since  ${\dd'}\colon {\overline{\lU}}^{(4)} \to (\fsl)^{\otimes 4}  $ is an $(\fsl)$-module map and ${\dd'}(P^0)= [ \rb_{12} + \rb_{13},\rb_{23} ] $. Therefore relations {\eqref{4i}} and {\eqref{5i}} are satisfied by $(\overline{r}, P^0)$, in the adjoint categorical  representation of $\st$. \qedhere}
}
\begin{Remark}\label{taut}
 {By the proof of the previous theorem a more general result holds. Looking at \eqref{mainseq}, let $t \in (\fsl) \tn (\fsl)$ be any inverse image, under $\pi \tn \pi$, of $r \in \sl \tn \sl$, where $r$ is in \eqref{r}. Let also $Q$ be any inverse image, under $\dd'\colon \overline{\lU}^{(3)} \to (\fsl)^{\tn 3}$, of $[t_{12}+t_{13},t_{23}]$.  Then $(t,Q)$ is an infinitesimal 2-Yang-Baxter operator in $\st$, with respect to the adjoint representation. }
\end{Remark}

\subsubsection{The left-hand-side of relation {\eqref{4i}} applied to $P^0$} 
\label{reliii}

A word on notation: to distinguish between the unital elements of $\Fo$ (i.e. the constant function of value $1$) and the identity in the enveloping algebra $\mathcal{U}(\fsl)$, from now on we will denote the former as $1_{\Fo}$, writing simply $1$ for the latter.

{Let us explicitly compute:}
$$\rb_{14}\t (P^0_{213} + P^0_{234}) +(\rb_{12} + \rb_{23} + \rb_{24})\t P^0_{314} - (\rb_{13}+\rb_{34})\t P^0_{214}.$$
We compute one term at a time, starting with 
$\frac{1}{16}\rb_{14} \t(P^0_{213}+P^0_{234})$ which explicitly is

\begin{equation*}
\begin{split}
 \big( 2\, \zp{\ep} & \tn 1\tn 1\tn\zp{\em} + 2\, \zp{\em}\tn 1\tn 1\tn\zp{\ep} - 4\, \zp{\eo}\tn 1\tn 1\tn\zp{\eo}\big) \, \t \\
 & \big( \zp{\eo}\tn\zp{\em}\tn x\tn 1 + x\tn\zp{\em}\tn\zp{\eo}\tn 1 - x\tn\zp{\eo}\tn\zp{\em}\tn 1 - \zp{\em}\tn\zp{\eo}\tn x\tn 1 \, + \\
 & + 1\tn\zp{\em}\tn\zp{\eo}\tn x + 1\tn\zp{\em}\tn x\tn\zp{\eo} - 1\tn\zp{\eo}\tn x\tn\zp{\em} - 1\tn\zp{\eo}\tn\zp{\em}\tn x \big) \, . 
\end{split}
\end{equation*}
The computations are done in the following order: in the first two lines we write the first term of $\overline{r}_{14}$ acting on $P^0_{213}$, in the third and fourth line we write the second term of $\overline{r}_{14}$ acting on $P^0_{213}$, then the fifth and sixth line refer to the third term of $\overline{r}_{14}$ acting on $P^0_{213}$; finally the same order is used for $\overline{r}_{14}$ acting on $P^0_{234}$, giving $12$ lines in total:
{\allowdisplaybreaks
\begin{equation*}
\begin{split}
 & \quad 2\, (\al{\ep}{\eo},-\ep)\tn\zp{\em}\tn x\tn\zp{\em} + 2\, \ep\t x\tn\zp{\em}\tn\zp{\eo}\tn\zp{\em} \, + \\
 & -2\, \ep\t x\tn\zp{\eo}\tn\zp{\em}\tn\zp{\em} - 2\, (\al{\ep}{\em},-2\, \eo)\tn\zp{\eo}\tn x\tn\zp{\em} \, + \\
 & + 2\, (\al{\em}{\eo},\em)\tn\zp{\em}\tn x\tn\zp{\ep} + 2\, \em\t x\tn\zp{\em}\tn\zp{\eo}\tn\zp{\ep} \, + \\
 & - 2\, \em\t x\tn\zp{\eo}\tn\zp{\em}\tn\zp{\ep} \, + \\
 & -4\, \eo\t x\tn \zp{\em}\tn\zp{\eo}\tn\zp{\eo} + 4\, \eo\t x\tn\eeo\tn\eem\tn\eeo \, + \\
 & +4\, (\al{\eo}{\em},-\em)\tn\eeo\tn x\tn\eeo \, + \\
 & +2\, \eep\tn\eem\tn\eeo\tn \em\t x + 2\, \eep\tn\eem\tn x\tn (\al{\em}{\eo},\em) \, + \\
 & -2\, \eep\tn\eeo\tn\eem\tn \em\t x \, + \\
 & + 2\, \eem\tn\eem\tn\eeo\tn \ep\t x + 2\, \eem\tn\eem\tn x\tn (\al{\ep}{\eo},-\ep) \, + \\
 & -2\, \eem\tn\eeo\tn x (\al{\ep}{\em},-2\eo) -2\, \eem\tn\eeo\tn\eem\tn\ep\t x + \, \\
 & -4\, \eeo\tn\eem\tn\eeo\tn\eo\t x + 4\, \eeo\tn\eeo\tn x\tn (\al{\eo}{\em},-\em) \, + \\
 & +4\, \eeo\tn\eeo\tn\eem\tn\eo\t x \, .
\end{split}
\end{equation*}}
We next evaluate the cocycle $\alpha$ and the action of $e_i$ on $x$. We denote as before $l.i$ the $i^{\rm th}$ term in the $l^{\rm th}$ line; then $2.2$ cancels with $11.2$, and $6.1$ with $10.1$. We can also sum $1.1$ with $7.2$ and $3.1$ with $9.2$. We get:
{\allowdisplaybreaks
\begin{equation*}
\begin{split}
& - 4\, (\d x,0)\tn\eem\tn x\tn\eem + 2\, x^2\tn\eem\tn\eeo\tn\eem \, + \\
& - 2\, x^2\tn\eeo\tn\eem\tn\eem -4\, \eem\tn\eem\tn x\tn (\d x,0) \, + \\
& + 2 \cdot\ufo\tn\eem\tn\eeo\tn\eep - 2\cdot\ufo\tn\eeo\tn\eem\tn\eep \, + \\
& - 4\, x\tn\eem\tn\eeo\tn\eeo + 4\, x\tn\eeo\tn\eem\tn\eeo \, + \\
& + 2\, \eep\tn\eem\tn\eeo\tn \ufo -2\, \eep\tn\eeo\tn\eem\tn \ufo \, + \\
& + 2\, \eem\tn\eem\tn\eeo\tn x^2 -2\, \eem\tn\eeo\tn\eem\tn x^2 \, + \\
& -4\, \eeo\tn\eem\tn\eeo\tn x + 4\, \eeo\tn\eeo\tn\eem\tn x \, .
\end{split}
\end{equation*}}
We refer to these terms as the term of type I; so the first term will be denoted as I.1, the second as I.2 etc. 

We proceed now with $\frac{1}{16}(\rb_{12} + \rb_{23} + \rb_{24})\t P^0_{314}$. Using the explicit expressions for $\overline{r}$ and $P^0$ this is written as
{\allowdisplaybreaks
\begin{equation*}
\begin{split}
& \big(2\, \eep\tn\eem\tn 1\tn 1 + 2\, \eem\tn\eep\tn 1\tn 1 - 4\, \eeo\tn\eeo\tn 1\tn 1 \, + \\
& \; + 2\tn\eep\tn\eem\tn 1 + \quad 2\tn\eem\tn\eep\tn 1 - 4\tn\eeo\tn\eeo\tn 1 \, + \\
& \; + 2\tn\eep\tn 1\tn\eem + 2\tn\eem\tn 1\tn\eep - 4\tn\eeo\tn 1\tn\eeo \big) \, \t \\
& \big( \eeo\tn 1\tn\eem\tn x + x\tn 1\tn\eem\tn\eeo - x\tn 1\tn\eeo\tn\eem  - \eem\tn 1\tn\eeo\tn x \big) \, . 
\end{split}
\end{equation*}}
We compute the terms using the same order as before: the first two lines refer to the action of the first term of $\overline{r}_{12}$, then the second (resp. third) term of $\overline{r}_{12}$ acting on $P^0_{314}$ gives the third and fourth (resp. fifth and sixth) lines, the seventh and eight line refer to the action of the first term of $\overline{r}_{23}$ and so on:
{\allowdisplaybreaks
\begin{equation*}
\begin{split}
& \quad \, 2 \, (\al{\ep}{\eo},-\ep)\tn\eem\tn\eem\tn x + 2\, \ep\t x\tn\eem\tn\eem\tn\eeo \, + \\
& -2\, \ep\t x\tn \eem\tn \eeo\tn\eem -2\, (\al{\ep}{\em},-2\eo)\tn\eem\tn\eeo\tn x \, + \\
& +2 (\al{\em}{\eo},\em)\tn\ep\tn\em\tn x + 2\, \em\t x \tn\eep\tn\eem\tn\eeo \, + \\
& -2\, \em\t x \tn \eep\tn\eeo\tn\eem \, + \\
& -4 \, \eo\t x\tn\eeo\tn\eem\tn\eeo + 4\, \eo\t x\tn \eeo\tn\eeo\tn\eem \, + \\
& +4 \, (\al{\eo}{\em},-\em)\tn\eeo\tn\eeo\tn x \, + \\
& -2\, x\tn\eep\tn (\al{\em}{\eo},\em)\tn\eem - 2, \eem\tn\eep\tn (\al{\em}{\eo},\em)\tn x \, + \\
& +2\, \eeo\tn\eem\tn (\al{\ep}{\em},-2\eo)\tn x + 2\, x\tn\eem\tn (\al{\ep}{\em},-2\eo)\tn\eeo \, + \\
& -2\, x\tn\eem\tn (\al{\ep}{\eo},-\em)\tn\eem - 2\, \eem\tn\eem\tn (\al{\ep}{\eo},-\ep)\tn x , + \\
& -4\, \eeo\tn\eeo\tn (\al{\eo}{\em},-\em)\tn x - 4 \, x\tn\eeo\tn (\al{\eo}{\em},-\em)\tn\eeo \, + \\
& +2\, \eeo\tn\eep\tn\eem\tn\em\t x + 2\, x\tn\eep\tn\eem\tn (\al{\em}{\eo},\em) \,+ \\
& -2\, \eem\tn\eep\tn\eeo\tn\em\t x \, + \\
& + 2\, \eeo\tn\eem\tn\eem\tn\ep\t x + 2\, x\tn\eem\tn\eem\tn (\al{\ep}{\eo},-\ep) \, + \\
& -2\, x\tn\eem\tn\eeo\tn (\al{\ep}{\em},-2\eo) -2\, \eem\tn\eem\tn\eeo\tn\ep\t x \, + \\
& -4\, \eeo\tn\eeo\tn\eem\tn\eo\t x +4\, x\tn\eeo\tn\eeo\tn (\al{\eo}{\em},-\em) \, + \\
& +4\, \eem\tn\eeo\tn\eeo\tn \eo\t x \, .
\end{split}
\end{equation*}}
Again we evaluate the cocycle $\alpha$ and the action of $e_i$ on $x$. After that there are several cancellations; in the usual notation these are: $2.2$ with $8.1$, $3.1$ with $7.2$, $5.1$ with $10.2$, $5.2$ with $15.2$, $6.1$ with $16.1$, $7.1$ with $11.2$, $8.2$ with $14.1$ and $10.1$ with $15.1$. Hence we get:
\begin{equation*}
\begin{split}
& -2\, (2\d x,\ep)\tn\eem\tn\eem\tn x + 2\, x^2\tn\eem\tn\eem\tn\eeo \, + \\
& -2\, x^2\tn\eem\tn\eeo\tn\eem + 2\cdot\ufo\tn\eep\tn\eem\tn\eeo \, + \\
& -2\cdot\ufo\tn\eep\tn\eeo\tn\eem + 2\, x\tn\eem\tn (2\d x,\ep)\tn\eem \, + \\
& +2\, \eem\tn\eem\tn (2\d x,\ep)\tn x + 2\, \eeo\tn\eep\tn\eem\tn \ufo \, + \\
& -2\, \eem\tn\eep\tn\eeo\tn \ufo + 2\, \eeo\tn\eem\tn\eem\tn x^2 \, + \\
& -2\, x\tn\eem\tn\eem\tn (2\d x,\ep) - 2\, \eem\tn\eem\tn\eeo\tn x^2 \, .
\end{split}
\end{equation*}
We denote these terms as the terms of type II, and refer to them as II.1, II.2 etc. 

We finally compute $\frac{1}{16}(\rb_{13}+\rb_{34})\t P^0_{214}$. The explicit form is:
\begin{equation*}
\begin{split}
& \big( 2\, \eep\tn 1\tn\eem\tn 1 + 2\, \eem\tn 1\tn\eep\tn 1 -4\, \eeo\tn 1\tn\eeo\tn 1 \, + \\
& \; + 2\tn 1\tn \eep\tn\eem + 2\tn 1\tn \eem\tn \eep - 4\tn 1\tn \eeo\tn\eeo \big) \, \t \\
& \big( \eeo\tn\eem\tn 1\tn x + x\tn\eem\tn 1\tn\eeo - x\tn\eeo\tn 1\tn\eem - \eem\tn\eeo\tn 1\tn x \big) \, .
\end{split}
\end{equation*}   
With the usual order of computation (first two lines given by the first term of $\rb_{13}$ acting on $P^0_{214}$, seventh and eight line given by the first term of $\rb_{34}$ acting on $P^0_{214}$ etc.), we have:
{\allowdisplaybreaks
\begin{equation*}
\begin{split}
& 2\, (\al{\ep}{\eo},-\ep)\tn\eem\tn\eem\tn x + 2\, \ep\t x\tn \eem\tn\eem\tn\eeo \, + \\
& -2\, \ep\t x\tn\eeo\tn\eem\tn\eem - 2\, (\al{\ep}{\em},-2\eo)\tn\eeo\tn\eem\tn x \, + \\
& +2\, (\al{\em}{\eo},\em) \tn \eem\tn\eep\tn x + 2\, \em\t x \tn \eem \tn \eep\tn\eeo \, + \\
& -2\, \em\t x\tn\eeo\tn\eep\tn\eem \, + \\
& -4\, \eo\t x\tn \eem\tn\eeo\tn\eeo + 4\, \eo\t x \tn \eeo\tn\eeo\tn\eem \, + \\
& +4\, (\al{\eo}{\em},-\em)\tn\eeo\tn\eeo\tn x \, + \\
& +2\, \eeo\tn\eem\tn\eep\tn \em\t x + 2\, x\tn\eem\tn\eep\tn (\al{\em}{\eo},\em) \, + \\
& -2\, \eem\tn\eeo\tn\eep\tn\em\t x \, + \\
& +2\, \eeo\tn\eem\tn\eem\tn\ep\t x + 2\, x\tn\eem\tn\eem\tn (\al{\ep}{\eo},-\ep) \, + \\
& -2\, x\tn\eeo\tn\eem\tn (\al{\ep}{\em},-2\eo) - 2\, \eem\tn\eeo\tn\eem\tn\ep\t x \, + \\
& -4\, \eeo\tn\eem\tn\eeo\tn\eo\t x + 4\, x\tn\eeo\tn\eeo\tn (\al{\eo}{\em},-\em) \, + \\
& +4\, \eem\tn\eeo\tn\eeo\tn\eo\t x \, .
\end{split}
\end{equation*}}
We evaluate the cocycle $\alpha$ and the action of $e_i$ on $x$. We have cancellations of $5.2$ with $11.2$ and of $6.1$ with $12.1$. We then get:
\begin{equation*}
\begin{split}
& -2\, (2\d x,\ep)\tn\eep\tn\eem\tn x + 2\, x^2\tn\eem\tn\eem\tn\eeo \, + \\
& -2\, x^2\tn\eeo\tn\eem\tn\eem + 4\, \eeo\tn\eeo\tn\eem\tn x \, + \\
& +2\, \eem\tn\eem\tn\eep\tn x + 2\cdot\ufo\tn\eem\tn\eep\tn\eeo \, + \\
& -2\cdot\ufo\tn\eeo\tn\eep\tn\eem -4\, x\tn\eem\tn\eeo\tn\eeo \, + \\
& +2\, \eeo\tn\eem\tn\eep\tn \ufo + 2\, x\tn \eem\tn\eep\tn\eem \, + \\
& -2\, \eem\tn\eeo\tn\eep\tn \ufo + 2\, \eeo\tn\eem\tn\eem\tn x^2 \, + \\
& -2\, x\tn\eem\tn\eem\tn (2\d x,\ep) + 4\, x\tn\eeo\tn\eem\tn\eeo \, + \\
& -2\, \eem\tn\eeo\tn\eem\tn x^2 -4\, \eeo\tn\eem\tn\eeo\tn x \, .
\end{split}
\end{equation*}  
We denote these terms as the terms of type III, and refer to them as III.1, III.2 etc.

We can now sum everything together, i.e. terms of type I plus terms of plus II minus terms of type III. We have the following cancellations: \\

\begin{tabular}{rcllrcllrcllrcl}
I.2 & with & II.3 \, ; &  & I.3 & with & III.3 \, ; & & I.7 & with & III.8 \, ; & & I.8 & with & III.14 \, ;\\
I.11 & with & II.12 \, ; & & I.12 & with & III.15 \, ; & & I.13 & with & III.16 \, ; & &I.14 & with & III.4 \, ; \\
II.1 & with & III.1 \, ; & & II.2 & with & III.2 \, ; & &II.10  & with & III.12 \, ; & &II.11 & with & III.13 \, .  
\end{tabular}\\

\noindent
Among what is left, we compute II.6 - III.10 to be 
$$4 \, x \tn \eem \tn (\d x,0) \tn \eem $$ 
and this cancel with I.1 in ${\overline{\lU}}^{(4)}$ by \eqref{irel}. Similarly, we compute II.7 - III.5 to be
$$4 \, \eem \tn \eem \tn (\d x,0) \tn x $$ 
which cancels with I.4, again in ${\overline{\lU}}^{(4)}$. The remaining terms are conveniently assembled into two groups as
\begin{equation*}
\mathrm{I.5 + I.6 + II.4 + II.5 - III.6 - III.7} \, ; \qquad  
\mathrm{I.9 + I.10 + II.8 + II.9 - III.9 - III.11} \, .
\end{equation*}
If we write these terms explicitly following the same order we see that:
\begin{equation}
\label{12t}
\begin{split}
\frac{1}{2\times 16} \,\big( \rb_{14}\t (P^0_{213} + P^0_{234}) +  & (\rb_{12} + \rb_{23} + \rb_{24})\t P^0_{314} - (\rb_{13}+\rb_{34})\t P^0_{214} \big) \,= \\
&\quad =  \ufo\tn\eem\tn\eeo\tn\eep - \ufo\tn\eeo\tn\eem\tn\eep \, + \\
&\quad  + \, \ufo\tn\eep\tn\eem\tn\eeo - \ufo\tn\eep\tn\eeo\tn\eem \, + \\
&\quad  - \, \ufo\tn\eem\tn\eep\tn\eeo + \ufo\tn\eeo\tn\eep\tn\eem \, + \\
&\quad  + \, \eep\tn\eem\tn\eeo\tn\ufo - \, \eep\tn\eeo\tn\eem\tn\ufo \, + \\ 
&\quad  +\, \eeo\tn\eep\tn\eem\tn\ufo \, - \, \eem\tn\eep\tn\eeo\tn\ufo \, + \\
&\quad  -\, \eeo\tn\eem\tn\eep\tn\ufo + \, \eem\tn\eeo\tn\eep\tn\ufo \, \\
&\quad ={-\frac{1}{8} \left (\ufo\tn \overline{ V} + \overline{V} \tn \ufo\right ),}
\end{split}
\end{equation}
{where $\overline{V}$ is defined in \ref{corrC}, below.
These terms do not vanish in general. However, since $\ufo$, appearing in \eqref{12t} either in the first or in the fourth factor of the tensor product, is $(\fsl)$-invariant (Remark \ref{crux}), the right-hand-side of \eqref{12t} vanishes in the adjoint categorical representation of $\st$, a result anticipated in Theorem \ref{adj2R}.}

\subsubsection{{The left-hand-side of relation {\eqref{5i}} applied to $P^0$}}
\label{reliv} 
Let us now explicitly compute:
$$\rb_{23}\t (P^0_{214} + P^0_{314}) - \rb_{14}\t (P^0_{423}+P^0_{123})\, .$$
Again we compute one term at time. We start with $\frac{1}{16}\rb_{23}\t (P^0_{214} + P^0_{314})$, which explicitly is
\begin{equation*}
\begin{split}
 \big( 2\tn\eep & \tn\eem\tn 1 + 2\tn\eem\tn\eep\tn 1 - 4\tn\eeo\tn\eeo\tn 1\big) \, \t \\
& \big( \eeo \tn \eem \tn 1 \tn x  + x \tn \eem \tn 1 \tn \eeo - x \tn \eeo \tn 1 \tn \eem  - \eem \tn \eeo \tn 1 \tn x \, + \\
& \; + \eeo \tn 1 \tn \eem \tn x + x \tn 1 \tn \eem \tn \eeo - x \tn \eeo \tn 1 \tn \eem - \eem \tn 1 \tn \eeo \tn x \big) \, .
\end{split}
\end{equation*}
We follow the usual order for computations, acting first with each factor of $\overline{r}_{23}$ on each factor of $P^0_{214}$, and then on $P^0_{314}$. We directly evaluate the action of the cocycle $\alpha$. We get:
{\allowdisplaybreaks
\begin{equation*}
\begin{split}
& \quad \, 2 \, \eeo \tn (0,-2\eo) \tn \eem \tn x + 2 \, x \tn (0,-2\eo) \tn \eem \tn \eeo \, + \\
& - 2 \, x \tn (-2\d x,-\ep) \tn \eem \tn \eem - 2 \, \eem \tn (-2\d x,-\ep) \tn \eem \tn x \, + \\
& - 2 \, x \tn \eem \tn \eep \tn \eem - 2 \, \eem \tn \eem \tn \eep \tn x \, + \\
& - 4 \, \eeo \tn (0,-\em) \tn \eeo \tn x - 4 \, x \tn (0,-\em) \tn \eeo \tn \eeo \, + \\
& - 2 \, x \tn \eep \tn \eem \tn \eem - 2 \, \eem \tn \eep \tn \eem \tn x \, + \\
& + 2 \, \eeo \tn \eem \tn (0,-2\eo) \tn x + 2 \, x\tn \eem \tn (0,-2\eo) \tn \eeo \, + \\
& -2 \, x \tn \eem \tn (-2\d x,\ep) \tn \eem - 2 \, \eem \tn \eem \tn (-2\d x,-\ep) \tn x \, + \\
& -4 \, \eeo \tn \eeo \tn (0,-\em) \tn x - 4 \, x \tn \eeo \tn (0,-\em) \tn \eeo \, .
\end{split}
\end{equation*}}
We have the following cancellations: $1.1$ with $8.1$, $1.2$ with $8.2$, $4.1$ with $6.1$ and $4.2$ with $6.2$. The remaining contribution is
\begin{equation*}
\begin{split}
& 4 \, x \tn (\d x,0) \tn \eem \tn \eem + 4 \, \eem \tn (\d x,0) \tn \eem \tn x \, + \\
& 4 \, x \tn \eem \tn (\d x,0) \tn \eem + 4 \, \eem \tn \eem \tn (\d x,0) \tn x 
\end{split}
\end{equation*}
where the four terms come respectively from the summation of $2.1$ with $5.1$, $2.2$ with $5.2$, $3.1$ with $7.1$ and $3.2$ with $7.2$. We denote them terms of type I.

We go ahead by computing $\frac{1}{16}\overline{r}_{14}\t (P^0_{423} + P^0_{123})$. This explicitly is 
\begin{equation*}
\begin{split}
 \big( 2\, \zp{\ep} & \tn 1 \tn 1\tn\zp{\em} + 2\, \zp{\em}\tn 1\tn 1\tn\zp{\ep} - 4\, \zp{\eo}\tn 1\tn 1\tn\zp{\eo}\big) \, \t \\
 & \big( 1 \tn \eeo \tn x \tn \eem + 1 \tn x \tn \eeo \tn \eem - 1\tn x \tn \eem \tn \eeo  - 1 \tn \eem \tn x \tn \eeo \, + \\
 & \; + \eem \tn \eeo \tn x \tn 1 + \eem \tn x \tn \eeo \tn 1 - \eeo \tn x \tn \eem \tn 1 - \eeo \tn \eem \tn x \tn 1 \big) \, .
\end{split}
\end{equation*}
We follow the usual order for computing the various terms, evaluating the action of the cocycle $\alpha$. We get:
{\allowdisplaybreaks
\begin{equation*}
\begin{split}
& -2 \, \eep \tn x \tn \eem \tn \eem - 2 \, \eep \tn \eem \tn x \tn \eem \, + \\
& +2 \, \eem \tn \eeo \tn x \tn (0,-2\eo) + 2 \, \eem \tn x \tn \eeo \tn (0,-2\eo) \, + \\
& -2 \, \eem \tn x \tn \eem \tn (-2\d x,-\em) - 2 \, \eem \tn \eem \tn x \tn (-2\d x, -\ep) \, + \\
& -4 \, \eeo \tn \eeo \tn x \tn (0,-\em) - 4 \, \eeo \tn x \tn \eeo \tn (0,-\em) \, + \\
& +2 \, (0,-2\eo) \tn \eeo \tn x \tn \eem + 2 \, (-2\d x,-\ep) \tn \eem \tn x \tn \eem \, + \\
& -2 \, \eem \tn x \tn \eem \tn \eep - 2 \, \eem \tn \eem \tn x \tn \eep \, + \\
& -4 \, (0,-\em) \tn \eeo \tn x \tn \eeo - 4\, (0,-\em) \tn x \tn \eeo \tn \eeo \, .
\end{split}
\end{equation*}}  
We have the following cancellations: $2.1$ with $8.1$, $2.2$ with $8.2$, $4.1$ with $5.1$ and $4.2$ with $5.2$. We are left with 
\begin{equation*}
\begin{split}
& 4 \, (\d x,0) \tn x \tn \eem \tn \eem + 4 \, (\d x,0) \tn \eem \tn x \tn \eem \, + \\
& + 4 \, \eem \tn x \tn \eem \tn (\d x,0) + 4 \, \eem \tn \eem \tn x \tn (\d x,0) 
\end{split}
\end{equation*}
where the four terms come respectively from the sum of $1.1$ with $6.1$, $1.2$ with $6.2$, $3.1$ with $7.1$ and $3.2$ with $7.2$. These are denoted terms of type II. \\

We can now sum the two contributions, i.e. terms of type I minus terms of type II. This amounts to:
\begin{equation*}
\begin{split}
& \quad \, 4 \, x \tn (\d x,0) \tn \eem \tn \eem + 4 \, \eem \tn (\d x,0) \tn \eem \tn x \, + \\
& + 4 \, x \tn \eem \tn (\d x,0) \tn \eem + 4\, \eem \tn \eem \tn (\d x,0) \tn x \, + \\
& - 4 \, (\d x,0) \tn x \tn \eem \tn \eem + 4 \, (\d x,0) \tn \eem \tn x \tn \eem \, + \\
& - 4 \, \eem \tn x \tn \eem \tn (\d x,0) + 4 \, \eem \tn \eem \tn x \tn (\d x,0) ,
\end{split}
\end{equation*}
where we can recognize four pairs of terms, each one being equal to zero in ${\overline{\lU}}^{(4)}$, by \eqref{irel}: $1.1$ with $3.1$, $1.2$ with $4.1$, $2.1$ with $3.2$ and $2.2$ with $4.2$. Hence we have:
\begin{equation}\label{canc}
\rb_{23}\t (P^0_{214} + P^0_{314}) - \rb_{14}\t (P^0_{423}+P^0_{123}) = 0 \, .
\end{equation}

\subsubsection{The correction term $C$}
\label{corrC}

{From our computations in \ref{reliii} and \ref{reliv}, we see that the pair $(\rb,P^0)$ satisfies all the relations for a free infinitesimal 2-{Yang-Baxter} operator in $\st=(\dd\colon \F_0 \to \F_1 \rtimes_\alpha \sl,\t)$, except for the remaining terms \eqref{12t} in relation {\eqref{4i}}. By considering $P=P^0+ \lambda C$, for  suitable $C\in\mathrm{ker}\big( \dd'\colon \overline{\lU}^{(3)} \to (\fsl)^{3 \otimes}\big)$ and $\lambda \in \C$, Proposition \ref{invi}, it is possible to cancel out these terms. }

{We start with some general observations. Given $X \in \sl$, we denote its lift by the obvious section $\sigma\colon\sl\to\fsl$ of $\pi\colon \fsl \to \sl$ \eqref{mainseq}  as $\overline{X}=\sigma(X)=(0,X)\in \fsl$. By \eqref{fslbr} we have:}
$$[\overline{X},\overline{Y}]=[(0,X),(0,Y)]=\big(\alpha(X,Y),[X,Y]\big), \textrm{ for each } X,Y \in \sl, $$
thence $\sigma$ is not a Lie algebra morphism. It is however very close to being so. Indeed, let $1_{\Fo}\in \Fo$ be the constant function of $x \in \mathbb{R}$, with value ``1'', thus $\mu \t 1_{\Fo}=0$, for each $\mu \in \fsl$, Remark \ref{crux}. Recalling the notation of \ref{Sinsmap}, in  $\lU^{(2)}$, we have, for each $X,Y \in \sl$: 
$$[\overline{X},  \overline{Y}] \tn 1_{\F_0}=\overline{[X,Y]} \tn 1_{\Fo} + \big(\alpha(X,Y),0\big) \tn {1_{\Fo}} .$$
Now by \eqref{defalfa}, certainly $\alpha(X,Y)=df$, for some $f \in \Fo$. Thus $\big(\alpha(X,Y),0\big)=\dd(f)$ for some $f \in \Fo$. Looking at the defining relations \eqref{irel} for $\overline{\lU}^{(n)}$, we have:
$$\big(\alpha(X,Y),0\big) \tn 1_{\Fo} =\dd(f) \tn 1_{\Fo}=f \tn \dd(1_{\Fo})=f \tn (0,0)=0, $$
since, for each $g \in \Fo$, $\dd(g)=(dg,0)$, where $d$ denotes exterior derivative. Therefore, for each $X,Y \in \sl$:
$$[\overline{X}, \overline{Y}] \tn 1_{\F_0}=\overline{[X,Y]} \tn 1_{\Fo}, \textrm{ in } \overline{\lU}^{(2)}. $$

By a completely analogous argument, one can prove the following lemma:
\begin{Lemma}\label{keyl} 
Fix integers $n \ge 2$ and $a \in \{1, \dots ,n\}$. Consider  arbitrary elements $Y_i \in \sl$, for $i \in \{1,\dots,n \} \setminus \{a\}$. For each $j \in  \{1,\dots,n \} \setminus \{a\}$, and each $X \in \sl$, we have the following equality in $\overline{\lU}^{(n)}$:
$$\overline Y_1 \tn \dots\tn [\overline X, \overline Y_j ] \tn \dots \tn \overline Y_{a-1} \tn 1_{\Fo} \tn \overline Y_{a+1} \tn \dots \tn \overline Y_n=\overline Y_1 \tn \dots\tn \overline{[ X, Y_j ]} \tn \dots \tn \overline Y_{a-1} \tn 1_{\Fo} \tn \overline Y_{a+1} \tn \dots \tn \overline Y_n \, .
$$
\end{Lemma}

\noindent
This result will be used time after time in the following discussion.

Let $r=\sum_i s_i\otimes t_i\in \sl \tn \sl$ be the infinitesimal Yang-Baxter operator in $\sl$, derived from the Cartan-Killing form; see \eqref{r}. It is a symmetric tensor,  and it further satisfies the 4-term relation: $[r_{12}+r_{13},r_{23}]=0$, explicitly:
$$\sum_{i,j} s_i \tn [t_i,s_j] \tn t_j + \sum_{i,j} s_{i} \tn s_j \tn [t_i,t_j]=0, \textrm{ in } \sl \tn \sl \tn \sl.$$
Given that $r$ is symmetric, several variants of the equation $[r_{12}+r_{13},r_{23}]$ hold. For example:
$$[r_{31}+r_{32},r_{12}]=0 \textrm{ and } [r_{21}+r_{23},r_{13}]=0. $$

Since $X \in \sl \mapsto  \overline{X} \in \fsl$ is not a Lie algebra map, the tensor: $$\overline{r} = \sum_i \overline{s_i} \tn \overline{t_i} \in (\fsl) \tn (\fsl),$$ made explicit in \eqref{rbar}, does not satisfy the 4-term relation. However, by the previous lemma:
\begin{Lemma}
Define
\begin{equation}
\label{defT}
T=T_{123}=\sum_i \overline{s_i} \tn \overline{t_i} \otimes 1_{\Fo} \in \overline{\lU}^{(3)} \, .
\end{equation}
Then $T_{123}=T_{213}$,  and moreover:  
$$ \bar{r}_{12} \t T_{234}+\bar{r}_{13}\t T_{234}=0 \,,  \textrm{ in } \overline{\lU}^{(4)} \, .$$
Explicitly:
$$\sum_{i,j} \overline{s_i} \tn [\, \overline{t_i},\overline{s_j} \,] \tn \overline{t_j} \otimes 1_{F_0} + \sum_{i,j} \overline{s_{i}} \tn\overline{ s_j} \tn [ \, \overline{t_i},\overline{t_j} \,]\otimes 1_{F_0} =0 \,,  \textrm{ in } \overline{\lU}^{(4)} .$$

\end{Lemma}
\proof{It follows from Lemma \ref{keyl} and the fact that $r=\sum_i s_i \tn t_i$ satisfies the 4-term relation. \qedhere} \\

Let us now define:
$$V=V_{123}=\sum_{i,j} s_i\tn [t_i,s_j] \tn t_j=[r_{12},r_{23}] \in \sl \tn \sl \tn \sl \, .$$
Explicitly:
\begin{equation*}
 V = 8\,\big(- \em\tn\eo\tn\ep + \eo\tn\em\tn\ep - \ep\tn\em\tn \eo + \ep\tn\eo\tn\em +\em\tn\ep\tn\eo - \eo\tn\ep\tn\em\big)  \, .
\end{equation*}
Also put:
$$\overline{V}=\overline{V}_{123}=\sum_{i,j} \overline{s_i}\tn \overline{[t_i,s_j]} \tn \overline{t_j}=\overline{[r_{12},r_{23}]} \in (\fsl) \tn (\fsl) \tn (\fsl) \, .$$
Explicitly:
\begin{equation*}
\begin{split}
{\frac{1}{8}} \overline{V} = & - \eem\tn\eeo\tn\eep + \eeo\tn\eem\tn\eep - \eep\tn\eem\tn\eeo + \\ 
& + \eep\tn\eeo\tn\eem +\eem\tn\eep\tn\eeo - \eeo\tn\eep\tn\eem \, .
\end{split}
\end{equation*}
In general, given a positive integer $n$, noting that $r$ is symmetric, for distinct $a,b,c \in \{1,\dots,n\}$, we put:
$$V_{abc} =[r_{ab},r_{bc}]\in \sl \tn \sl \tn \sl \, .$$
By the 4-term relation and the symmetry of $r$, we have cyclic invariance: $$V_{abc}=V_{bca}=V_{cab} \,,$$ and, 
by anti-symmetry of the Lie bracket, we also have: 
$$V_{abc}= - V_{cba}=-V_{bac}=-V_{acb} \,.$$ 

Recalling the right-hand-side of equation \eqref{12t}, consider now, noting Lemma \ref{keyl}:
$$W=W_{1234}=\overline{V} \otimes {1_{\F0} }=\sum_{i,j} \overline{s_i}\tn \overline{[t_i,s_j]}\tn \overline{t_j} \otimes 1_{\Fo}=\sum_{i,j} \overline{s_i}\tn [\overline{t_i},\overline{s_j}] \tn \overline{t_j} \otimes 1_{\Fo}\in \overline{\lU}^{(4)} \, .$$
Explicitly:
\begin{equation*}
\begin{split}
{\frac{1}{8}} W = & - \eem\tn\eeo\tn\eep \tn 1_{\Fo} + \eeo\tn\eem\tn\eep\tn 1_{\Fo} - \eep\tn\eem\tn\eeo\tn 1_{\Fo} + \\ 
& + \eep\tn\eeo\tn\eem\tn 1_{\Fo} +\eem\tn\eep\tn\eeo\tn 1_{\Fo} - \eeo\tn\eep\tn\eem\tn 1_{\Fo} \, .
\end{split}
\end{equation*}
From the symmetries of $V$, hence of $\overline{V}$, we have cyclic invariance in the first three indices, namely: 
\begin{equation}\label{s1}
W_{abcd}=W_{bcad}=W_{cabd},
\end{equation}
 and also  the anti-symmetry: 
\begin{equation}\label{s2}
W_{abcd}=-W_{bacd}=-W_{cbad}=-W_{acbd}.
\end{equation}

Equation \eqref{12t} can be written as:
\begin{equation}
\label{new12t}
\rb_{14}\t (P^0_{213} + P^0_{234}) +  (\rb_{12} + \rb_{23} + \rb_{24})\t P^0_{314} - (\rb_{13}+\rb_{34})\t P^0_{214} = -4\, (\overline{V}\tn\ufo + \ufo\tn\overline{V}) = -4 \, (W_{1234} + W_{2341}).
\end{equation}
Next lemma motivates linear combinations of tensors $T_{ijk}$ as natural candidates for the correction term $C$. 
\begin{Lemma}\label{tech1} Let $n$ be a positive integer.
For $a,b,c,d$, distinct indices in $\{1,\dots,n\}$, we have, in  $\overline{\lU}^{(n)}$:
$$\overline{r}_{ab} \t T_{bcd}=W_{abcd} \, , \qquad \overline{r}_{ad} \t T_{bcd}=0 \, .$$
\end{Lemma}
\proof{The first equation follows from Lemma \ref{keyl}, the second from the fact that the action of $\fsl$ on the constant functions in $\Fo$ is trivial, Remark \ref{crux}. \qedhere}

\begin{Remark}
There are several variants of these equations obtained by  using $T_{abc}=T_{bac}$ and $\rb_{ab}=\rb_{ba}$; for example: $$\bar r_{ba} \t T_{cbd}=\bar r_{ab} \t T_{bcd}=W_{abcd} .$$
These will be used without further explanation. 
\end{Remark}

\noindent
We can now present the correction term. Define $C\in \overline{\lU}^{(3)}$ as (where $2_{\Fo}=2\times 1_{\Fo}$, a constant polynomial in $x$):
\begin{equation}\label{defC}
C=C_{123}=2 T_{231}-T_{123}-T_{312}=\sum_i 2_{F_0}  \otimes \overline{s_i} \tn \overline{t_i}  -\sum_i  \overline{s_i} \tn \overline{t_i} \tn 1_{F_0}    -\sum_i  \overline{s_i} \tn 1_{F_0}  \otimes  \overline{t_i} \, . 
\end{equation}
We next compute the left-hand-side of relations {\eqref{2i}}-{\eqref{5i}} on $C$. From the symmetry of $r$ we easily get that $C_{123}+C_{231}+C_{312}=0$ and $C_{123}=C_{132}$. Since $\dd(f)=(\d f, 0),$ for each $f \in \Fo$, we have:
$${\dd'}(C)=\sum_i \dd(2_{F_0})  \otimes \overline{s_i} \tn \overline{t_i}  -\sum_i  \overline{s_i} \tn \overline{t_i} \tn \dd(1_{F_0}   ) -\sum_i  \overline{s_i} \tn\dd( 1_{F_0} ) \otimes  \overline{t_i} =0. $$

\begin{Lemma}
 We have  (compare with relation {\eqref{4i}}):
 $$\rb_{14}\t (C_{213} + C_{234}) +(\rb_{12} + \rb_{23} + \rb_{24})\t C_{314} - (\rb_{13}+\rb_{34})\t C_{214} =  W_{1234} + W_{2341} = \overline{V} \tn {1_{\Fo}} +1_{\Fo} \tn \overline{V}.$$
\end{Lemma}
\proof{ By using the explicit formula  \eqref{defC} for $C$, Lemma \ref{tech1} and \eqref{s1} and \eqref{s2}, the left-hand-side is:
\begin{equation*}
\begin{split} 
\rb_{14} & \t (2T_{132}- T_{213}-T_{321}  + 2T_{342} - T_{234}-T_{423}) +(\rb_{12} + \rb_{23} + \rb_{24} )\t ( 2T_{143}- T_{314}-T_{431}  ) + \\ 
& 
\qquad\qquad\qquad\qquad\qquad\qquad
\qquad\qquad\qquad\qquad\qquad\qquad\qquad 
- (\rb_{13}+\overline{r}_{34})\t ({2}T_{142}- T_{214}-T_{421} ) = \\
 & =  2W_{4132}-W_{4123}+2W_{1432}-W_{1423}+2W_{2143}-W_{2134}
- W_{2314}-W_{2341}+2W_{2413}-W_{2431} + \\
& \qquad\qquad\qquad\qquad\qquad\qquad
\qquad\qquad\qquad\qquad\qquad\qquad\quad 
- 2W_{3142}+W_{3124}-2W_{3412}+W_{3421} =\\
& = W_{3124}+W_{3421}=W_{1234}+W_{2341}=\overline{V} \tn {1_{\Fo}} +1_{\Fo} \tn \overline{V}.
\end{split}
\end{equation*}\qedhere}

\begin{Lemma}
We have  (compare with relation {\eqref{5i}}):
 $$\rb_{23}\t (C_{214} + C_{314}) - \rb_{14}\t (C_{423}+C_{123}) = 0.$$
\end{Lemma}
\proof{ As before, we use the explicit formula  \eqref{defC} for $C$, and Lemma \ref{tech1}, to compute:
\begin{align*}
\rb_{23} & \t (C_{214} + C_{314}) - \rb_{14}\t (C_{423}+C_{123}) = \\
 &=
  \rb_{23}\t (2T_{142}-T_{214 }-T_{421 } + 2T_{143}-T_{314}-T_{431 }) - \rb_{14}\t (2T_{234}-T_{423 }-T_{342 }+2T_{231}-T_{123 }-T_{312}) = \\
 &=-W_{3214}-W_{3241} - W_{2314}-W_{2341}+ W_{1423} + W_{1432}+W_{4123} +W_{4132}=0 \, .
 \end{align*}
\qedhere} \\

The two previous lemmas, and  remaining calculations on $C$, together with \eqref{12t} and \eqref{canc}, and the remaining calculations on $P^0$, lead us to the following theorem, which is the main result of this paper:
\begin{Theorem}
\label{main2r}
Consider the differential crossed module $\st=(\partial \colon \Fo \to \fsl ,\t)$, associated to the string Lie 2-algebra, embedded inside  the exact sequence:
$$
0 \lra \mathbb{C} \stackrel{i}{\lra} \Fo \stackrel{\dd}{\lra} \fsl \stackrel{\pi}{\lra} \sl \lra 0.
$$  We denote by $\{\em,\eo,\ep\}$ the standard basis of $\sl$ satisfying \eqref{com}, and by $r=\sum_i s_i\tn t_i \in \sl\tn\sl$ the infinitesimal {Yang-Baxter operator} in $\sl$, derived from the Cartan-Killing form; explicitly:
$$  r = 2 \, \em\tn\ep + 2 \, \ep\tn\em - 4 \, \eo\tn\eo \, =\sum_{i} s_i \tn t_i.
$$
Also, let  $x$ be the formal variable in $\Fo$. Consider the following elements $\overline{r}\in (\fsl)\tn(\fsl)$ and $P^0,C \in  \mathfrak{U}^{(3)}$:
\begin{align*}
\overline{r} & = 2\, \zp{\ep} \tn \zp{\em} + 2 \, \zp{\em} \tn \zp{\ep} - 4 \, \zp{\eo} \tn \zp{\eo} =\sum_i \overline{s_i} \tn \overline{t_i}\,,\\
P^0  & = 16 \, \big(\zp{\em}\tn\zp{\eo}\tn x + \zp{\em}\tn x\tn\zp{\eo} - \zp{\eo}\tn x\tn \zp{\em} - \zp{\eo}\tn\zp{\em}\tn x \big)\,, \\
C & = \sum_i 2_{F_0}  \otimes \overline{s_i} \tn \overline{t_i}  -\sum_i  \overline{s_i} \tn \overline{t_i} \tn 1_{F_0}    -\sum_i  \overline{s_i} \tn 1_{F_0}  \otimes  \overline{t_i} \, .
\end{align*}
Then the  pair: $$(\rb,P)=\big(\rb, P^0+4 C\big)$$ is a free infinitesimal 2-Yang-Baxter operator in the differential crossed module $\st$. Moreover $(\pi \tn \pi) (\overline{r})=r$.
\end{Theorem}

{By combining this theorem with Theorem \ref{qws}, and mainly \eqref{def2Knizhnik-Zamolodchikov}, we can see that, given a categorical representation of $\st$ in a chain-complex of vector spaces, then we have a flat 2-connection in the configuration space $\C(n)/S_n$, categorifying the $\mathfrak{sl}(2,\C)$-Knizhnik-Zamolodchikov connection.  As  explicit examples of  categorical representation of $\st$, we can consider the adjoint representation, and its tensor powers, considering the tensor product of categorical representations defined in \cite{CFM11}.}
\section{Acknowledgements}
We would like to thank Friedrich Wagemann for correspondence on the string Lie 2-algebra. 
L.S. Cirio acknowledges support by the National Research Fund, Luxembourg, AFR project 1164566.
J. Faria Martins was  partially supported by CMA/FCT/UNL, under the project PEst-OE/MAT/UI0297/2011. This work was partially supported by FCT (Portugal) through the projects
PTDC/MAT/098770/2008 
and PTDC/MAT/101503/2008. 


\begin{thebibliography}{qq}

\bibitem {AC} Abad C.A.; Crainic M.: Representations up to homotopy of Lie algebroids. Journal f\"{u}r die reine und angewandte Mathematik Volume 2012, Issue 663, Pages 91-126.

\bibitem {AF} Altschuler D.; Freidel L.: On universal Vassiliev invariants, Comm. Math. Phys. Volume 170, Number 1 (1995), 41-62.

\bibitem {Ar}  Arnol'd V.I.: The cohomology ring of the colored braid group. Math. Notes 5, 138-140 (1969); translation from Mat. Zametki 5, 227-231 (1969).

\bibitem {Art}  Artin E.. Theory of braids, Annals of Math. (2) 48 (1947), 101 - 126.


\bibitem {ACJ} Aschieri P.; Cantini L.; Jur\v{c}o B.: Nonabelian bundle gerbes, their differential geometry and gauge theory.  Comm. Math. Phys.  254  (2005),  no. 2, 367--400.

\bibitem {B1}  Baez J.C.: Higher Yang-Mills Theory, arXiv:hep-th/0206130. 

\bibitem {BC} Baez J.C.; Crans A.S.: Higher-dimensional algebra. VI. Lie 2-algebras. 
Theory Appl. Categ. 12 (2004), 492--538 (electronic).

\bibitem  {BSCS}  Baez J.C.; Stevenson D.;  Crans A.S.; Schreiber U.: From loop groups to 2-groups. Homology, Homotopy and Applications, Vol. 9 (2007), No. 2, pp.101-135. 

\bibitem{BaH} Baez J.C.; Huerta J.: An Invitation to Higher Gauge Theory, Gen. Relativity Gravitation 43 (2011), no. 9, 2335-2392. 


\bibitem {BL} Baez J.C.; Lauda A.D.: Higher-dimensional algebra. V. 2-groups.  Theory Appl. Categ.  12  (2004), 423--491 (electronic)

\bibitem {BNeu} {Baez J.C.; Neuchl M.:
Higher dimensional algebra. I: Braided monoidal 2-categories. Adv. Math. 121, No. 2, 196-244 (1996). }
\bibitem{BS1}  Baez J.C.; Schreiber U.: Higher Gauge Theory: 2-Connections on 2-Bundles. arXiv:hep-th/0412325. 

\bibitem {BS2}  Baez J.C.; Schreiber U.: Higher Gauge Theory.  Categories in algebra, geometry and mathematical physics, 7-30, Contemp. Math., 431, Amer. Math. Soc., Providence, RI, 2007.  

\bibitem{BWC} {Baez J.C.; Wise D.K.; Crans A.S.: Exotic statistics for strings in 4d
BF theory. Adv. Theor. Math. Phys. 11 (2007) 707-749.}



\bibitem  {BN} Bar-Natan D.: On the Vassiliev knot invariants. Topology 34, No. 2, 423-472 (1995).

\bibitem {BM} Barrett J.W.; Mackaay M.:  Categorical representations of categorical groups. Theory Appl. Categ. 16 (2006), No. 20, 529--557 (electronic). 

\bibitem  {Bi} Birman J. S.: Braids, Links, and the Mapping Class Groups. Ann. Math. Studies, No. 82. Princeton, NJ: Princeton University Press, 1976. 

\bibitem{BiBr}  Birman J.S.;  Brendle T.E.:  Braids: A Survey. Handbook of knot theory, 19-103, Elsevier B. V., Amsterdam, 2005.	

\bibitem {BrMe} Breen L.; Messing W.: Differential geometry of gerbes.  Adv. Math.  198  (2005),  no. 2, 732--846.

\bibitem  {Br} Brown K.: Cohomology of groups. Graduate Texts in Mathematics, 87. Springer-Verlag, New York-Berlin, 1982.

\bibitem  {BHS} Brown R., Higgins P.J., Sivera R.: {Nonabelian algebraic topology.} Filtered spaces, crossed complexes, cubical homotopy groupoids. Tracts in Mathematics 15. European Mathematical Society 2010. 

\bibitem{CKS} Carter S.; Kamada S.;  Saito M.: Surfaces in 4-spaces. Encyclopaedia of Mathematical Sciences 142. Low-Dimensional Topology 3. Springer-Verlag, Berlin  (2004).
      
\bibitem {CV} Chmutov S. V.; Varchenko A. N.: Remarks on the Vassiliev knot invariants coming from sl2. Topology 36 (1997), no. 1, 153–178.


\bibitem{CFM11} Cirio L.S.; Faria Martins J.: Categorifying the Knizhnik-Zamolodchikov connection. Differential Geometry and its Applications, Volume 30, Issue 3, June 2012, 238-261.

\bibitem {D} Dold A.: Lectures on algebraic topology. Reprint of the 1972 edition. Classics in Mathematics. Springer-Verlag, Berlin, 1995. 

\bibitem {Dr} Drinfel'd V.G.: On quasitriangular quasi-Hopf algebras and a group closely connected with Gal($\overline{{\mathbb{Q}}}/{\mathbb{Q}})$. Leningr. Math. J. 2, No.4, 829-860 (1991); translation from Algebra Anal. 2, No.4, 149-181 (1990). 

\bibitem {EML} Eilenberg S.; MacLane S.: Determination of the second homology and cohomology groups of a space by means of homotopy invariants. Proc. Nat. Acad. Sci. U. S. A. 32, (1946). 277-280.

\bibitem{E}  Elgueta J.: Representation theory of 2-groups on Kapranov and Voevodsky's 2-vector spaces. Advances in Mathematics
Volume 213, Issue 1, 1 August 2007, Pages 53-92 

\bibitem {FMM} Faria Martins J.; Mikovi\'{c} A.: Lie crossed modules and gauge-invariant actions for 2-BF theories. Advances in Theoretical and Mathematical Physics, Volume 15, Number 4 (August 2011), 1059-1084. 

\bibitem {FMP1} Faria Martins J.; Picken R.:  On two-dimensional holonomy. Transactions of the American Mathematical Society, 362 (2010), 5657-5695. 	

\bibitem{FMP2} Faria Martins J.; Picken R.: Surface holonomy for non-abelian 2-bundles via double groupoids. Advances in Mathematics Volume 226, Issue 4, 1 March 2011, Pages 3309-3366


\bibitem {FMP3} Faria Martins J.; Picken R.: The fundamental Gray 3-groupoid of a smooth manifold and local 3-dimensional holonomy based on a 2-crossed module,  Differential Geometry and its Applications,
Volume 29, Issue 2, March 2011, 179-206. 


\bibitem{ger66} Gerstenhaber M.: On the deformation of ring and algebras. II. Ann. of Math.  84 (1966), 1--19. 

\bibitem {GP} Girelli F.; Pfeiffer H.: Higher gauge theory---differential versus integral formulation.  J. Math. Phys.  45  (2004),  no. 10, 3949--3971.

\bibitem {Gu}  {Gurski N.: Loop spaces, and coherence for monoidal and braided monoidal bicategories. Adv. Math. 226 (2011), no. 5, 4225-4265.} 

\bibitem {KP} Kamps K. H.; Porter T.: 2-groupoid enrichments in homotopy theory and algebra.  $K$-Theory  25  (2002),  no. 4, 373--409.

\bibitem {Ka} Kassel C.: Quantum groups. Graduate Texts in Mathematics, 155. Springer-Verlag New York, 1995.


\bibitem{KT}  Khovanov M.; Thomas R. {Braid cobordisms, triangulated categories, and
flag varieties. Homology, Homotopy and Applications, Vol. 9 (2007), No. 2, pp.19-94.}

\bibitem {KZ} Knizhnik V.G.; Zamolodchikov A.B.: Current Algebra and 
 and  Wess-Zumino Model in  Two-Dimensions, Nucl. Phys. B 247: 83-103.

\bibitem {Ko} {Kontsevich M.: Vassiliev's knot invariants. I. M. Gel'fand Seminar, 137--150, Adv. Soviet Math., 16, Part 2, Amer. Math. Soc., Providence, RI, 1993.}

\bibitem  {Kh1} Kohno T.: Monodromy representations of braid groups and Yang-Baxter equations. Annales de l'institut Fourier, 37 no. 4 (1987), p. 139-160.

\bibitem {Kh2} Kohno T.: Conformal field theory and topology. Translations of Mathematical Monographs. Iwanami Series in Modern Mathematics. 210. Providence, RI: American Mathematical Society (AMS), 2002.

\bibitem {LM} Le T.Q.T.; Murakami J.: Representation of the category of tangles by Kontsevich's iterated integral. Comm. Math. Phys.
168, (1995), no. 3, 535-562.

\bibitem{MP} Mackaay M.; Picken  R.F.: Holonomy and parallel transport 
for abelian gerbes.  Adv. Math.  170  ({2}00{2}),  no. {2}, {2}87-219.
 
\bibitem {ML} Mac Lane S. Cohomology theory in abstract groups. III. Operator Homomorphisms of Kernels. Annals of Mathematics 50 (3) (1949), 736--761.


\bibitem{Sch} Scheunert M.: Graded tensor calculus, J. Math. Phys. 24 (1983), no. 11, 2658-2670.

\bibitem {SW1} Schreiber U.; Waldorf K.: Smooth Functors vs. Differential Forms. Homology Homotopy Appl. 13 (2011), no. 1, 143-203. 

\bibitem{SW2} Schreiber U.; Waldorf K.: Connections on non-abelian gerbes and their holonomy. arXiv:0808.1923v1 [math.DG]. 

\bibitem{wag06} Wagemann F.: On Lie algebra crossed modules. Comm. Algebra 34 (2006), no. 5, 1699--1722.

\bibitem {W} {Witten E.: Quantum field theory and the Jones polynomial.
Comm. Math. Phys. 121 (1989), no. 3, 351--399.}

\bibitem  {Wo}  Wockel C.:  Principal 2-bundles and their gauge 2-groups. Forum Mathematicum. Volume 23, Issue 3, Pages 565-610

\end{thebibliography}
\end{document}